\documentclass[twocolumn]{aastex631}

\begin{document}

\title{The ATLAS Virtual Research Assistant}

\correspondingauthor{H.~F. Stevance}
\email{hfstevance@gmail.com}

\author[0000-0002-0504-4323]{H.~F.~Stevance}
\affil{Astrophysics sub-Department, Department of Physics, University of Oxford, Keble Road, Oxford, OX1 3RH, UK}
\affil{Astrophysics Research Centre, School of Mathematics and Physics, Queen's University Belfast, BT7 1NN, UK}
\affil{University of Sheffield, Astrophysics Research Cluster, Hicks Building, Broomhall, Sheffield S3 7RH}

\author[0000-0001-9535-3199]{K. W. Smith} 
\author[0000-0002-8229-1731]{S.~J.~Smartt} 
\affil{Astrophysics sub-Department, Department of Physics, University of Oxford, Keble Road, Oxford, OX1 3RH, UK}
\affil{Astrophysics Research Centre, School of Mathematics and Physics, Queen's University Belfast, BT7 1NN, UK}
\author{S.~J.~Roberts} 
\affil{Department of Engineering Science, University of Oxford}

\author[0000-0002-9986-3898]{N.~Erasmus}
\affil{South African Astronomical Observatory, Cape Town, 7925, South Africa}
\affil{Department of Physics, Stellenbosch University, Stellenbosch, 7602, South Africa}

\author[0000-0002-1229-2499]{D.~R.~Young}
\affil{South African Astronomical Observatory, Cape Town, 7925, South Africa}
\affil{Astrophysics Research Centre, School of Mathematics and Physics, Queen's University Belfast, BT7 1NN, UK}

\author[0000-0003-3068-4258]{A. Clocchiatti}
\affiliation{Instituto de Astrof\'{\i}sica, Pontificia Universidad Cat\'olica de Chile, Vicu\~na Mackenna 4860, Macul, Santiago, Chile}
\affiliation{Millennium Institute of Astrophysics MAS, Nuncio Monse\~nor S\'otero Sanz 100, Of. 104, Providencia, Santiago, Chile}



\begin{abstract}
We present the Virtual Research Assistant (VRA) of the ATLAS sky survey which performs preliminary eyeballing on our clean transient data stream. The VRA uses Histogram Based Gradient Boosted Decision Tree Classifiers trained on real data to score incoming alerts on two axes: ``Real" and ``Galactic". The alerts are then ranked using a geometric distance such that the most ``Real" and ``Extra-galactic" receive high scores; the scores are updated when new lightcurve data is obtained on subsequent visits. To assess the quality of the training we use the Recall at rank K, which is more informative to our science goal than general metrics (e.g. accuracy, F1-Scores). We also establish benchmarks for our metric based on the pre-VRA eyeballing strategy, to ensure our models provide notable improvements before being added to the ATLAS pipeline. 
Then, policies are defined on the ranked list to select the most promising alerts for humans to eyeball and to automatically remove the bogus alerts. In production the VRA method has resulted in a reduction in eyeballing workload by 85\% with a loss of follow-up opportunity $<$0.08\%. It also allows us to automatically trigger follow-up observations with the Lesedi telescope, paving the way to automated methods that will be required in the era of LSST. Finally, this is a demonstration that feature-based methods remain extremely relevant in our field, being trainable on only a few thousand samples and highly interpretable; they also offer a direct way to inject expertise into models through feature engineering. 
\end{abstract}

\keywords{Sky surveys(1464), Transient detection (1957), Astrostatistics (1882), Interdisciplinary astronomy (804), Astroinformatics (78)}


%
%

\section{Introduction}
\label{sec:intro}

The first two decades of the 21st century have seen a revolution in astronomers' ability to survey the sky on a large scale and in the time domain, with facilities such as Pan-STARRS
(Panoramic Survey Telescope and Rapid Response System; \citealt{keiser2002,Chambers2016}), PTF (Palomar Transient Factory; \citealt{law2009}), ASAS-SN (All-Sky Automated
Survey for Supernovae; \citealt{shappee2014}), ZTF (Zwicky
Transient Facility; \citealt{bellm2019}), BlackGem \citep{blackgem}, GOTO (Gravitational-wave Optical Transient Observer; \citealt{steeghs2022, dyer2024}) and ATLAS
(Asteroid Terrestrial Last-Alert System; \citealt{tonry2018, smith2020}).
These wide-field sky surveys have allowed astronomers to routinely find new transient events which range from the common -- a few thousand examples -- thermonuclear (type Ia) and core-collapse Supernovae (CCSNe) (e.g. \citealt{perley2020}, Srivstav et al. in prep.), to rarer -- a few hundred -- Tidal Disruption Events (TDEs; \citealt{gezari2021}) and Superluminous supernovae (SLSNe; \citealt{galyam2019}), as well as recently discovered optical counterparts of both gamma-ray bursts 
(GRBs; \citealt{2013ApJ...769..130C}) 
and Fast X-ray transients (FXTs; \citealt{gillanders2024}). Until the advent of ZTF and ATLAS, the counterparts to high energy transients had typically been found by focused follow-up but now that the whole sky can be scanned every 24 to 48 \,hrs, the optical afterglows are frequently found either without a high energy trigger \citep{2025MNRAS.537.2362P}, or through post-hoc association 
\citep[e.g.][]{2017ApJ...850..149S}. 

Once transients are found in sky surveys, follow-up observations can be carried out and the science exploitation phase begins.
These additional observations often require  facilities with larger apertures and/or specialised instruments that are in high demand (e.g. Liverpool Telescope \citealt{steele2004}; the PESSTO and ePESSTO programs on the New Technology Telescope, \citealt{smartt2015}; or X-shooter on the Very Large Telescope; \citealt{vernet2011}). 
After overcoming the technical challenge of rapidly observing large areas of the sky with a rapid cadence (a few days), the field of transient astronomy  transitioned from a target-limited regime to a resource-limited regime, where the number of transients far outweighed the availability of follow-up facilities. 
There began the new challenge of data curation and prioritization:
How can we select the most promising or interesting alerts in a vast stream without overwhelming the science teams with data to manually eyeball?
This usually starts with basic cuts (e.g. based on signal-to-noise), followed by cross-matching to astrophysical catalogues \citep{Young_sherlock_2023}, real-bogus classification using Convolutional Neural Networks (e.g. \citealt{weston2024, killenstein2021}) and finally eyeballing performed by humans to determine which alerts are indeed real and which need further attention. 
Even in surveys where this final step is handled with the help of citizen scientists (e.g. \citealt{killenstein2024}), they only contribute a small amount of reported discoveries.  
In ATLAS, eyeballing requires between 200 and 400 alerts a day; a new step of automation was therefore required. 

At this stage, early photometric transient classification -- which uses as little lightcurve information as possible to infer likely spectroscopic classes -- may seem like an attractive option. 
Some promising examples can be found in the works of \cite{muthukrishna2019} (RAPID),  a Recurrent Neural Network aiming to classify 12 types of explosive transients as early as 2 days since trigger,  and that of \cite{gagliano2023} who present a multi-modal neural network using shallow learning on the image stamps and additional features to classify supernovae as early as 3 days after alert.
Although preliminary classification can help prioritizing transients for follow-up, these algorithms can only perform successfully in a stream that has been cleaned of other contaminants such as galactic transients and left-over bogus detections, as they were trained on clean data sets (PLAsTiCC \citealt{plasticc}, and the ZTF Bright Transient Survey data, respectively). 

A different strategy is therefore required to clean the stream, one more suited to the task of triaging  when little light curve information is known and many contaminants are present. A succesful example of this is {\tt BTSBot}, which flags potential bright extragalactic transients that are candidate for follow-up  \citep{btsbot}.
Its classification is simpler than in the photometric transient classifiers (binary Vs multi-class) but effective and adapted to the task of filtering data in a stream composed of many types of astrophysical events and left-over bogus. 

In this paper we present a different approach to data curation in an impure transient stream.
We call it the Virtual Research Assistant (VRA) because the strategy implemented in our design follows that of our human eyeballers. In Section \ref{sec:design} we summarise the design of the VRA, place it in the context of the ATLAS pipelines, and establish benchmarks we will use to assess the success of our algorithms. In Section \ref{sec:scoring} we present our data sets and the training of our models. In Section \ref{sec:performance} we describe how the combined performance of our models and policies are evaluated before being launched in production, and then report the in-production performance of the VRA. Further discussions can be found Section \ref{sec:disc} and we conclude in Section \ref{sec:concl}.
In addition to this manuscript, we point the reader to the Manual \citep{stevance_2025_14944209} for further details on earlier prototypes of the VRA and for up-to-date information on the current VRA version and eyeballing policies. 
All the data and code used for the training and analysis presented here can be found in the VRAv1 Code and Data release \citep{stevance_2025_14906192}.


%
%
%
\section{Overview}
\label{sec:design}

\subsection{ATLAS transient searches} 
\label{ref:atlas_summary}
The ATLAS sky survey \citep{tonry2018} is composed of four 0.5m telescopes: two in Hawaii, one in Chile, one in South Africa.
Two filters are used for observations: the cyan (c) filter (420-650 nm) used during dark time and the orange (o) filter (560-820 nm) used during bright time. 
In survey mode ATLAS performs four 30 second exposures of tile or sky pointing, each separated by roughly 15 minutes, a strategy motivated by the main science case of ATLAS (the discovery and follow-up of Near-Earth Objects; \citealt{Heinze2021}). 
The individual frames are detrended and calibrated (astrometric and photometric)
on site for each unit and the data are transferred to Hawaii. 
Difference imaging with respect to the ATLAS wallpaper is carried out after which all sources with significance greater than 5$\sigma$ are cataloged. 
These detection catalogs and the reduced and calibrated images are transferred to Transient Servers in Queen's University Belfast 
\citep{smith2020}. The catalog files of the difference image detections contain about   $\mathcal{O}(10^7)$ source, all of which are ingested into a relational database. 
Before the development of the VRA, the  transient alert processing 
\citep[as summarised in][]{smith2020} was as follows:

\begin{enumerate}
\item \textbf{Quality cuts}: three or more good quality, co-spatial, detections at significance of $5\sigma$ or greater within 1 night are required to define an object. 

\item \textbf{Astrophysical cross-matching} with {\sc sherlock} \citep{Young_sherlock_2023}. Known variable stars are removed from the stream and contextual information is added, such as potential host cross-matching, angular distance to potential hosts, redshift (if known) (see Sherlock documentation for further details). 

\item \textbf{Real/Bogus classification} using a Convolutional Neural Network \citep{weston2024}. The $20\times20$ central pixels of each difference image receives a score between 0 (Bogus) and 1 (Real). Below a 0.2 threshold the alerts are not sent for human eyeballing, they are directly labeled as garbage (but not deleted from the database).

\item \textbf{Humans} eyeball the data to classify them into four broad categories: ``Good" (extra galactic transient), ``Attic"/``Galactic", ``Garbage", ``Proper Motion" (alerts due to stars moving between the time of observation and the date at which the wallpapers were constructed). To do this, humans have access to a broad range of (multimodal) data: Stamps of the wallpaper, observation and difference imaging; lightcurve; contextual information (see e.g. Figure \ref{fig:missed_day1}). 
\end{enumerate}

\begin{figure}[ht!]
\centering
\includegraphics[width=8cm]{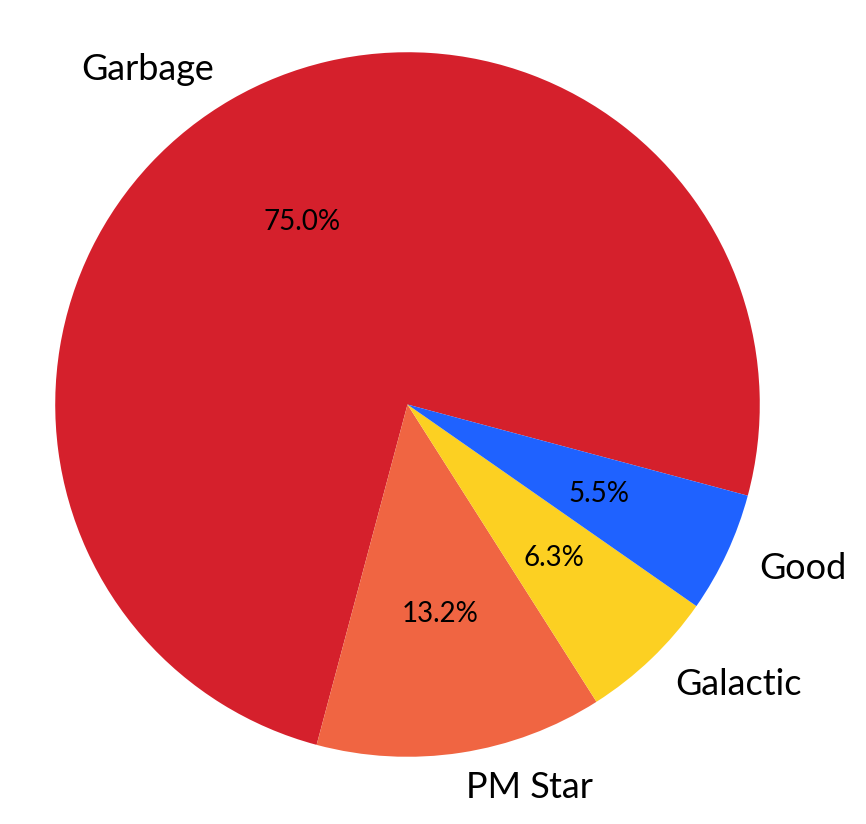}
\caption{Alert type distribution in the ATLAS eyeball list between 27th of March and 13th of August 2024 for a total of 40,802 alerts, all eyeballed by humans and predating the introduction of the first VRA prototype in production \label{fig:pie_chart}. See the label description in Section \ref{sec:labels}.}
\end{figure}

We can gauge the workload of the eyeballers using the first data set gathered for VRA training between 27th of March and 13th of August 2024. Over that period a total of 40,802 objects were presented to humans for scanning, averaging nearly 300 objects per day.  As we can see in Figure \ref{fig:pie_chart}, nearly 90\% of those objects were labeled as Garbage or Proper Motion, the rest being nearly evenly split between the Attic (Galactic) and Good (Extragalactic) categories. 

\begin{figure}[ht!]
\centering
\includegraphics[width=8.5cm]{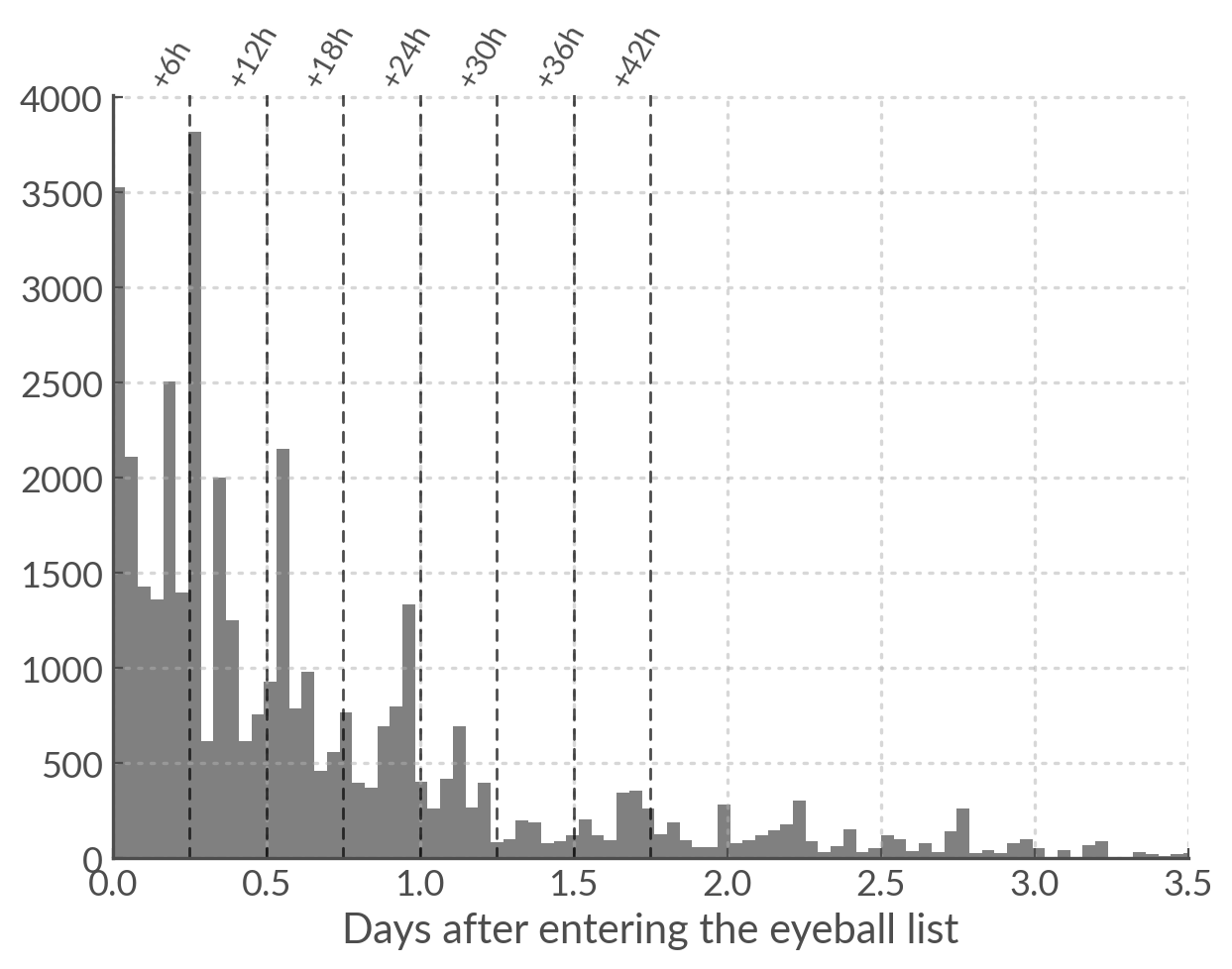}
\caption{Histogram of the time delay between an alert entering the eyeball list and a human making a classification between 27th March 2024 and 13th August 2024.
\label{fig:when_decisions_made}}
\end{figure}

Figure \ref{fig:pie_chart} encapsulates the problem we will address in this paper, the real, extragalactic transient sources are still only a few percent (5.5\% over this period) of the objects that a human on duty will scan through manually. 
Further automation of the eyeballing process is challenging, at least in part because  human eyeballers are able to triage the objects with very little lightcurve information.  As we can see in Figure \ref{fig:when_decisions_made}, a large portion of the objects (90\%) is labeled in less than 48h.
Hence human scanning is quick and effective but the ratio of good to reject objects makes it an inefficient use of scientist time.

\subsection{Scope and benchmark}
\label{sec:benchmark}
The primary goal of this project is to minimise eyeballer workload without compromising extragalactic transient identification.  
Henceforth, we shall define the objects that make it through the basic filtering, as ``alerts" for which we aim to make the process more efficient. 
A ``simple" way (conceptually, albeit not necessarily technically) to do this is to order the eyeball list with alerts that are most likely to be ``Good" at the top. 
If we can provide a ranking that is robust enough to guarantee complete recovery of the ``Good" alerts in the top X\% of the list (X to be determined after training), we can then crop the bottom of the eyeball list. 
In essence, that is the strategy implemented with the Real/Bogus score, where all alerts below 0.2 are automatically labelled as Garbage. 
In practice, however, the RB score ordering is not very effective beyond the initial crop. 

\begin{figure*}[ht!]
\centering
\includegraphics[width=8.2cm]{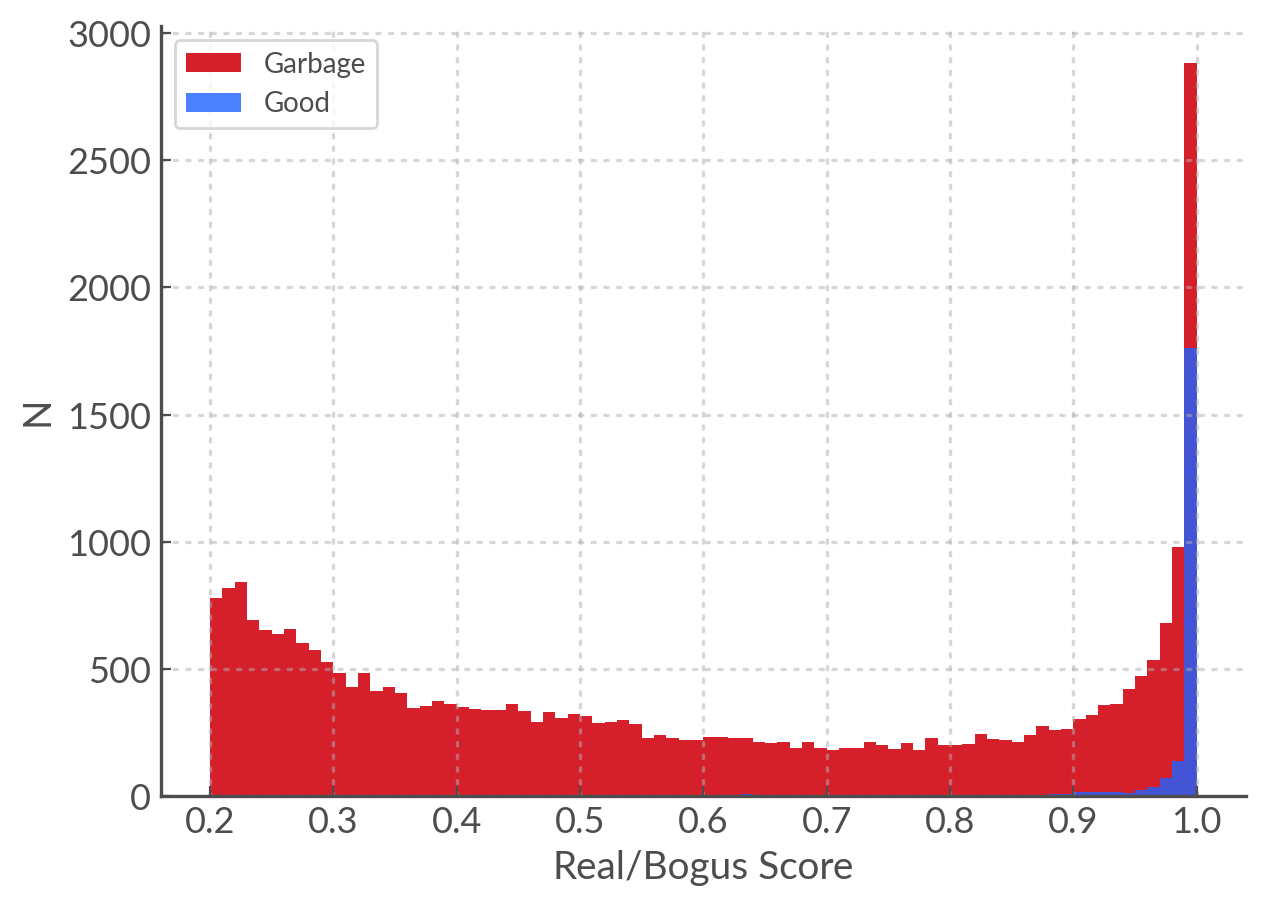}
\includegraphics[width=6cm]{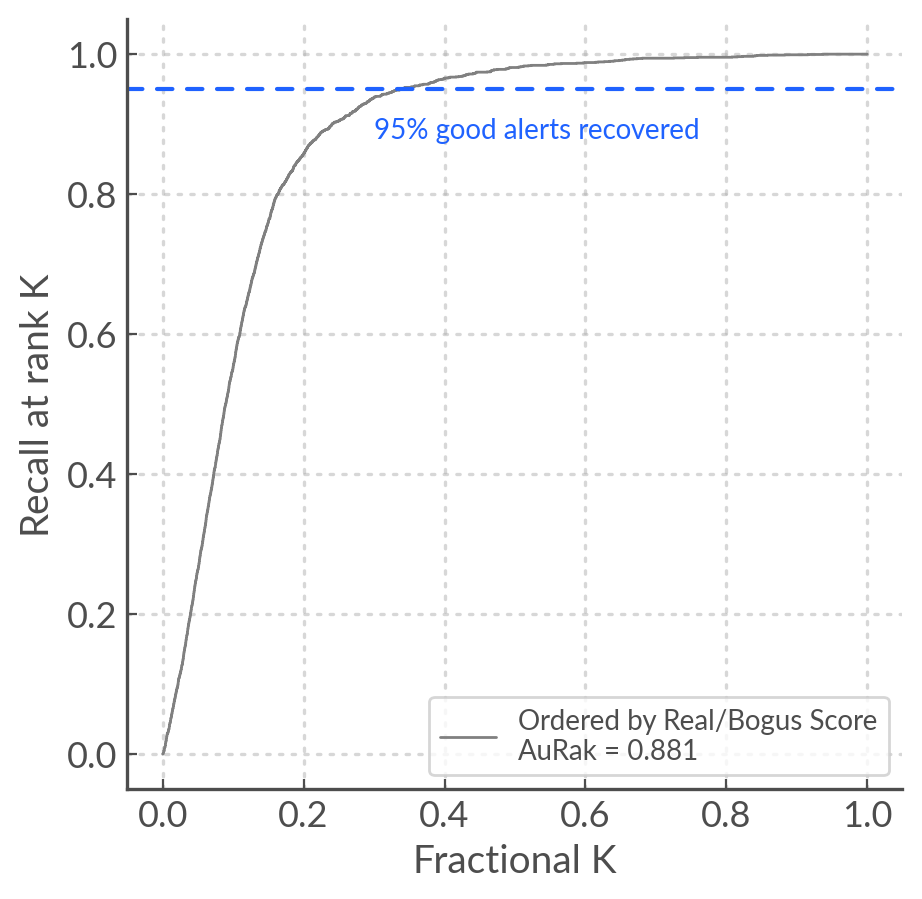}
\caption{\textbf{Left:} Distribution of the Real/Bogus score for the human-labeled "garbage" and "good" alerts over the period 27th March - 16th August 2024. \textbf{Right:} Recall at rank K for the data set ordered by real/bogus score. When ordering by real/bogus score the eyeballers would have to, on average, eyeball the top 35\% of the list to recover 95\% of the good objects (amounting to 5.3\% of the list). To recover 99\% of the good objects, 60\% of the list (ordered by RB score) must be eyeballed.  
\label{fig:rb_score_ordering}}
\end{figure*}

As can be seen in Figure \ref{fig:rb_score_ordering} (left panel), the distribution of the RB scores for the garbage alerts has a secondary peak at an RB score of 1 (the primary peak at 0 is not shown as the plots only show scores for eyeballed alerts with RB score$>0.2$), which leads to confusion in the high RB score regions. 
In the right panel of Figure \ref{fig:rb_score_ordering} we also show the Recall at rank K (R@K) obtained from using the RB score to rank the eyeball list. 
The R@K is defined as:

\begin{equation}
    {\rm R@K} = \frac{\rm N\, relevant \,alerts\,in\,top\,K}{\rm N\, relevant\,alerts}
\end{equation}

Since our scientific focus is on extra-galactic transients (``Good" list) these are the objects considered relevant for R@K calculations. 
Our goal is to create models that result in an R@K curve that is steeper that that in Figure \ref{fig:rb_score_ordering} and ideally reaches 100\% recall closer to the top of the list (currently only beyond a fraction of 0.8, or 80\% down the list).  
Inspired by the AUROC (Area Under the Receiver Operator Characteristic) metric which is classically used when training machine learning (ML) models, we define the Area under the Recall at rank K (AuRaK) as one of our model evaluation metrics. 
For the full data set ordered by RB score, we obtain an AuRaK = 0.88.
Any model we create should exceed this value and show a steeper rise, otherwise they would provide no improvements compared to the current strategy.

The other issues with ordering and selection by RB score is that a simple, binary, RB score does not discriminate between galactic and extra-galactic transients, and it does not capture the new information provided by new lightcurve points obtained at on subsequent observations (whether it be a detection or a non-detection).  We need to create a system that can update the ranking of an alert when new data is gathered. 

\subsection{A Transient Agnostic Score Space}
One of the first steps in the design of the VRA eyeballing system was to perform interviews with the most experienced members of the eyeballing team to ask what questions they ask themselves prior to making decisions and what pieces of data they use to answer them. 
At this stage of eyeballing, there are only three important questions to be answered:
\begin{enumerate}
    \item Is it Real?
    \item Is it Galactic?
    \item Is it Fast? 
\end{enumerate}

The transient classification, even a broad version of it (e.g. is it a type II vs a type Ia) is not a major consideration in the first few days of an alerts because in most cases the data is simply \textit{insufficient} to make a reliable, informed, decision.
Therefore a transient specific classifier is not adapted to the task at hand.

\begin{figure}
    \centering
    \includegraphics[width=1\linewidth]{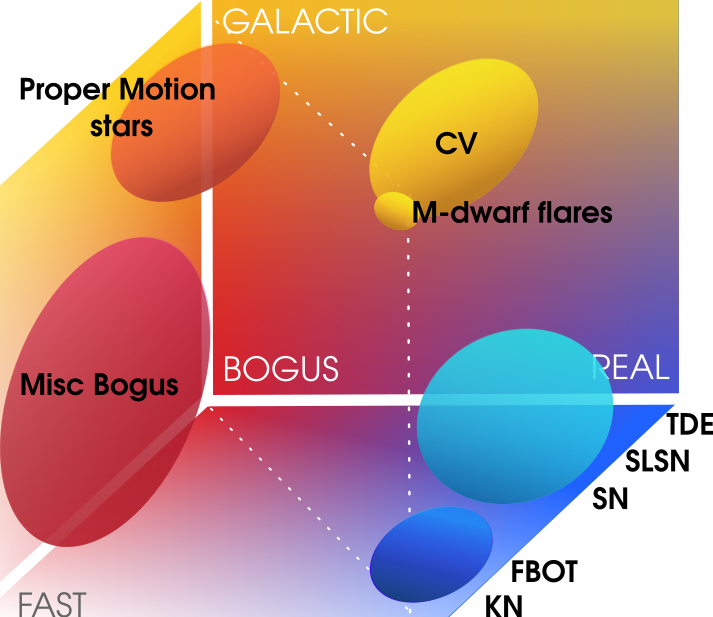}
    \caption{A toy representation of our Transient agnostic score space, defined by three dimensions: Real (x), Galactic (y), Fast (z). The latter relates to the timescale on which the transient lightcurve evolves - it is not used in this iteration of the VRA and is a more subjective quantity (see Discussion \ref{sec:disc}). Nonetheless we can conceive a 3D space where all types of transients and bogus alerts can be found. Cataclysmic Variables (CV) for example are fast (they can rise by several magnitudes within a day) and galactic transients; Kilonovae and Fast Blue Optical Transients (FBOTs) are also rapidly evolving but extra-galactic in nature. Tidal Disruption Events (TDEs), Superluminous supernovae (SLSNe) are extr-galactic too but slower evolving than supernovae. As for the bogus alerts, artifacts from high proper motion stars are galactic in nature whilst other miscellaneous bogus events (e.g. from trailing) may show a variety of behaviour. In the specific case of ATLAS (see Appendix \ref{sec:app_feature_distribs}) most of our miscellaneous bogus alerts are highly correlated with the Galactic plane - this may not be a behaviour that extends to other surveys. }
    \label{fig:score_space}
\end{figure}

Additionally, the three questions highlighted above conveniently  define a Score Space within which all varieties of transients and boguses live (see Figure \ref{fig:score_space}). 
Since it is not specific to one class of transients we call it ``transient agnostic"\footnote{Although we note that our implementation in the ATLAS VRA, because it will be assessed using metrics that favour extra-galactic transients, will be biased towards performing well on extra-galactic transients (see Sections  \ref{sec:scoring} and \ref{sec:policy_eval})}.
Overall the VRA is designed to follow a similar strategy to the human eyeballers: we create scoring algorithms which take data from the stream and provide a Real Score ($p_{\rm real}$) and a Galactic Score ($p_{\rm gal}$).
In the end the idea of assigning a fast score ($p_{\rm fast}$) was deprioritised for this version of the VRA since performance was found to be satisfactory (see Section \ref{sec:policy_eval}), but future iterations of the ATLAS VRA (or other bots in future surveys) may want to use all three axes. 

Once alerts have been placed in Score Space, we must then rank them from most relevant to least relevant in the eyeball list. 
The advantage of a transient agnostic score space is that ordering the alerts by relevance can be adjusted for different science cases without having to modify (retrain) the scoring algorithms. 
Those looking for galactic alerts candidates would want alerts near the ($p_{\rm real}=$1, $p_{\rm gal}$=1, $p_{\rm fast}=$1) coordinate to be ranked highest, whilst supernova astronomers would favour alerts in the regions corresponding to ($p_{\rm real}=$1,  $p_{\rm gal}$=0).

The details of how the scoring algorithms are trained can be found in Section \ref{sec:scoring}.

\subsection{Eyeballing policies}
\label{sec:eyeball_policies}
Using the real and galactic scores we calculate two properties: the VRA score\footnote{Note it is often called the rank in our internal codes and databases} which ranges from 0 to 10 and used to rank alerts from least to most Real/Extragalactic; the Galactic flag, Boolean (True or False) which identifies alerts as being likely to be Galactic.

The VRA score measures the distance to the (1,0) coordinate in score space, normalises it, makes shorter distances yield a high score and multiplies the result by 10 to obtain a score between 0 and 10. 
It is calculated as follows:
\begin{equation}
\label{eq:vra_rank}
   \mathrm{VRA}_{\rm score} =    10 \times \frac{\sqrt{f^2+1} - \sqrt{(1-p_{\rm real})^2 + (f\times p_{\rm gal})^2}}{\sqrt{f^2+1}}
\end{equation}

 where $f$ is a scaler applied to the galactic axis to better separate the ``Garbage" and ``PM" distributions from real events (see left panel Figure \ref{fig:ss_ratk_day1}). 
 The scaler $f$ is a parameter that allows us to tune the separation between distributions in score space. 
 In the current version of the VRA $f=0.5$, which allows greater separability between the bogus classes (``Garbage", ``PM") and the real classes (``Galactic", ``Good") - this is most easily seen in Figure \ref{fig:ss_ratk_day1}. 
 The intuition between this choice is as follows: for our science goals, we are more tolerant of confusion between the galactic and extra galactic transients than confusion between real alerts (of any kind) and bogus alerts. 
 For a discussion on the choice of $f$ see Section \ref{sec:disc}.
 If an alert has a cross-match to the Transient Name Server (TNS; \citealt{tns2021}) we automatically upgrade its rank to 10. 
The eyeballers are tasked with inspecting objects with scores $>7$.

 In addition, we calculate a Galactic flag which measures the distance to the (1,1) coordinate of Score Space (with a scalar $f=0.9$) and returns True if that distance is $<$0.4.
 Objects that do not meet the VRA score threshold of 7 but do get flagged as potentially galactic are moved to a separate ``Galactic Candidate" list to be eyeballed with lower priority.
Finally, for objects with $D<100$Mpc, the VRA scores are used to flag high rank objects for automated follow-up (see Section \ref{sec:concl}) but eyeballers are still tasked with looking at all incoming objects.
At the end of each ingest cycle a slackbot is triggered and presents the eyeballers with three tiers of eyeballing priorities: Fast Track (immediate), Extra-Galactic (within the hour), Galactic (within 24 to 48h).


\subsection{Auto-garbaging policies}
\label{sec:garbage_policies}
Using the VRA score calculated with eq. \ref{eq:vra_rank} we select alerts for auto-garbaging if they meet the following criteria (at time of writing): if on day 1 their rank VRA$_{\rm score}<1$; if on the second visit maximum VRA$_{\rm score}<2$; if on the third visit and beyond the mean VRA$_{\rm score}<3$

The general form of these policies was chosen when the VRA was first added to production in August 2024, and further evaluation of the policies such as described in Section \ref{sec:policy_eval} led to the specific values presented here. 
It is worth re-emphasising that ``Garbage" is a label in the ATLAS  Transient Server Database
and a list in which the alerts are moved. \textit{Data are not deleted from the database}.

Finally, note that all policies are subject to change over time as new versions of the VRA may be trained or policies revisited.
For up-to-date information regarding the VRA policies and version please see the most up-to-date version of the Technical Manual \citep{stevance_2025_14944209} or the VRA website\footnote{\url{https://heloises.github.io/atlasvras/index.html}}.
 
\subsection{Monitoring}
\label{sec:monitoring}
The operations of the VRA are monitored weekly.
Every Friday a report is sent to a slack channel summarising the following information:
 The number of objects which entered the eyeball list (RB score $>0.2$); the number of objects that were eyeballed by humans; the number of objects reported to TNS
 ; a pie-chart showing the distribution of labels for the past week; the number of potential VRA misses.

Potential misses are defined as alerts which would have not met the VRA rank threshold but whose rank was raised to 10 by a cross-match to TNS. 
To further monitor potential misses we also have deployed a bot which cross-matches the ``Garbage" list items of the past week to the TNS. 
TNS items can still land in the ``Garbage" list if their RB score was lower than 0.2 or if they met the VRA auto-garbaging policies before they were reported to TNS. 

Finally, a prugatory sentinel runs every day to flag any alerts that have not been eyeballed or auto-garbaged but are more that 15 days old  because the day $N$ models are only trained with data up to day 15 (see Section \ref{sec:models}) and we do not trust scores predicted out of distribution. 
We have not found this workload to be substantial (only a handful of objects) so we have not found the need to add additional eyeballing or garbaging policies. 
These are then eyeballed by a human to make a final decision. 

These bots send slack alerts and record their reports to csv files which can be inspected at a later date. 
These regular checks have been crucial to development and will allow us to monitor the VRA for a decrease in performance in the future, which could occur as a result of data-drift and could call for a retraining of the scoring algorithms. 

%
%
\section{Real/Galactic Scoring}
\label{sec:scoring}

Placing alerts in our Score Space defined in Figure \ref{fig:score_space} is done by using two binary classifiers: A Real/Bogus and a Galactic/Extra-Galactic classifier.
Additionally, we differentiate between alerts that are newly added to the eyeball list (day 1) and those which are receiving additional light-curve information on subsequent days by creating \textit{day 1} and \textit{day $N$} models. 
This is motivated by the fact that, although a majority of our alerts are classified on day 1 (Figure \ref{fig:when_decisions_made}), this is partially skewed because many alerts are obviously bogus and do not require further data to make a decision. 
When establishing whether an alert is extra-galactic or galactic, waiting for additional light curve data is not uncommon in eyeballing, and the VRA must be able to use this new information. 
The \textit{day $N$} models are trained on data ranging from day 2 to day 15 and include additional features (see Table \ref{tab:features} and Section \ref{sec:data}) to capture informative ligthcurve evolution that will help our Real and Galactic classifiers.
Overall we have four binary classifiers: Real, \textit{day 1}; Galactic \textit{day 1}; Real \textit{day  N}; Galactic \textit{day N}

Our classifiers are trained using data taken from the stream and labeled by our eyeballers. The labels and notable caveats are described in Section \ref{sec:labels};  the data set and features are described in detail in section \ref{sec:data}; then the training of our models is presented in Section \ref{sec:models}.

\subsection{The labels}
\label{sec:labels}

There are four alert classification categories, which we use as follows:
\begin{itemize}
    \item \textbf{``Garbage":} This is a broad category that encompasses most types of bogus alerts such as trails, star spikes, bad psf, detector issues, bad subtractions in crowded fields or associated with bright galaxy cores. This category is used to provide samples with label $p_{\rm real}=0$.
    
    \item \textbf{``PM or Proper Motion":} This is used to separate the bad subtractions that are specifically suspected to have occurred as a result of the drift of a star compared to its position in the ATLAS wallpaper. This category provides samples with labels $p_{\rm real}=0$, $p_{\rm gal}=1$.

    \item \textbf{``Attic/Galactic":} The Attic is a list in the ATLAS transient server (see \citealt{smith2020}) used to store \textit{real} alerts that do not belong in our ``Good" list. This contains mostly galactic events (Cataclysmic variables, stellar flares, stellar variability) and we use this category to create training samples with labels $p_{\rm real}=1$, $p_{\rm gal}=1$.

    \item \textbf{``Good":} This list is dedicated to the extra-galactic transient alerts (SN, TDE, SLSN, FBOTs, but not AGNs) and the samples drawn from this are given labels  $p_{\rm real}=1$, $p_{\rm gal}=0$. 
    
\end{itemize}

As with any real sample, the labeling is not pure and there are known areas of confusion or mislabeling. 
The ``PM" category is a more recent addition to the web server and although it predates the start of our data gathering for the VRA project some eyeballers would place examples of ``PM" stars in the garbage. 
Another area of confusion arises in the ``Attic" which contains duplicate Good objects and some AGNs in low numbers. Re-eyeballing of the data during development allowed us to find some of these alerts and remove them from the $p_{\rm gal}=1$ labels (some contamination may remain, see \citealt{stevance_2025_14944209} for details).


\subsection{Data and Features}
\label{sec:data}

The data used to train the models presented in this paper were gathered from the eyeball list between 27th of March 2024 and 22nd January 2025.
These do not reflect the full extent of the bogus properties in the ATLAS stream and are solely intended to train a model that works downstream of previous automation steps
(points 1 to 3 see Section \ref{ref:atlas_summary}). 

The VRA underwent several rounds of prototyping and the data is divided into sub-data sets. 
The first sub-data set was gathered between 27th March 2024 and 13th August 2024, during which no VRA prototypes were actively participating in eyeballing. 
These data best reflect the eyeballer workload and decision speed, although we note that it is limited in time to a period of four and half months during which the galactic center is very visible to the Chilean and South African ATLAS units. 
The second sub-data set was gathered between 18th August 2024 and 22nd January 2025.
This is the first dataset that is impacted by the VRA, which is reflected in the large fraction of data that are labelled as ``Auto-garbage". 

\begin{figure}
    \centering
    \includegraphics[width=1\linewidth]{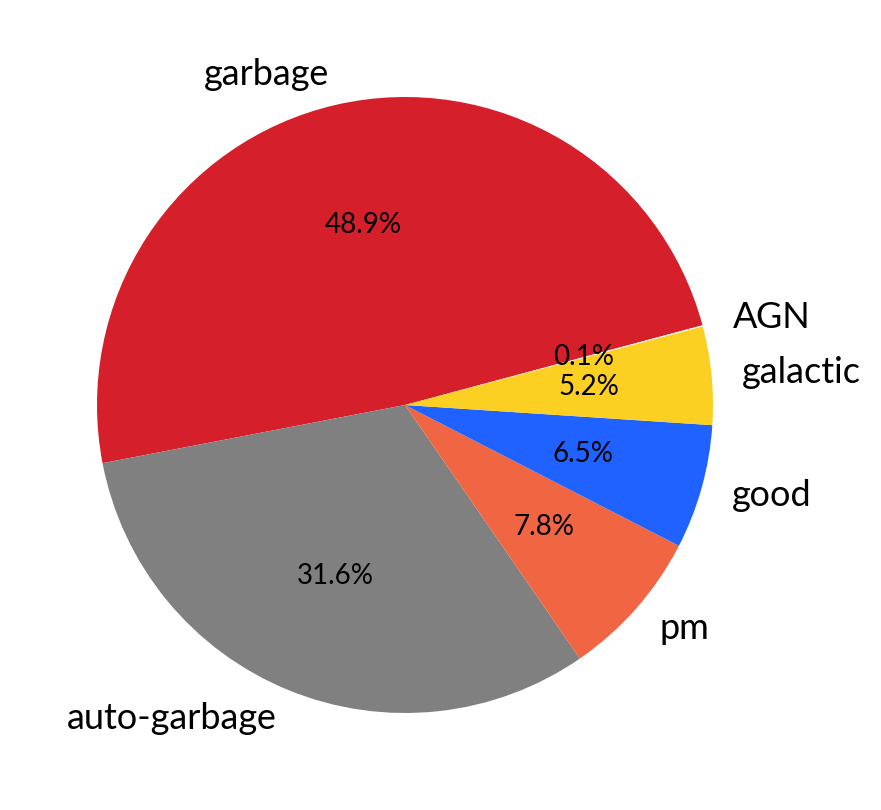}
    \caption{ Alert type distribution in our full data sets spanning 27th March 2024 to 22nd January 2025. Some of these alerts were re-eyeballed during development as their human labels were discrepant with their location in Score Space. A few Active Galactic Nuclei (AGNs) were found in the ``Galactic" (``Attic") alerts and marked as such.  \label{fig:pie_chart_full_data}}
\end{figure}

The full data set used here contains 75,129 alerts with a label distribution shown in Figure \ref{fig:pie_chart_full_data}.
The raw JSON data and cleaned data frames are available alongside the codes used to clean the data set and make the features described in the following sections \citep{stevance_2025_14906192}. 
The list of features used by the day 1 and day $N$ models are summarised in Table \ref{tab:features}, and the feature distributions are shown in Appendix \ref{sec:app_feature_distribs}.

\begin{table*}[!h]
    \centering
    \caption{Features used by the day 1 and day $N$ models. The column names are those used in the code and data release. }
    \label{tab:features}
    \begin{tabular}{l| l| l| l}
       \textbf{Model} & \textbf{Column name} & \textbf{Type} & \textbf{Description} \\
       \hline  	
       day 1 + $N$ & {\tt Nnondet\_std} &  float & Standard deviation of the number of non detections between detections \\
       day 1 +  N & {\tt Nnondet\_mean}  & float &  Mean of the number of non detections between detections \\
       day 1 + $N$ & {\tt magdet\_std}   & float &   Standard deviation of the magnitude of the historical detections \\
       day 1 + $N$ & {\tt DET\_Nsince\_min5d}  & float &	Number of detections between phase -5 days and day 1\\ 
        day 1 + $N$ & {\tt NON\_Nsince\_min5d} 	 & float &	Number of detections between phase -5 days and day 1\\ 
       day 1 + $N$ & {\tt DET\_mag\_median\_min5d} & float &  Median magnitude of the detections between phase -5 d and day 1.\\
       day 1 + $N$ & {\tt log10\_std\_ra\_min5d} 	 & float &	log 10 of the Standard deviation of the RA in the detections from phase -5 days   \\
       day 1 + $N$ & {\tt log10\_std\_dec\_min5d}  & float &	log 10 of the Standard deviation of the Dec in the detections from phase -5 days \\
       day 1 + $N$ & {\tt ra}       &  float &  RA \\
       day 1 + $N$ & {\tt dec} & float & Dec \\
       day 1 + $N$ & {\tt rb\_pix} & float & Real/Bogus Score from the CNN \cite{weston2024}\\
       day 1 + $N$ & {\tt z} 	& float &	Spectroscopic redshift (if known, else NaN)\\ 
       day 1 + $N$ & {\tt photoz} 	& float &	Photometric redshift (if known, else NaN) \\ 
       day 1 + $N$ & {\tt ebv\_sfd} 	& float & Extinction $E(B-V)$ calculated using {\tt dustmaps} SFD\\
       day 1 + $N$ & {\tt log10\_sep\_arcsec} & float & log 10 of the projected separation between the best matched source (in arcsec) \\
       day 1 + $N$ & {\tt SN} & bool & \textbf{[PRUNED]} (sherlock classification) if SUPERNOVA \\
       day 1 + $N$ & {\tt NT} & bool & \textbf{[PRUNED]} (sherlock classification) if NUCLEAR TRANSIENT\\
       day 1 + $N$ & {\tt ORPHAN} & bool & \textbf{[PRUNED]} (sherlock classification) if ORPHAN\\
       day 1 + $N$ & {\tt CV} & bool &  (sherlock classification) if CATACLYSMIC VARIABLE \\
       day 1 + $N$ & {\tt UNCLEAR} & bool & \textbf{[PRUNED]} (sherlock classification) if UNCLEAR \\
       day $N$ only & {\tt max\_mag} & float & Maximum (median) magnitude seen since phase -5 d \\
       day $N$ only & {\tt max\_mag\_day} & float & Day of the maximum magnitude \\
       day $N$ only & {\tt DET\_N\_total} & float & Number of detections since phase -5 d \\
       day $N$ only & {\tt NON\_N\_total} & float & Number of non detections since phase -5 d \\
       day $N$ only & {\tt DET\_mag\_median} &  float &  Median magnitude of the detections since phase -5 d.   \\
       day $N$ only & {\tt NON\_mag\_median} & float &  Median magnitude of the non detections since phase -5 d \\
       day $N$ only & {\tt DET\_N\_today}  & float &  \textbf{[PRUNED]} Number of detections seen today \\
       day $N$ only & {\tt NON\_N\_today}  & float &  \textbf{[PRUNED]} Number of detections seen today \\
        \hline
    \end{tabular}
\end{table*}

\subsubsection{Long term lightcurve history (-100 days)}

One of the characteristics that eyeballers look for in the alert lightcurves are historical detections, as they can indicate recurrent activity or outbursts (real) or regular bad subtractions or artifacts at that location (bogus).

\begin{figure*}[ht!]
\centering
\includegraphics[width=15cm]{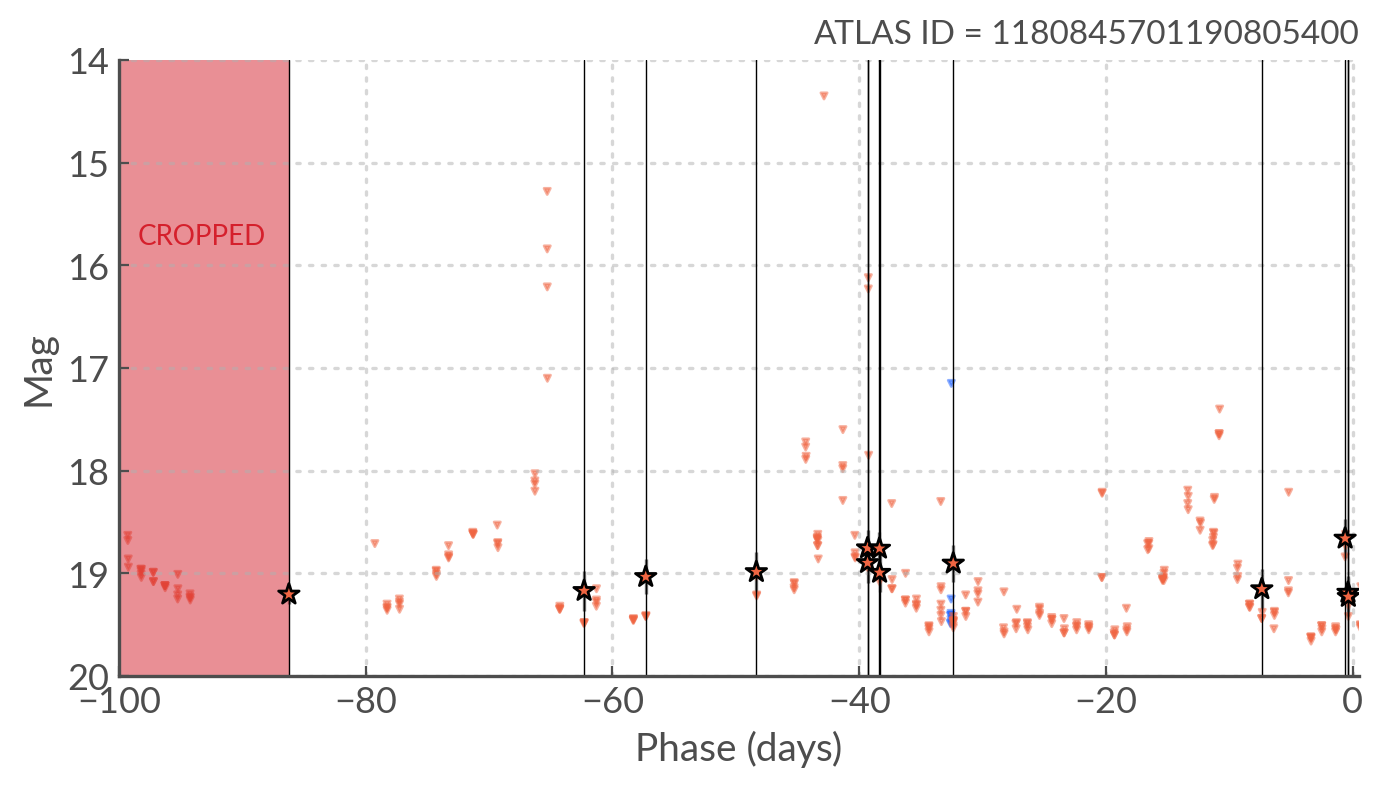}
\caption{Visualisation of the lightcurve history used to create the VRA features. The star markers show detections (any filter) and the triangles show the 5-sigma limiting magnitude in the ATLAS frame of each non-detections. The time axis is given with respect to the time of first alert (t=0). Each detection is also highlighted with a vertical line to show how historical detections segment the lightcurve - it is within these segments that we count the number of non detections to then calculate the mean and standard deviation for the {\tt Nnondet\_std} and  {\tt Nnondet\_mean} features. Also in red we show that every measurement before the first detection in our -100 day window is cropped (red overlay).
\label{fig:lc_history}}
\end{figure*}

To attempt to capture these behaviors we calculate three features: The mean and standard deviation of the number of non detections between each historical detection ({\tt Nnondet\_mean}, {\tt Nnondet\_std}), and the standard deviation of the magnitude of these historical detections ({\tt magdet\_std}).
Here a historical detection is defined as any detection that occurred within -100 days of entering the eyeball list. 
This includes lone detections that would not pass the quality cuts which require a minimum of three detections within a single night.  
To calculate the features we first crop every data point before the first historical detections, within our chosen time window of -100 days, in order to anchor our count of the non detections. 
To illustrate this process with show in Figure \ref{fig:lc_history} an example of a garbage alert with spurious detections.
The mean and standard deviation of the number of non-detections between each historical detection are 19.1 and 27.4. 

For feature calculations we ignore the filter information, which means that orange and cyan magnitudes are considered together when calculating e.g. the standard deviation.
Also it is worth stating that magnitudes being a logarithmic transformation of the flux, taking the standard deviation of a series of magnitudes is not the correct way to formally assess their variability. 
Nonetheless we use such features as they are informative and fast to compute, but we emphasise that \textit{they are not strictly physical measurements and they should not be used outside of this context} - certainly not if deriving astrophysical quantities.

\subsubsection{Recent lightcurve history (-5 days)}
The recent lightcurve history is defined as the data captured within 5 days of the alert entering the eyeball list. 
This is of particular interest as real transients that were rising but faint in preceding days may have shown one or two detections but fell short of our requirement for alerts to have three out of four 5$\sigma$ detections in one night to enter the data stream. 
We chose 5 days before alert as a cutoff since given the ATLAS cadence it usually corresponds to an additional one to two previous visits which from eyeballing experience is where lower signal detections begin to be visible for rising extra-galactic transients.

We record three recent light curve history features: the number of detections ({\tt DET\_Nsince\_min5d}), the number of non detections ({\tt NON\_Nsince\_min5d}) seen in the 5 days preceding and including the first alert, and the median magnitude of the detections over that period ({\tt DET\_mag\_median\_min5d}).

\subsubsection{New lightcurve information for the day $N$ models}
\label{sec:new_lc_features}
For the day $N$ models, we calculate four additional lightcurve features. 
These are designed to try and capture the new lightcurve information gathered by new telescope visits for sources that were ambiguous and not classified by eyeballers on day 1.  The four additional features  are the total number of detections ({\tt DET\_N\_total}) and non detections ({\tt NON\_N\_total}) since -5 days, the minimum (maximum brightness) magnitude value recorded so far ({\tt max\_mag}) and the day (phase) on which that magnitude value was recorded ({\tt max\_mag\_day}). 
Again here all filters were considered together and there is no distinction between a cyan maximum and an orange maximum. 
We do not record which filter the maximum was recorded in as it would not be informative for VRA classification since the orange and cyan filters alternate based on the phase of the moon rather than  characteristics related to the transients themselves. 

We also do not attempt the fit the lightcurves to find the peak magnitude. 
Previous work has shown that the simpler feature of the minimum magnitude and the day at which is occurred was surprisingly informative \citep{btsbot} and we also find that to be the case (see Figures \ref{fig:permpinp_day1} and \ref{fig:permpinp_dayN}).

We only calculate these features until a maximum phase of 15 days after the initial alert. 

\subsubsection{Sky location and extinction}

Some of the most important contextual features used are the on-sky positions (Right Ascension and Declination) and the exctinction ($E(B-V)$; {\tt ebv\_sfd}) - see Figures \ref{fig:permpinp_day1} and \ref{fig:permpinp_dayN}).
This is because they define whether the source is associated with a crowded area of the sky (Galactic plane) that is prone to bad subtractions (see Figure \ref{fig:RADEC}).
High values of extinction reduce the likelihood of an alert being an extra-galactic source. In the extreme, the highest values of foreground extinction (e.g. $E(B-V)\gtrsim1$) preclude extragalactic sources simply due to the limit they place on the absolute magnitude.

\subsubsection{Scatter on the sky}
Another important piece of information leveraged by the eyeballers is the scatter in the RA ({\tt log10\_std\_ra\_min5d}) and Dec ({\tt log10\_std\_dec\_min5d})
of the individual exposures. 
Typically a well localised alert, with little scatter in its position centroid, will be associated with a real alert. 
The converse is not necessarily true however, as real transients can and do occur in images that show jitter or trailing. 
 We apply a log10 transformation to these features to obtain a more symmetrical distribution and since these values are never 0 there is no risk of getting undefined values.

\subsubsection{{\sc Sherlock} cross-matching}
Finally we also record features relating to cross-matching with astrophysical catalogs. 
This cross-matching is already performed upstream with {\sc Sherlock} \citep{Young_sherlock_2023}, which is a contextual classifier that uses boosted decisions trees to perform rapid cross-matching to existing astrophysical catalogues for transient surveys used in ATLAS and the Lasair data broker \citep{lasair}.
It adds features to the data stream related to host/source cross-matching (such as angular distance, redshift), as well as a classification based on the features in the cross-match.

For our purposes we use four of these pre-calculated features: the logged separation in arcseconds to the cross-matched source ({\tt log10\_sep\_arcsec}), the spectroscopic redshift ({\tt z}) , the photometric redshift {\tt (photoz}), and finally the CV (Cataclysmic Variable) flag. 
This flag indicates that a known classified source is within 0.5 arcsec of our transient candidate.

In this final version of the VRA we do not use the SN (suspected supernova), ORPHAN (no host), NT (nuclear transient) or UNCLEAR {\sc Sherlock} flags because they induce confusion into the models (see Figures \ref{fig:permpinp_day1} and \ref{fig:permpinp_dayN}).
That is due in part to the fact that many Galactic stars are tagged as extended sources in some catalogs, leading both {\sc Sherlock} and our models to confuse galactic transients with potential supernovae. 

\subsection{The Models}
\label{sec:models}


\subsubsection{Histogram Based Gradient Boosted Classifiers}
 To place our alerts in score space we create two classifiers: a real/bogus and a galactic/extragalactic  (four if we include their \textit{day 1} and \textit{day $N$} variants).
We chose to use a Gradient Boosting method \citep{friedman2001} called Histogram Based Gradient Boosted Decision Trees ((H)GBDT).

Light curve data in its raw state is a time-series, which is not handled well by such feature-based classifiers. 
This is why we devised an array of features to capture long and short term light-curve behaviour. 
Although there exists other forms of ML methods such as neural networks (especially recurrent neural networks) that could use the raw light-curve as input, there are two reasons why we did not chose to use such models. 
Firstly, neural networks are data-hungry. To be trained effectively they require data sets of order 50$\times$ the number of parameters in the model \citep{ALWOSHEEL2018167}. Although we appreciate that this is somewhat a `rule of thumb', other studies indicate that the GBDT family can provide strong baselines across a wide range of data set sizes \citep{2023arXiv230502997M}.
The fact that (H)GBDT models can perform well with only a few thousand examples allows us to proceed without the need for data augmentation to artificially increase the number of data samples, as is often the case when using neural networks with large numbers of parameters.

Secondly, the light-curve information is not rich on the timescales considered here, and it is difficult to represent astrophysical time series in classic ML and statistic tools. 
For example they are not built to handle non-detection information (a non-detection is \textit{not} the same as no observation). 
On the whole we prefer to engineer our own features based on our understanding of the light-curve information and our goals.

\begin{figure}
\centering
\includegraphics[width=8cm]{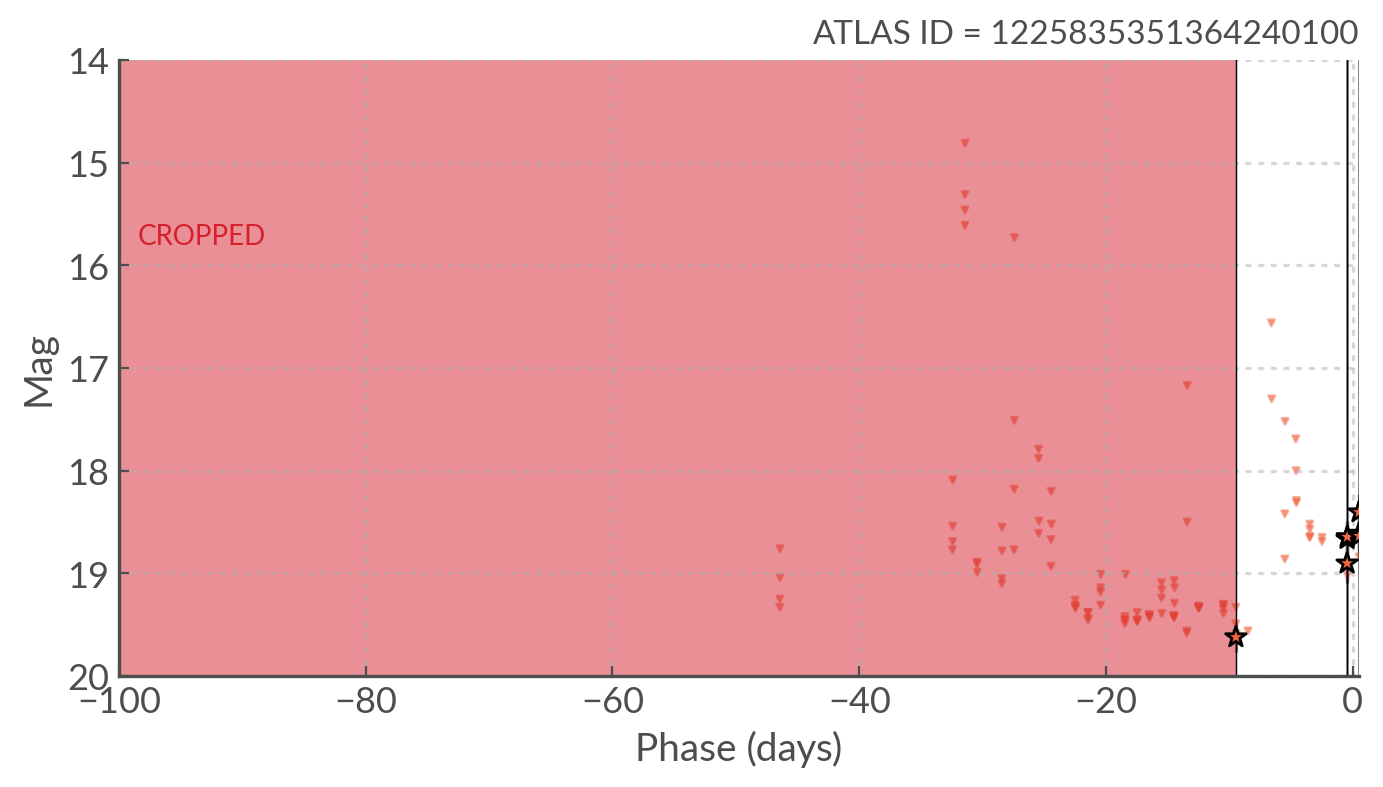}
\caption{Long term history of AT2024lwd (ATLAS\_ID=1225835351364240100) as recorded by the VRA feature calculator
\label{fig:24lwd_hist}}
\end{figure}

Finally, the use of feature based methods allows for easier interpretability of the models which makes diagnosing potential issues easier (see later our discussion of a AT2024lwd in Section \ref{sec:performance} and Figures \ref{fig:24lwd_hist}, \ref{fig:24lwd}) 

We use the {\tt scikit-learn} \citep{scikit-learn} implementation of the (H)GBDT, which is based on LightGBM \citep{lightGMB}.
There are several advantages to the histogram based approach, of main importance to us is the native handling of null values. 
In most machine learning models null values need to be imputed, in the (H)GBDT the feature values are discretised into histograms with (typically) 256 bins, 255 of which are used for numerical values and the final bin for null values\footnote{see manual page for (H)GBDT in  \href{https://scikit-learn.org/stable/modules/generated/sklearn.ensemble.HistGradientBoostingClassifier.html}{scikit-learn}}. 
Another advantage (and the main reason this improvement on the original algorithm was devised) is their speed.
This only becomes a concern when dealing with tens of thousands of training samples, which is barely the case here but could become important in the future. 

\subsubsection{day 1 Models}
We first split the data into a training set and a validation set, at ratios of 0.85 and 0.15 of the full data set.
The training set is then balanced by sub-sampling overly represented classes.
As mentioned above, the data set considered here is a combination of two sub data sets, the train/validation split and re-sampling of which were performed individually.
Additionally, the resampling was performed before parts of the data sets were re-eyeballed. 
The full historical details of data gathering and re-eyeballing can be found in the online data release \citep{stevance_2025_14906192} and the technical manual \cite{stevance_2025_14944209}.
Overall the training set is not fully balanced, as we can see in Table \ref{tab:n_classes_day1models}, but it is not largely dominated by bogus alerts as is the full data set and the validation set.

\begin{table}[!h]
    \centering
    \caption{Day-1 Models training and validation set label distribution. }
    \label{tab:n_classes_day1models}
    \begin{tabular}{l| l| l}
       \textbf{Label}  & \textbf{Training} & \textbf{Val.} \\ 
\hline
     Good [$p_{\rm real}$=1; $p_{\rm gal}$=0]& 4,249 & 745 \\
     Galactic [$p_{\rm real}$=1; $p_{\rm gal}$=1] & 2,899 & 457 \\
    Proper Motion [$p_{\rm real}$=0; $p_{\rm gal}$=1] & 2,736 & 905 \\
     Garbage [$p_{\rm real}$=NaN; $p_{\rm gal}$=0]  & 2,921 & 5,380 \\
    Auto-garbage [$p_{\rm real}$=NaN; $p_{\rm gal}$=0] & 1,600 & 3,571 \\
    \hline
       \textbf{TOTAL} & \textbf{14,405} & \textbf{11,058}\\
    \end{tabular}
\end{table}

Both the $p_{\rm real}$ and $p_{\rm gal}$ scorers were trained with an {\tt l2\_regularization} parameter of 10, a {\tt class\_weight} parameter set to `balanced' and a {\tt random\_seed} = 42. 
The $p_{\rm real}$ scorer was trained with a {\tt learning\_rate} of 0.1 (default) whilst the $p_{\rm gal}$ scorer {\tt learning\_rate} was set to 0.2. 
In an earlier prototype, hyper-parameter optimization using grid search was conducted on the {\tt learning\_rate} and {\tt l2\_regularization} value. 
We found very little difference in performance between most (reasonable) values of these parameters.
The performance gain were too marginal to justify the time and computational cost of rerunning the hyper-parameter search and follow-up tests on the AuRaK and eyeballing policies and subsequent trainings of the VRA. 
The details of our hyper-parameter searches during development can be found in \cite{stevance_2025_14944209} and the code si available in the data release \citep{stevance_2025_14906192}. 

\begin{figure*}
\centering
\includegraphics[width=10cm]{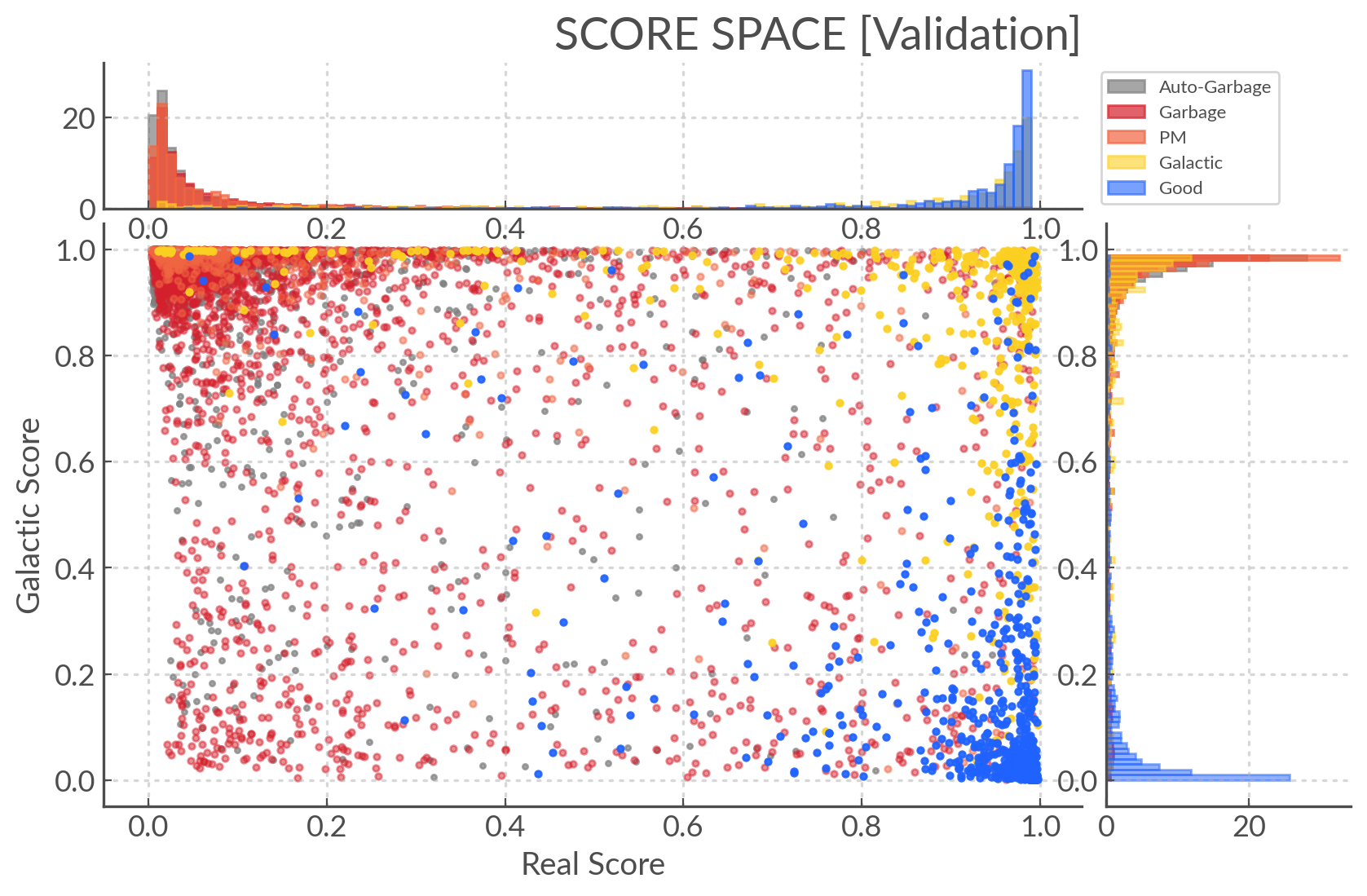}
\includegraphics[width=6.5cm]{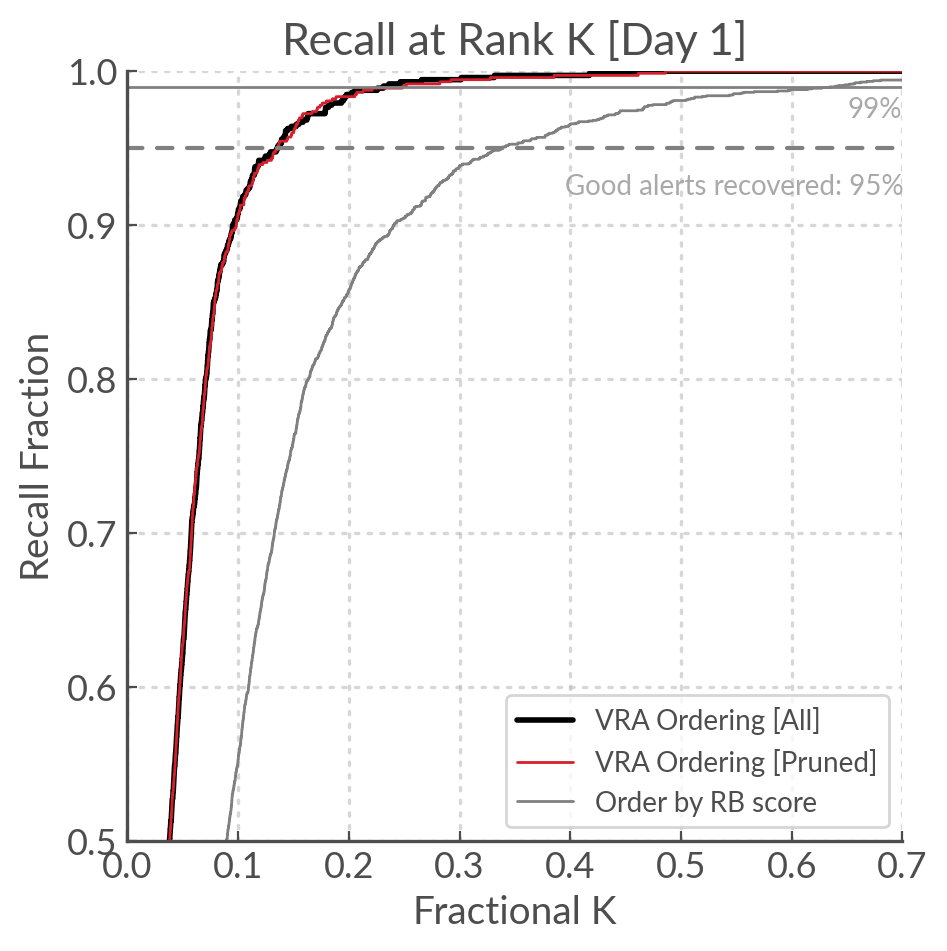}
\caption{Score space and R@K of our validation data set for the day 1 models, colour coded according to their human-given labels. The score space is that obtained with the models trained excluding the pruned features (see Table \ref{tab:features}). The R@K plot shows both the R@K curves obtained for the models trained on all the features and those trained excluding the pruned features. For comparison with our benchmark in Figure \ref{fig:rb_score_ordering} we show the R@K curve obtained when ordering by the RB scores calculated by the CNN. 
\label{fig:ss_ratk_day1}}
\end{figure*}

In Figure \ref{fig:ss_ratk_day1} we show the score space and R@K curve obtained with our day 1 models. 
We can see good separation of the real classes (``Good" and ``Galactic") from the bogus classes (``PM" and ``Garbage"), and the extra-galactic transients (``Good") distribution is well confined to the bottom-right corner of the score space.
The ``Galactic" samples are also  well separated and found in the top right corner of the plot where we expect.
The Garbage samples naturally concentrate towards the top of score space, which is unsurprising as ``Garbage" alerts are more likely to occur in crowded fields (as can be seen in Figure \ref{fig:RADEC}, they partially track the galactic plane).

We can also compare our day 1 models performance in ordering the alerts using Eq. \ref{eq:vra_rank} compared to our benchmark defined in Section \ref{sec:benchmark} (right panel of Figure \ref{fig:ss_ratk_day1}).
We get 95\% (99\%) recall of the ``Good" objects within the top 15\% (25\%) of the list. 
To achieve the same recovery of ``Good" alerts when ordering with the RB score from the CNN real/bogus classifier, we would have to scan 35\%($>$60\%) of the list. 
The AuRaK has also increased from 0.88 for our benchmark to 0.951 with our day 1 models. 

\begin{figure}
\centering
\includegraphics[width=7cm]{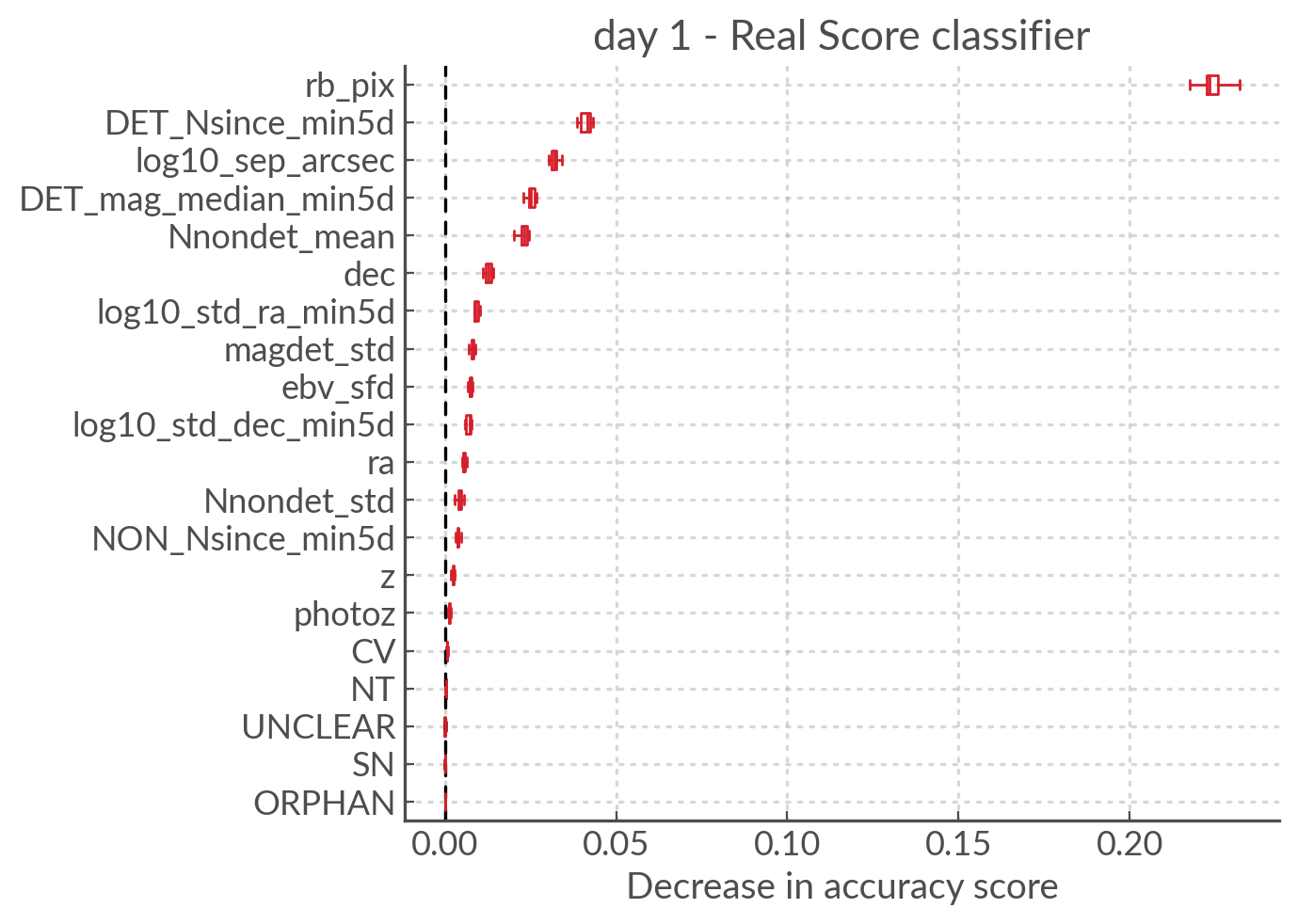}
\includegraphics[width=7cm]{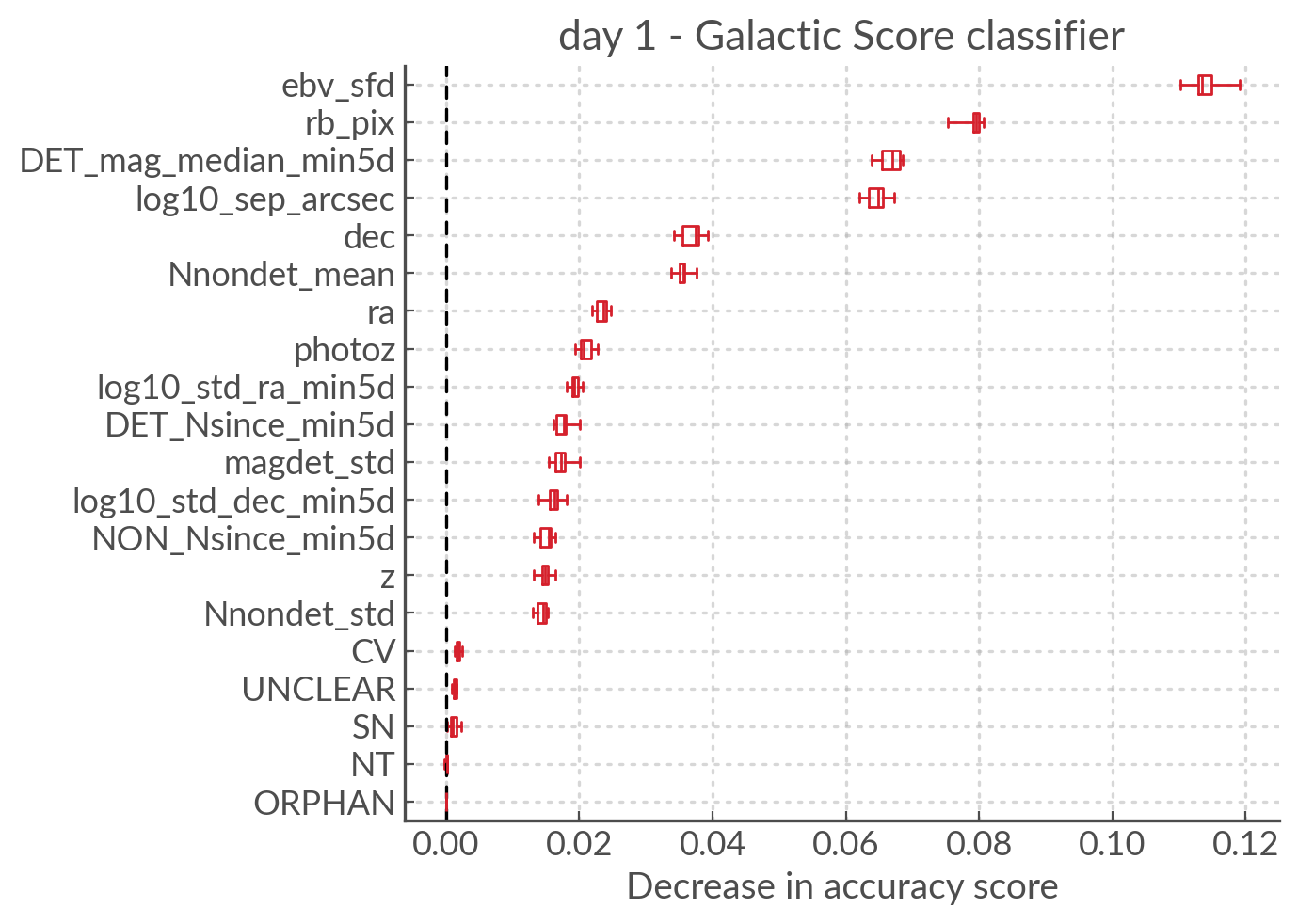}
\caption{Permutation importance of the day 1 model features. 
\label{fig:permpinp_day1}}
\end{figure}

Then, we can quantify how informative each of our features are using a technique called Permutation Importance, which consists in scrambling one feature at a time and retraining a model to evaluate the decrease in prediction accuracy. 
In the present case we performed 10 iterations (for each feature) using the {\tt scikit-learn} implementation; the results for both our Real and Galactic scoring models are presented in Figure \ref{fig:permpinp_day1}. 
We can see that the {\sc Sherlock} flags SN, UNCLEAR, ORPHAN and NT do not result in reduced accuracy for either the Real or the Galactic scoring model.
This is because many of the {\sc Sherlock} features used to create those flags are already given to the model (such as the separation in arcsecond), and the main feature that is omitted (wether the cross-matched source is a galaxy) is one that causes confusion and leads to an over-abundance of galactic sources being flagged as SN. 
These features therefore either represent a source of confusion or a duplication of information; as a result the (H)GBDT algorithm learns to ignore them. 

We note that a low score on the permutation importance graphs shown in Figure \ref{fig:permpinp_day1} can also result when a feature is only available in a small portion of our samples, such as the redshift.
It is obvious from an astrophysical stand-point that redshift is a useful quantity, but since it is only available for roughly 13\% of our alerts, a metric like the permutation importance does not necessarily reflect how informative that feature is for those samples. 

That is why before pruning features we must verify that omitting them does not affect our science metric, in this case the R@K. 
In Figure \ref{fig:ss_ratk_day1} we plot both the R@K curves for the models trained on all features (black) and those trained omitting the SN, UNCLEAR, ORPHAN and NT categories (red). 
As we can see the curves are virtually identical, confirming that these features are not informative and can be removed from the training process.

\subsubsection{day $N$ Models}

For the models that update the scores when new light curve information is recorded (whether they be detections or non-detections), we need a data set with samples of varying light curve completeness. 
To create this sample we take all the alerts in our training and validation sets for the day 1 models, and we consider their lightcurve between day 2 and day 15.
For each ATLAS visit in that time range, we calculate the new lightcurve features described in Section \ref{sec:new_lc_features} (also see Table \ref{tab:features}). 
This means that a single object will be represented several times in this data set, increasing the number and distribution of alert types in our training and validation sets, as we can see in Table \ref{tab:n_classes_dayNmodels}.
The class balance is a little skewed towards the "PM" labeled data but overall it is sufficiently balanced that we do not consider resampling.

\begin{figure*}
\centering
\includegraphics[width=9cm]{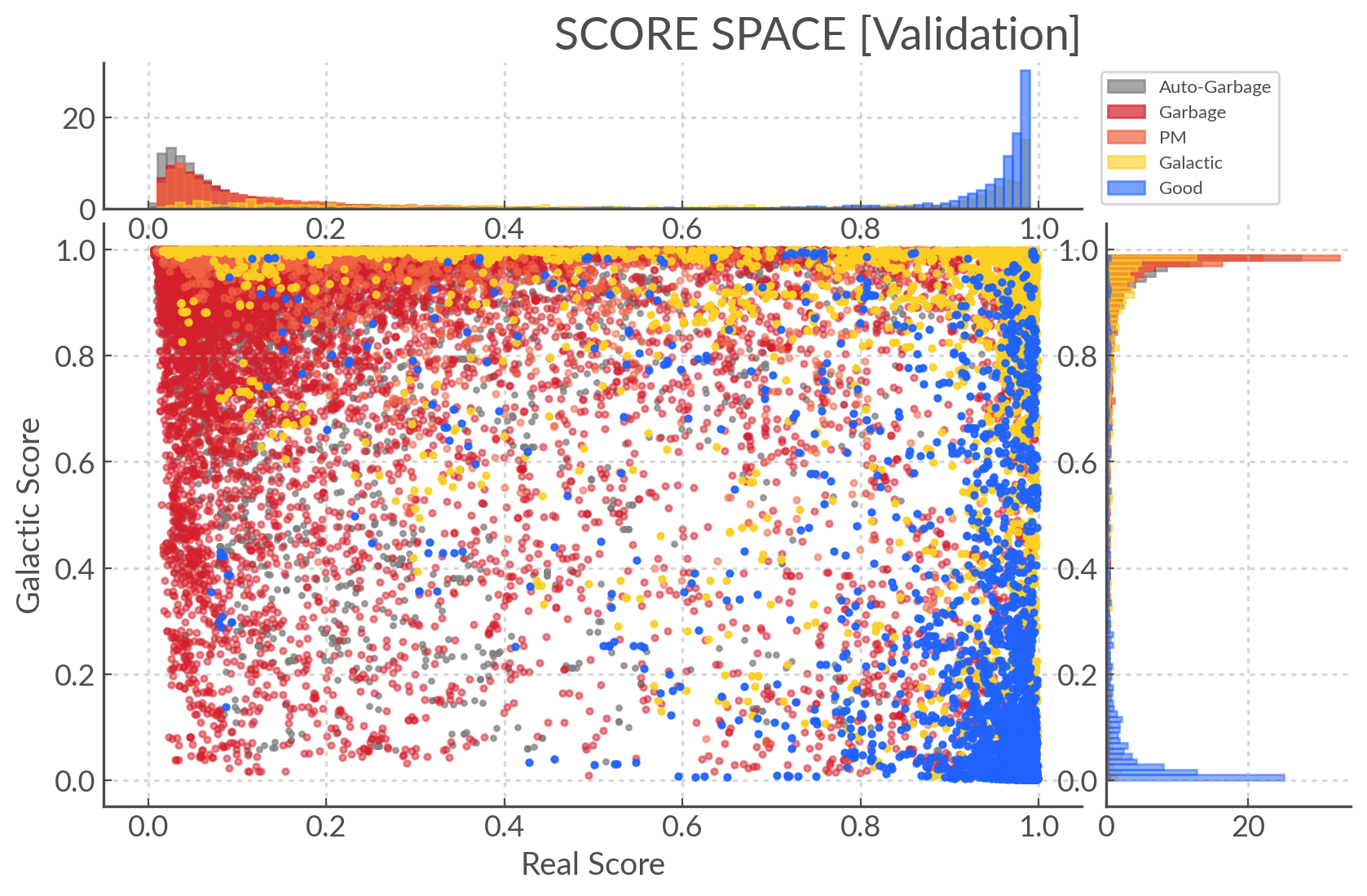}
\includegraphics[width=6cm]{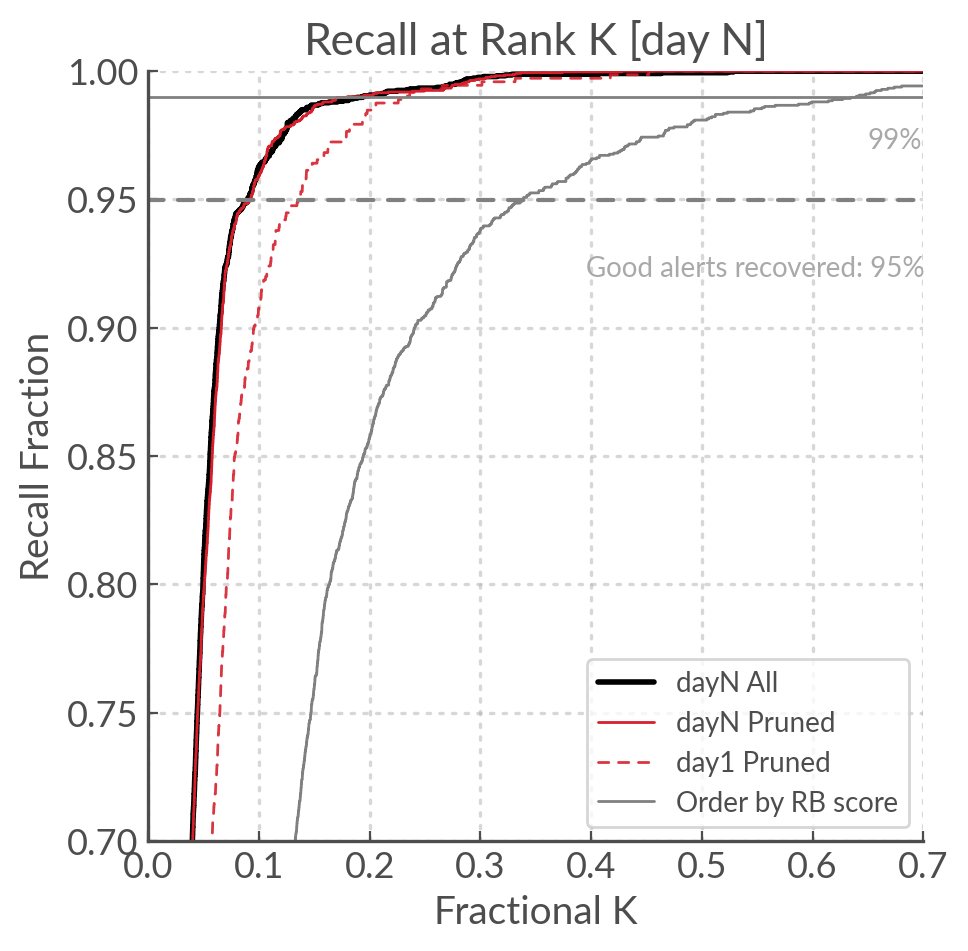}
\caption{Score space and R@K of our validation data set for the day $N$ models, colour coded according to their human-given labels. The score space is that obtained with the models trained excluding the pruned features. The R@K plot shows both the R@K curves obtained for the day $N$ models (solid lines) trained on all the features (black lines) and those trained excluding the pruned (red lines) features. We also show the R@K curves obtained with our day 1 model (dashed line). For comparison with our benchmark in Figure \ref{fig:rb_score_ordering} we show the R@K curve obtained when ordering by the RB scores calculated by the CNN.\label{fig:ss_ratk_dayN}}
\end{figure*}

In the left panel of Figure \ref{fig:ss_ratk_dayN} we show the score space for all the predictions on our validation set. 
The $p_{\rm gal}$ scorer performs better in the day $N$ models than the day 1 models with steeper inclines for the ``Good" and ``Galactic" distributions.
Although the ``Garbage" and ``PM" distributions look like they stretch across most of the 2D space, but we can see when looking at the marginalised $p_{\rm real}$ distribution that they are very similar to those in the day 1 score space plot in Figure \ref{fig:ss_ratk_day1}, but due to the larger number of samples even with medium transparency the scatter plots is inevitably crowded. 
We chose to plot individual points rather than, say, a kernel density approximation because these would hide the outliers in the ``Good" and ``Galactic" classes which are found in unexpected areas of the plot and informative during development.

In the right panel of \ref{fig:ss_ratk_dayN} we can see the R@K curve for the day $N$ models. 
The R@K plot here is not as directly interpretable as that of the day 1 models, because all samples are ordered together and not separated by visits or day $N$\footnote{since our cadence is not 24h visit N and day $N$ are not the same value}. 
Separating by visits would not be a useful comparison as the eyeball list  on any given day is mostly composed of new targets.
Therefore, sorting a handful of objects that are all on their third visit does not tell us how these rankings would perform in a complete eyeball list. 
A more direct evaluation of how our models and our eyeballing policies perform together can be found in Section \ref{sec:policy_eval}. 
The main take away from the R@K plot for day $N$ models is the comparison to day 1 models: If they did not perform better on the whole than day 1 models, it would mean that the features we have extracted on the new light curve data are uninformative. 
On the contrary we see that day $N$ models substantially outperform day 1 models in placing ``Good" objects at the top of the list, with a 95\% (99\%) recall achieved in the top 5\% (20\%) of the list.
Their AuRaK is 0.966.
      
This indicates that the new lightcurve features are helping with classification and that having a separate type of models which updates the scores after an object has already entered the eyeball list is useful.

\begin{table}[!h]
    \centering
    \caption{Day-N Models training and validation set label distribution.  \label{tab:n_classes_dayNmodels} }
    \begin{tabular}{l| l| l}
       \textbf{Label}  & \textbf{Training} & \textbf{Val.} \\ 
        \hline
        Good [$p_{\rm real}$=1; $p_{\rm gal}$=0]& 18,798 & 3,331 \\
        Galactic [$p_{\rm real}$=1; $p_{\rm gal}$=1] & 17,695 & 2,972 \\
        Proper Motion [$p_{\rm real}$=0; $p_{\rm gal}$=1]  & 15,655 & 5,537 \\
        Garbage [$p_{\rm real}$=NaN; $p_{\rm gal}$=0] & 14,812 & 31,393 \\
        Auto-garbage [$p_{\rm real}$=NaN; $p_{\rm gal}$=0]  & 11,711 & 26,262 \\
        \hline
        \textbf{TOTAL} & \textbf{78,671} & \textbf{69,495}\\

    \end{tabular}
\end{table}

\begin{figure}
\centering
\includegraphics[width=7cm]{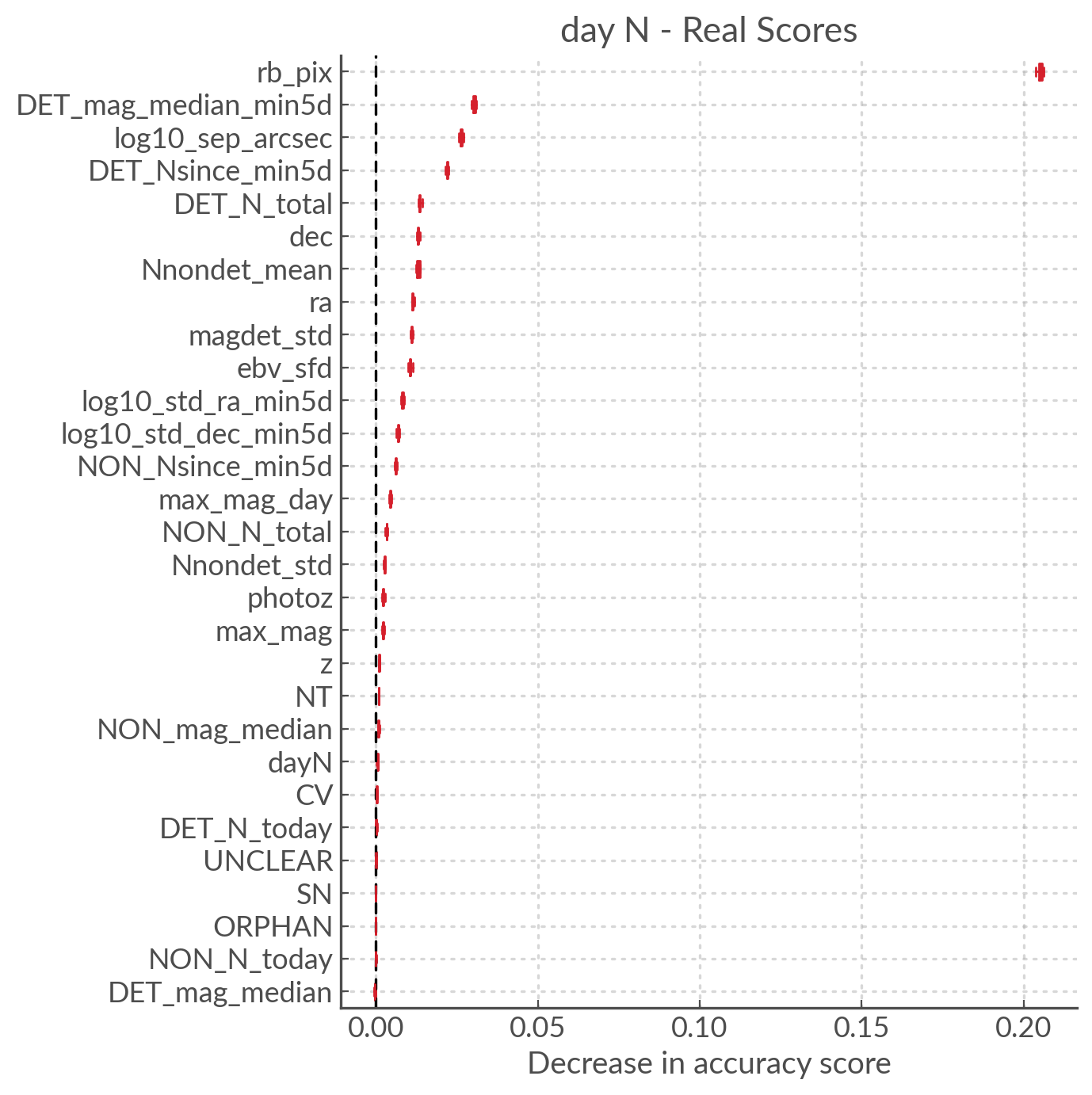}
\includegraphics[width=7cm]{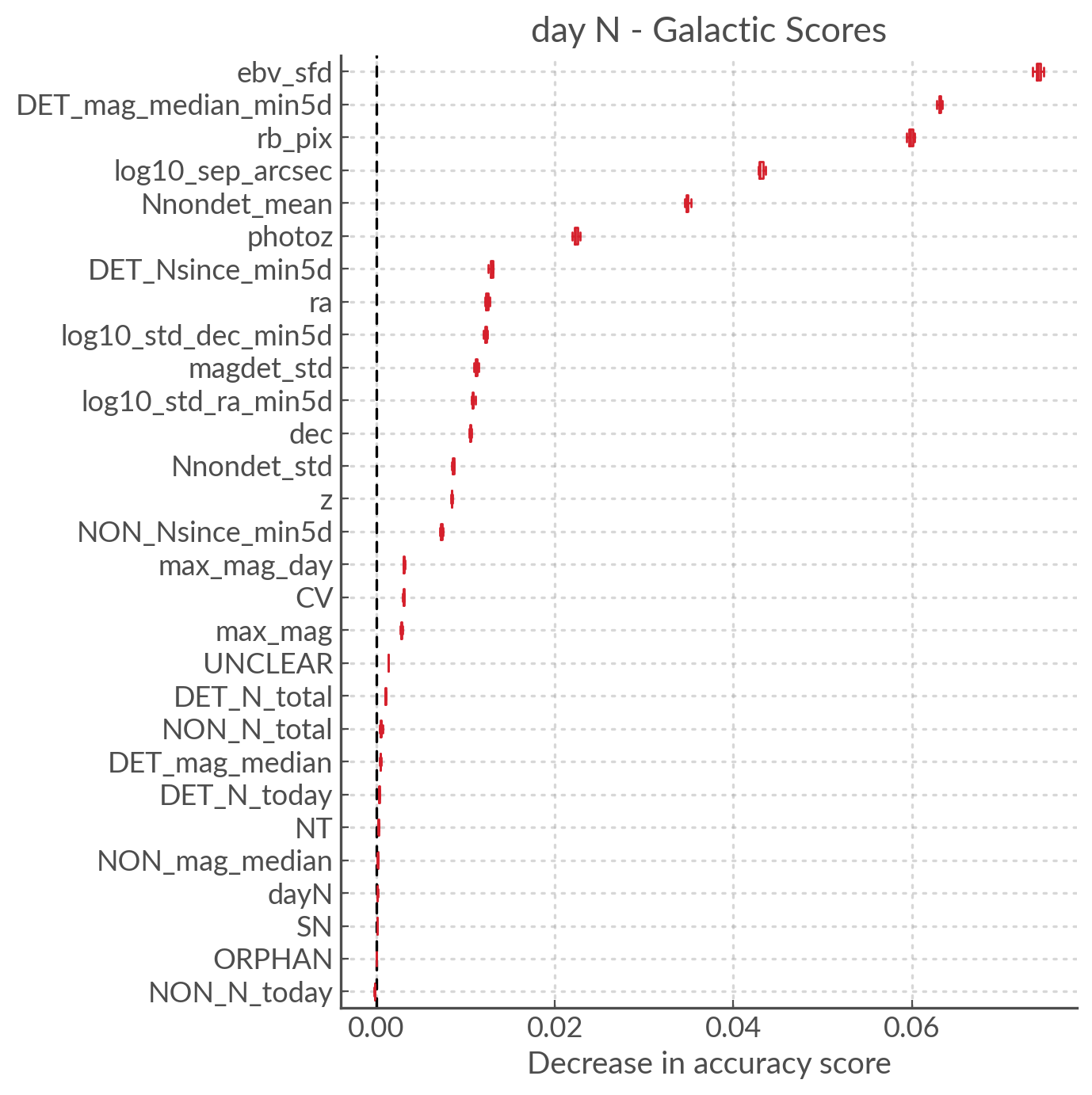}
\caption{Permutation importance for the day $N$ models
\label{fig:permpinp_dayN}}
\end{figure}

We can also calculate the permutation importance of each feature (also using 10 repeats) -- see Figure \ref{fig:permpinp_dayN} -- for which we include the {\sc Sherlock} flags to check that, as for the day 1 models, they do not provide new or useful information compared to the other features. 
Based on Figure \ref{fig:permpinp_dayN} we prune SN, UNCLEAR, ORPHAN and NT, as well as {\tt DET\_N\_today} and {\tt NON\_N\_today}.
The pruned models perform equally well (see right panel - Figure \ref{fig:ss_ratk_dayN}) and these features are removed from the final models. 
The low importance of {\tt DET\_N\_today} and {\tt NON\_N\_today} can be once again interpreted as the result of features duplicating information.
{\tt DET\_N\_today} and {\tt NON\_N\_today} are the number of detections and non detections (respectively) seen on a particular visit, but since we already calculate the total number of detections and non detections seen in total since -5 days (see Table \ref{tab:features} for the full list of features and their descriptions), this information is already included in the total count. 
Although {\tt DET\_mag\_median} and {\tt NON\_mag\_median} - the median magnitude of the detections and non detections on the day - are lower in the permutation importance, we found that removing them leads to more change in the R@K curve and a longer tail in the ``Good" object $p_{\rm gal}$ distribution so we did not prune these features. 

For HGBDT models, pruning is not strictly necessary (performance will not decrease if uninformative features are included), but we chose to remove the most uninformative features to limit how much data processing has to be done in production when running the VRA to limit code bloat.

%
\section{VRA Performance}
\label{sec:performance}

\begin{figure}
\centering
\includegraphics[width=9cm]{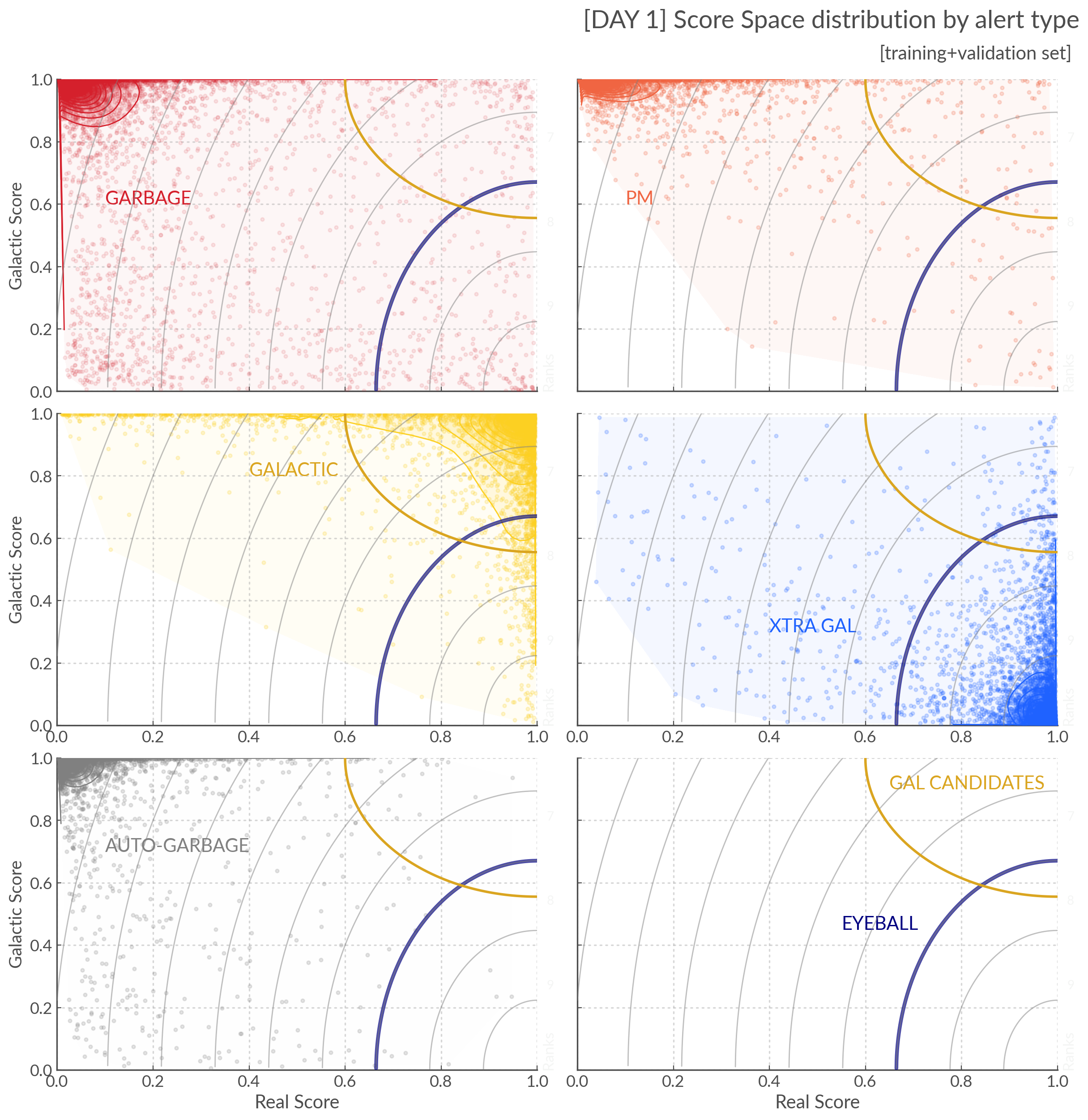}
\caption{Visualisation of the extragalactic eyeballing policies (blue line) and galactic candidate policies (yellow line) compared to the distribution of our alert samples. All VRA ranks from 1 to 9 are also shown (grey solid lines).
\label{fig:ss_withranks}}
\end{figure}

As described in Sections \ref{sec:eyeball_policies} and \ref{sec:garbage_policies}, we have policies to rank the most promising extragalactic candidates, move the most likely galactic transients into a separate eyeball list, and auto-garbaging policies to clear to bulk of the bogus detections.
We can visualize these policies against our model predictions in Figure \ref{fig:ss_withranks}.
The ranking strategy has already been evaluated using the R@K curve and AuRaK, but the threshold below which human scanners are not asked to eyeball has not been considered. 
In this section we evaluate how the models and policies interact with each other before implementing a model in production. 
\subsection{Policy Evaluation}
\label{sec:policy_eval}

\subsubsection{day 1}
\begin{figure*}
\centering
\includegraphics[width=16cm]{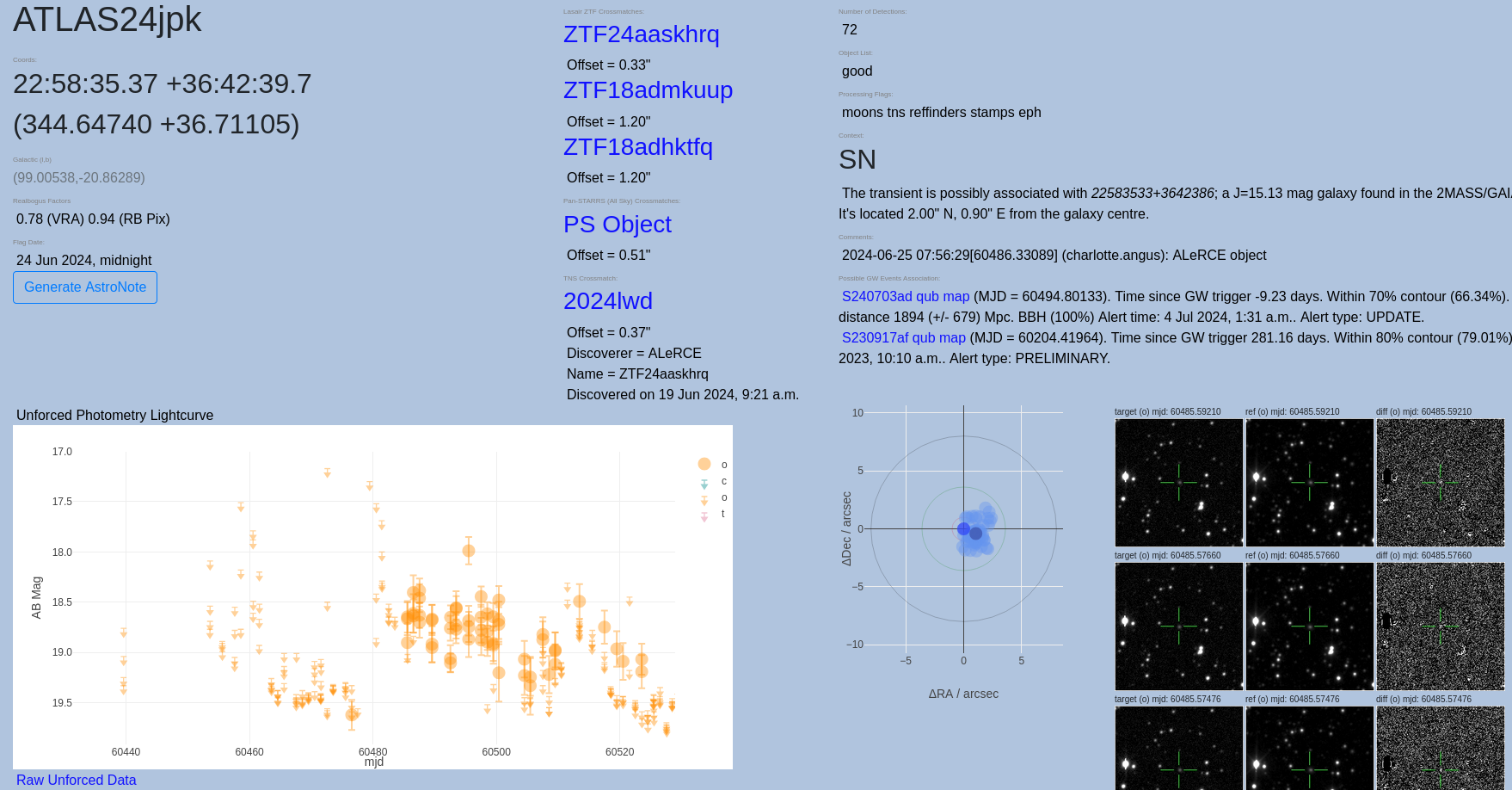}
\caption{Transient web server page for AT2024lwd which is mistakenly labeled as garbage by the VRA day 1 models during policy evaluation.
\label{fig:missed_day1}}
\end{figure*}

First we apply the rankings and policies described in Section \ref{sec:design} on our day 1 validation data set.
We find that 90\% of the ``Good" objects are eyeballed on day 1, and only 3 objects (0.32\%) are auto-garbaged. 
The first is AT2024ugz  which is very faint and has a lower RB score (0.87).
The second is AT2024aayb  and it received a very low RB score (0.28). 
The final object to be auto-garbaged on day 1 is AT2024lwd,  with $p_{\rm real}=0.082$	and  $p_{\rm gal}=0.9902$	(for a VRA$_{\rm score} = 0.67$) it is predicted by our classifiers to be a bogus alert correlated with the Galaxy.
When inspecting the web server (see Figure \ref{fig:missed_day1}) it is not immediately obvious why this alert would receive these predictions: the lightcurve is clear, the RB score is quite high (although not 1.0) and the on-sky location is not particularly near the galactic plane (b=-20 degrees).

Because we are using feature-based algorithms, our first port-of-call to understand this misclassification is to investigate the features. 
We find that five of the features calculated for AT2024lwd have values that are much more consistent with the Galactic and/or Garbage population ({\tt Nnondet\_std}, {\tt Nnondet\_mean}, {\tt log10\_std\_ra\_min5d}, {\tt log10\_std\_dec\_min5d} and {\tt ebv\_sfd}), (see Appendix \ref{sec:app_feature_distribs}, Figure \ref{fig:24lwd}).

The light curve history features {\tt Nnondet\_std} and {\tt Nnondet\_mean} are high because the rise of the transient happened to coincide with a waxing moon, as we can see in Figure \ref{fig:missed_day1}. A single detection in the days before the alert was followed by a streak of non-detections as the sky grew brighter faster than the transient.
Transients rising with the moon is not uncommon but a single detection (out of four frames) followed by 10 days of non detections is a very specific failure mode (at least in our experience - since deploying the VRA this case has not arisen).
In combination with an unusually large scatter in the Ra and Dec measurements, and a sky location that coincides with a slightly elevated extinction, the mis-classification is as a galactic bogus alert can be understood as a combination of unlikely events. 
No action is taken at this stage, but should similar cases be uncovered by cross-matching between the garbage and the TNS, we would review which features are most decisive in silencing these alerts and which additional training may need to be performed.

Overall on day 1, 69\% of the eyeball list is auto-garbaged, 5\% is sent to the Galactic Candidate eyeball list, 9\% remains in the extra-galactic candidate eyeball list and 18\% have been neither eyeballed nor auto-garbage, they await further data. We say that they are in \textit{purgatory}. 

\begin{figure}
\centering
\includegraphics[width=9cm]{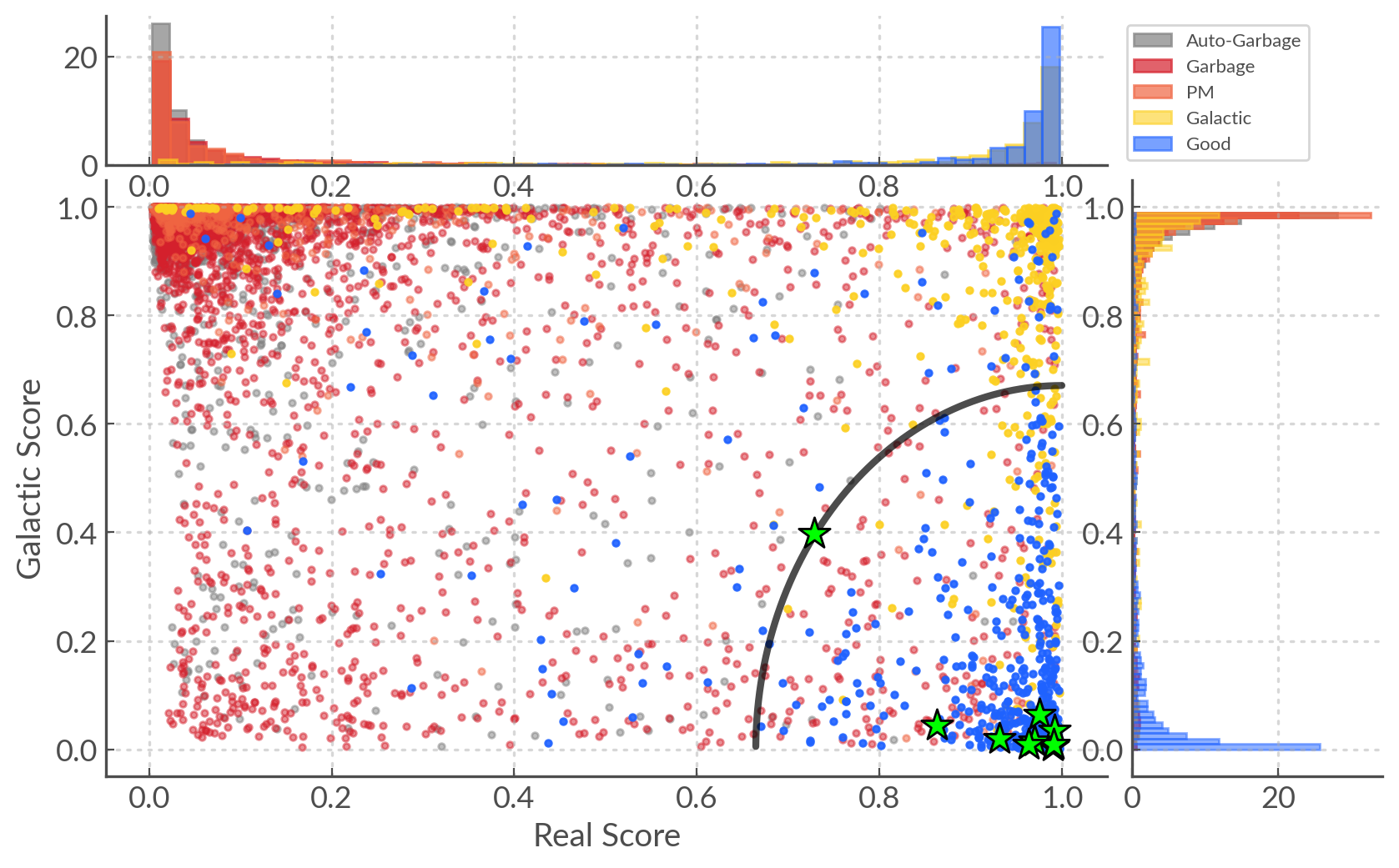}
\caption{Position of the key transients  (lime) listed in Table \ref{tab:key_transients} on the day 1 score space. The extragalactic candidate eyeballing threshold is show as a black line.
\label{fig:ss_keytransients}}
\end{figure}

\begin{table}[!h]
    \centering
    \caption{Key Transients and their day 1 models scores and ranks.}
    \label{tab:key_transients}
    \begin{tabular}{l | l l l l}

        \textbf{ Transient} & $\textbf{p}_{\rm real}$ & $\textbf{p}_{\rm gal}$ & \textbf{VRA$_{\rm score}$}  & VRA class\\
        \hline
        AT2018cow & 0.990 & 0.006 & 9.904  & Extra Gal. \\
        AT2018kzr & 0.931 & 0.019 & 9.380  & Extra Gal. \\
        SN2023zaw & 0.975 & 0.064 & 9.637  & Extra Gal. \\
        SN2023ufx & 0.992 & 0.035 & 9.827  & Extra Gal.\\
        AT2024eju & 0.728 & 0.400 & 6.987  & Purgatory\\
        SN2024atk & 0.991 & 0.008 & 9.910  & Extra Gal.\\
        SN2020kyg & 0.970 & 0.015 & 9.723  & Extra Gal.\\
        SN2020aedm & 0.963 & 0.008 & 9.670 & Extra Gal. \\
        SN2022ilv & 0.863 & 0.043 & 8.762  & Extra Gal. \\

    \end{tabular}
\end{table}

For the day 1 models we take an extra evaluation step that consists in looking at how some rare types of Extra-galactic transients that are particularly relevant to our science team are scored and ranked. 
These are also an interesting test set since they span several years of ATLAS data going back to 2018. Should data drift be a major issue we should see older objects perform significantly worse. 
The chosen objects are listed in Table \ref{tab:key_transients} and their predictions are shown in score space alongside the validation set in Figure \ref{fig:ss_keytransients}.
As we can see, all fall well within our eyeball threshold of 7 (see Section \ref{sec:design}) except for AT2024eju which has a rank 0.03 lower (rank 6.987).
AT2024eju  is qualitatively different to the rest of the transients in Table\,\ref{tab:key_transients}: it had a high RB score (0.99), but was only detected on one night and was initially thought to be a galactic source by a human scanner. 
However it was the optical counterpart to the fast x-ray transient EP20240315a as described in  \citep{gillanders2024}. This required the information from this external source (a 3 arcmin localization radius) to recognize it as an extragalactic alert. 
For such a borderline alert to have a VRA$_{\rm score}$ within 0.03 of our eyeballing threshold is a good indication that our policies are robust despite removing nearly 70\% of the stream on day 1.  

Finally we do not find that older objects perform any worse than more recent ones.  
Although this is not a sufficient test to claim that data drift will be a negligible issue, it is encouraging.

\subsubsection{After 4 visits}
To evaluate how the day $N$ models and later auto-garbaging policies perform, we take the samples that are left in purgatory and apply the corresponding policies before once again separating the alerts into extragalactic eyeballing, galactic eyeballing and purgatory. 
We repeat these successive steps for a total of 4 visits, which given the ATLAS cadence corresponds to between 4 and 15 days (our cut-off) after initial alert on average.

\begin{figure}
\centering
\includegraphics[width=9cm]{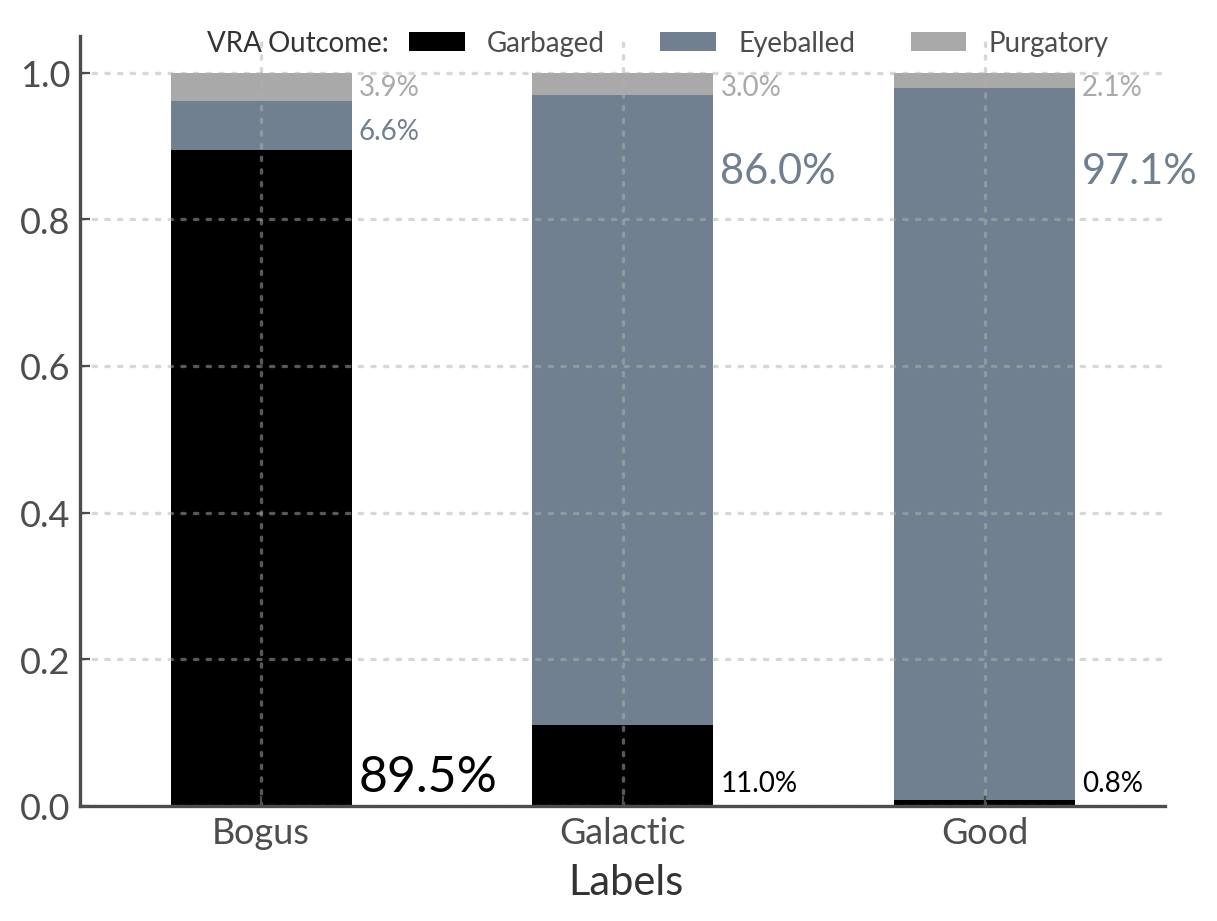}
\caption{VRA outcome for the ``Good", ``Galactic" and ``Bogus" alerts (which include ``Garbage", ``PM" and ``Auto-Garbage") after four visits. The \textit{eyeballed} outcome encompasses both alerts which are eyeballed as extragalactic and galactic candidates.
\label{fig:policy_alllabels}}
\end{figure}

We can see see in Figure \ref{fig:policy_alllabels} the fate of our data split by types (combining the ``Auto-garbage", ``Garbage" and ``Proper Motion" into one bin).
We can see that 97 \% of the Good objects are eyeballed (either through the Extra-galactic or the galactic eyeball list). 
The 2\% that remain in purgatory were checked by eye. 
Many are rather faint sources that were found by other surveys first and would have been eyeballed in production because their VRA$_{\rm score}$ would have been automatically raised to 10. 
All are alerts that we are happy can be eyeballed with a delay of 15 days to add to the good list for completeness: None would have been high priority follow-up targets where a 15 day latency would mean a missed opportunity. 
More details can be found in the code release, particularly in the ``Policy\_evaluation" jupyter notebook.
Additionally we have a good recovery of the Galactic transient events, with only 11\% being discarded. 

\begin{figure}
\centering
\includegraphics[width=9cm]{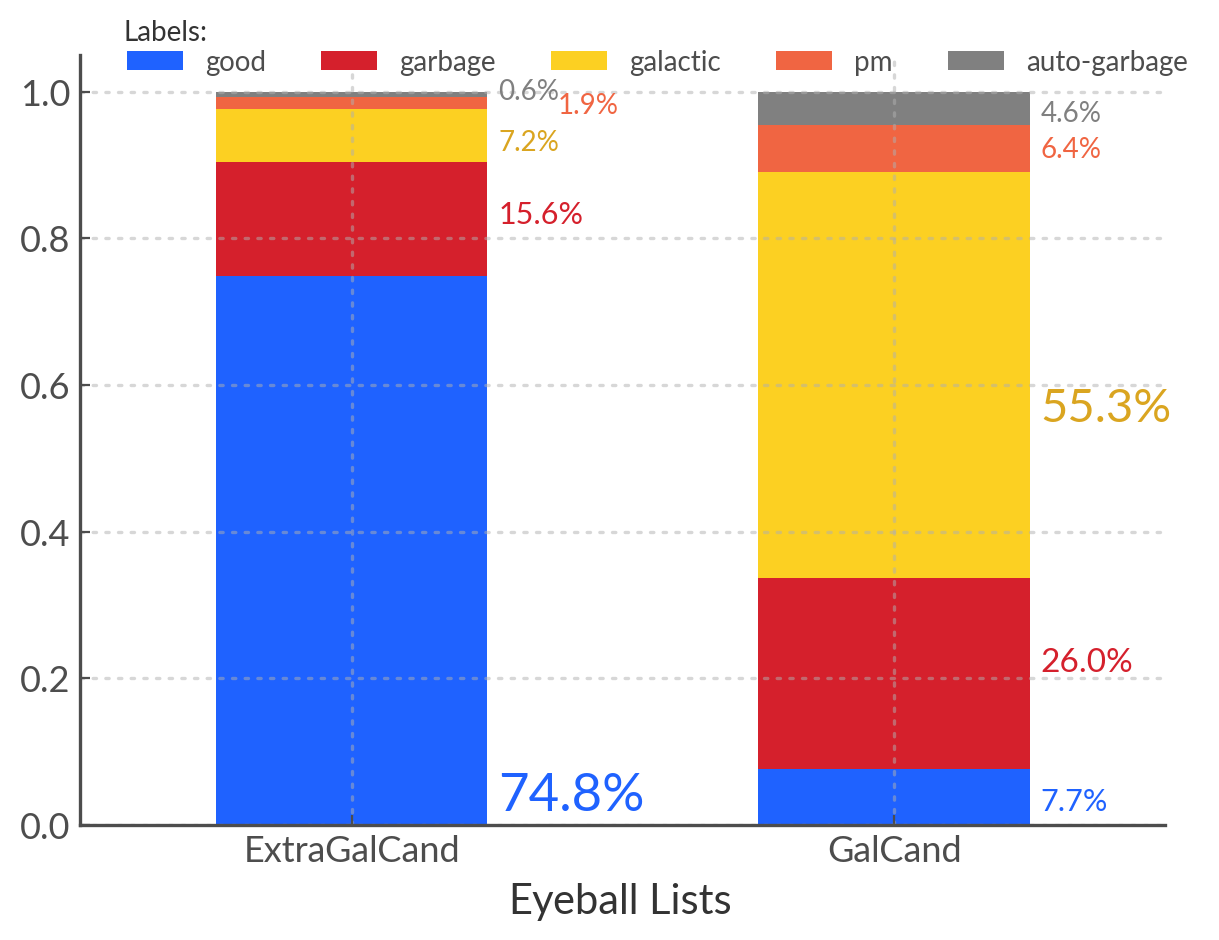}
\caption{Mixture of alert types in the extra-galactic candidate and galactic candidate eyeball lists (total after four visits). 
\label{fig:policy_eyeballlists}}
\end{figure}

Another important consideration is the composition of the Extragalactic and Galactic eyeball list.
As we can see in Figure \ref{fig:policy_eyeballlists}, the cummulative Extragalactic eyeball list (after 4 visits) is composed of $>$80\% real transients, $75\%$ of those being Extra-galactic objects we are targeting. 
The Galactic candidate list contains a higher fraction of bogus alerts but still $<40\%$.
As for the ``Good" objects in the Galactic eyeball list they were visually inspected and many were mislabeled CVs or interesting fast extragalactic transients which look similar.

Overall we consider our policies and how they interact with the models to be satisfactory. 
Based on our validation set, we can expect that over the course of a week 80.2\% of the incoming alerts will be auto-garbaged, 3.7\% will be left in purgatory and 16\% will be sent to human scanners for eyeballing. 
Of the 3.7\% left in purgatory during our tests 16 were labeled as ``Galactic" and 14 ``Good" objects. 
As mentioned above we visually checked those 14 objects and found no concerning outliers.

\subsection{In production performance}
\label{sec:prod}

\begin{figure}
\centering
\includegraphics[width=9cm]{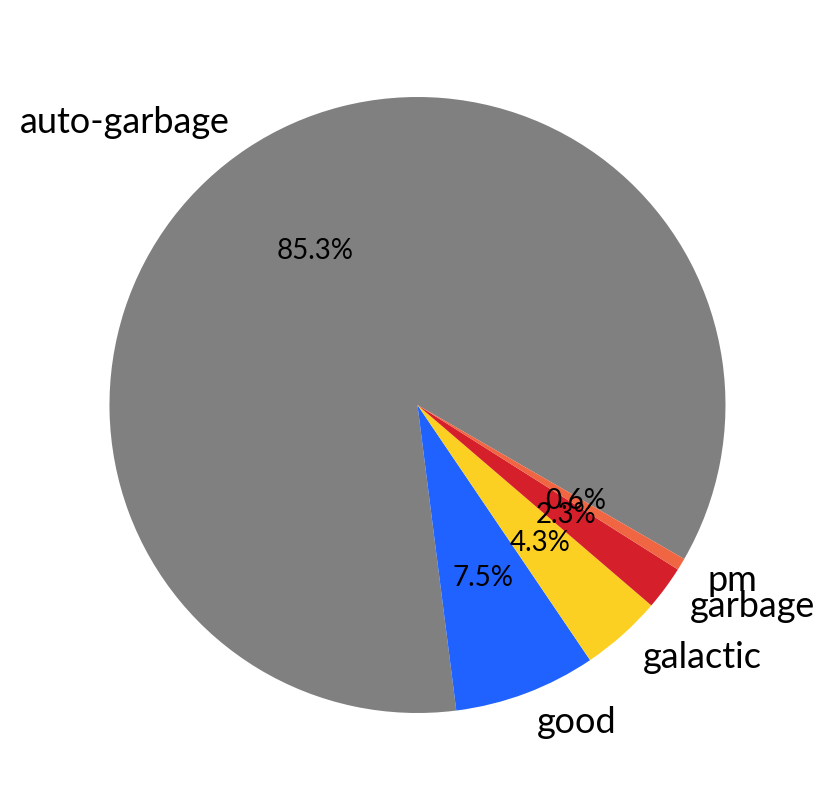}
\caption{Alert type distribution over the period  2025-04-04 to 2025-06-10. See the label description in Section \ref{sec:labels}.
\label{fig:prod}}
\end{figure}

We now present the performance of the VRA between 2025-04-04 and 2025-06-10, during which 16,938 alerts entered the eyeball list. As we can see in Figure \ref{fig:prod} the VRA auto-garbaged 85\% of the alerts over that period.
Our policy evaluation estimated that 80.2\% of objects would be handled by the VRA; the slightly better in-production performance is the result of acute hardware or weather events that are not taken into account in our tests. 
In this case some ATLAS units were subject to significant trailing in the images on the week starting 2025-05-23 leading to 7,516 alerts entering the eyeball list on that week alone. 
During such an event the VRA auto-garbages a higher fraction of alerts (91.5\% on that particular week), raising the average for the month.  

The TNS crossmatch to the garbage over this period found transients events miss-labeled as garbage but these were the result of human error. 
There were 164 ``potential misses" - alerts which did not meet the VRA score threshold initially but whose VRA score was raised to 10 when a cross-match to TNS was detected. 
Of these, 34 would not have risen above our threshold within the 15 day period covered by the models. 
In eyeballing these 34 events, we found that 11 were duplicates, 2 were CVs and 1 is suspected bogus. 
Of the 20 real events left, only 1 may have been the object of follow-up - SN2025hkm. 
The VRA Scores for that event rose and stabilized about 6.6 for over a week, still below our threshold likely due to a slightly low RB score (0.67).

This inidicates that future versions of the VRA could benefit from an additional feature that specifically counts the number of \textit{successive} detections, or potentially for objects with VRA scores above 6 for a few days in a row we could trigger the CNN again to update the RB score as the object is now brighter. 

We leave this for future enhancements; as the current potential loss rate is only 0.006\% of all the alerts entering the eyeball list, and 0.079\% of all the Extra Galactic events.

%
%

\section{Discussion}
\label{sec:disc}

\subsection{Raw Vs Forced photometry}
All of our light-curve features are based on the raw (difference) photometry measured on the night.
As a result we have to handle a mixture of detections and non-detections, and we are vulnerable to the effects of the waning and waxing moon (which affect our detection limits and can turn a detection into a non-detection).
An obvious solution to this problem would be to use the Forced Photometry light-curves instead. 
There is however a technical limitation and computational cost in production which makes this unfeasible at this stage.
The Forced Photometry is not calculated for ATLAS alerts entering the eyeball list unless they meet a specific quality criteria.
This is to ensure that the load on the compute servers result in the highest priority transients being processed fast and the forced photometry speed for those is not compromised. 
Consequently, at the point at which the VRA runs in the stream, forced photometry is not available in the majority of cases, and our algorithms are trained on the data readily available in production. 

We use this opportunity to highlight that this is an example of ``Data First" design in applied ML. 
The growing literature on ML in astronomy often suffers from a ``Model First" approach, where models are applied to a problem with no clear benchmark of success and the data provided to the algorithms is optimistic (if not at times unrealistic) compared to what we would expect in real life setting. 
This is where proofs of concepts can fail to lead to practical science solutions, because the limitations of the data ignored at the design stage are not easily overcome. 

Focusing back to the case of the VRA, its primary job is to reduce the false positive rate in the data flow without causing time delays or loss of opportunity. A next iteration would be to explore additional features that harness the forced photometry in the \textit{day $N$} models since 70\% of the stream is cleaned on day 1, lowering the cost to compute the forced photometry. 
This would allow analytic fits, extracting gradients and model comparison.

\subsection{The Fast axis}
Related to the discussion of using unforced photometry of detections or forced photometry at a known position, is that of scoring alerts on the Fast Axis imagined in Figure \ref{fig:score_space}.
One way to create transient agnostic scoring for this axis would be to evaluate the gradient of the light curve and apply a normalization factor. 
Without the forced photometry, the calculation of this gradient is compromised by the uncertainty introduced by non-detections. 

There are other simpler alternatives such as using the {\tt max\_mag} feature or creating a boolean flag where all alerts with day 1 magnitude below, say $m = 16$ are considered fast. 
These methods are strongly biased towards the Cataclysmic Variable population but could highlight nearby FBOTs such as AT2018cow, or shock breakout events. AT2018cow was 
discovered and recognised due to its very rapid rise to bright absolute magnitude 
\citep{2018ATel11727....1S,2018ApJ...865L...3P}.

On the whole since what is considered ``fast" is a less objective classification as real or galactic, and since creating a useful transient agnostic scoring method would not be feasible with the data available at the time of scoring, we omitted the Fast axis from our score space \textit{for this specific use case}.
In larger streams (e.g. LSST or future surveys) having an additional dimension to rank and prioritize alerts could provide sufficient benefits to justify adding complexity to the ranking method.

\subsection{Feature importance}

In Figures \ref{fig:permpinp_day1} and \ref{fig:permpinp_dayN} we showed the permutation importance of our features for the \textit{day 1} and day $N$ models. 
In this section we highlight a few take aways from this analysis\footnote{Although note that looking at individual features is an incomplete form of interpretation because the models use combinations of features to make decisions. }.

The first notable feature is the {\tt rb\_pix} (RB score).
As expected it has the largest influence on the $p_{\rm real}$ score and, perhaps surprisingly, the second (third) largest on the $p_{\rm gal}$ scores of the day 1 (day  $N$) models.
The relation between {\tt rb\_pix} and alert type is clear when we look at the {\tt rb\_pix} distribution by types (see Figure \ref{fig:rbpix_distrib}).

Another key feature is the extinction ({\tt ebv\_sfd}) which is the most predictive feature for the $p_{\rm gal}$ scores. This is not surprising since high extinction is correlated with the plane of the galaxy where more galactic events may occur, and makes extra-galactic transients fainter and less likely to be observed. 

Since a correlation with the galactic plane is important in determining whether an alert is galactic or extra-galactic, it may seem surprising that we do not use galactic coordinates for our scoring algorithm. 
In fact in earlier prototypes of the VRA we did test using the galactic latitude as a feature instead of RA and Dec but found that our $p_{\rm gal}$ classifier performed much worse (see \citealt{stevance_2025_14944209}).
There are two reasons for this: As we can see in Figure \ref{fig:RADEC} there is a strong correlation between the galactic plane and the bogus alerts; then there is the effect of the Galactic center, which can only be accounted for with the 2D coordinates (Ra and Dec or galactic latitude and longitude). 
At the time of those tests the conclusion was to keep Ra and Dec coordinates as they are already in the stream, and try including $E(B-V)$ as a feature which had not yet been tested and was then found to be very informative. 

The final features we will discuss are those related to redshift, {\tt z} and {\tt photoz} as we find the results from the permutation importance surprising. 
We expected {\tt z} to be consistently more important than {\tt photoz} since spectroscopic measurements of redshift are more reliable than photometric measurements. 
It could have been due to the larger availability of photometric redshift measurements, however we find on our data that 2905 samples have spectroscopic measurements, and 2748 have a photo z, so greater availability of one measurement over the other is ruled out as a cause.
We do not have a firm explanation to explain this discrepancy but we put forward a few hypetheses which could be tested in a future version. 
A first possibility is that the value of redshift is less important than the fact that there is a redshift at all - we could test this by turning the {\tt photoz} and {\tt z} features into boolean flags (individually and then together). 
A second possibility is that the relevance of the redshift split points in the decision trees are minimally affected by the errors on the redshift. Testing this is more difficult; we could try artificially adding noise to the {\tt photoz} and {\tt z} features and see how this affects their position in the permutation analysis.

\subsection{Choosing the policies}

Calculating the VRA score, galactic flag and applying our eyeballing and garbaging policies requires seven values to be set: Two scalar values $f$ for Equation \ref{eq:vra_rank} (one used when calculating the VRA score ($f=0.5$), one used when calculating the galactic flag ($f=0.9$)); an extragalactic candidate eyeballing threshold ($>7$); a distance to the (1,1) coordinate in score space to set the galactic flag to True ($<0.4$); currently three auto-garbaging policies for the first, second and third visit.

For our calculation of the VRA$_{\rm score}$ in this version of the VRA we tested $f$ values ranging from 0.4 to 1\footnote{Tests on earlier VRA prototypes with $f={0.1, 0.25, 0.5, 0.75, 1.0}$ can be found in \citealt{stevance_2025_14944209}}. We found that values of 0.4, 0.5 and 0.6 gave very similar results and all had an AuRaK of 0.951\footnote{See code release \citep{stevance_2025_14906192}}.
We chose $f=0.5$  because it has the convenient interpretation of weighing the ``Real" axis twice as much as the ``Galactic" axis.

To calculate the galactic flag, the choice of $f=0.9$ and distance $<0.4$ \textit{did not} go through a systematic search. Instead, the values were chosen to conservatively cover the galactic distribution without encroaching too much on the bogus distributions based on the visualisations in Figure \ref{fig:ss_withranks}.
One could create a larger grid search for the extra-galactic and galactic policy values and rerun all our policy diagnostics to optimise these values. 
This was not done because from the earlier grid searches we noted that optimised parameters are usually very close to ones chosen by visual inspection of the plots, and given current performance such a systematic search of parameters was not considered necessary. 
In the future this may take place as we review live performance of the current VRA over the next few months. 

Finally the auto-garbaging policies were created during the first live implementation of the VRA in August 2024. 
The in-production VRA scores were recorded for all alerts in the eyeball list and the objects were eyeballed as usual. 
This provided us with a first real test set, and the distribution of the VRA scores in this set was used to establish conservative eyeballing policies one after the other. 
In future iterations of the VRA the garbaging logic remained the same but the values changed slightly based on the policy evaluation presented in Section \ref{sec:policy_eval}.
For a history of the garbaging policies see the Technical Manual \citep{stevance_2025_14944209}.

\section{Summary and Conclusions}
\label{sec:concl}
The ATLAS Virtual Research Assistant is a bot which performs preliminary eyeballing to rank and prioritise alerts for the human eyeballers. 
It has reduced eyeballing workload by 85\% with no loss of follow-up opportunity.

It uses Histogram Based Gradient Boosted Classifiers to predict a ``Real" ($p_{\rm real}$) and ``Galactic" score  ($p_{\rm gal}$) for each alert and the scores are updated after each new visit by the ATLAS telescopes, up to 15 days after first entering the eyeball list.
The $p_{\rm real}$ and  $p_{\rm gal}$ values are then used to calculate the VRA score (see Equation \ref{eq:vra_rank}) which ranges from 0 (bogus) to 10 (real and extra-galactic).
We also calculate a ``Galactic" flag based on the distance to the $p_{\rm real}=1, p_{\rm gal}=1$ coordinate. 
Auto-garbaging policies are applied to remove the alerts most likely to be bogus from the eyeball list, and a VRA score threshold is used to select the alerts to be visually inspected by our team.

VRAs with similar strategies could be very useful to other sky surveys such as GOTO or BlackGem \citep{blackgem} to limit the reliance on citizen scientists or offer volunteers classification tasks that are more rewarding and engaging. 
Although the ATLAS VRA \textit{should not be run} as is on data from another survey, it could be worth exploring transfer learning techniques \citep{domingezSanchez2019,gupta2025} so that the VRAs of other teams can be trained using smaller training sets.
There may be issues with the differences in survey cadence and magnitude limits affecting the lightcurve features, and the strong dependence on the {\tt rb\_pix} value which is specific to our real-bogus classifier. 
If these effects render transfer-learning impossible, the advantage of the VRA design is that a relatively small sample (a few thousand) can provide good results in production. 

The success of the ATLAS VRA demonstrates that our field has not fully leveraged the potential of  feature-based machine learning methods, and we encourage our colleagues to not dismiss these without experimentation as they provide several advantages.
First, they can be trained with only a few thousand (sometimes a few hundred) samples which means that we did not have to rely on synthetic data.
Additionally feature-based methods are easier to interpret and provide us with a direct way to inject our expertise into the models (for a discussion see Appendix \ref{sec:imkeepingitstephen}).
(H)GBDT in particular have native support for categorical features and null values (without imputing).

Finally, the performance of the VRA has allowed us to introduce (since December 2024) an automated trigger mechanism for the 1-m Lesidi Telescope and the Mookodi instrument \citep{worters2016, erasmus2024}, as part of the South African Astronomical Observatory's “Intelligent Observatory" \citep{potter2024, erasmus2024spie}. 
Automated triggers have already resulted in classification (e.g. SN 2025arc \footnote{\url{https://www.wis-tns.org/object/2025arc}}), and our criteria are still being refined to increase the number of eligible alerts whilst minimizing unnecessary trigger. 
These tests are important precursors to the automated triggering system that need to be deployed on the LSST stream to shorten follow-up latency on instruments such as SOXs \citep{soxs}.

Our next focus will be to adapt the ATLAS VRA to data brokers such as Lasair and Fink \citep{lasair2024, fink2021}. We welcome discussions and collaboration from other survey teams should they wish to use our design to curate their data stream.


\begin{acknowledgments}
We are grateful to the referee for their helpful comments and suggestions. 
HFS is supported by Schmidt Science. 
KWS and SJS are supported by the Royal Society.
This work has made use of data from the Asteroid Terrestrial-impact Last Alert System (ATLAS) project. The Asteroid Terrestrial-impact Last Alert System (ATLAS) project is primarily funded to search for near earth asteroids through NASA grants NN12AR55G, 80NSSC18K0284, and 80NSSC18K1575; byproducts of the NEO search include images and catalogs from the survey area. This work was partially funded by Kepler/K2 grant J1944/80NSSC19K0112 and HST GO-15889, and STFC grants ST/T000198/1 and ST/S006109/1. The ATLAS science products have been made possible through the contributions of the University of Hawaii Institute for Astronomy, the Queen’s University Belfast, the Space Telescope Science Institute, the South African Astronomical Observatory, and The Millennium Institute of Astrophysics (MAS), Chile.
This work made use of observations made at the South African Astronomical Observatory (SAAO) which is supported by the South African National Research Foundation (NRF) .

\end{acknowledgments}

%

\vspace{5mm}
\facilities{ATLAS, SAAO}


\software{atlasapiclient \citep{atlaspapiclient},
          atlasvras \citep{heloise_2025_14983116},
          pandas \citep{reback2020pandas, mckinney-proc-scipy-2010},
          numpy \citep{harris2020array},
          sklearn \citep{sklearn_api},
          matplotlib \citep{Hunter:2007}
          }



\newpage
\appendix

\section{Feature Distributions}
\label{sec:app_feature_distribs}

In this appendix we show plots of the distributions of all our features split into five types -- ``Good", ``Galactic", ``Proper Motion", ``Garbage" and ``Auto-Garbage" (as previously defined) -- and add supplementary information that is not discussed in the main text. 

\subsection{Contextual Features}
\subsubsection{RB score from the CNN}
\begin{figure*}[ht!]
\centering
\includegraphics[width=12cm]{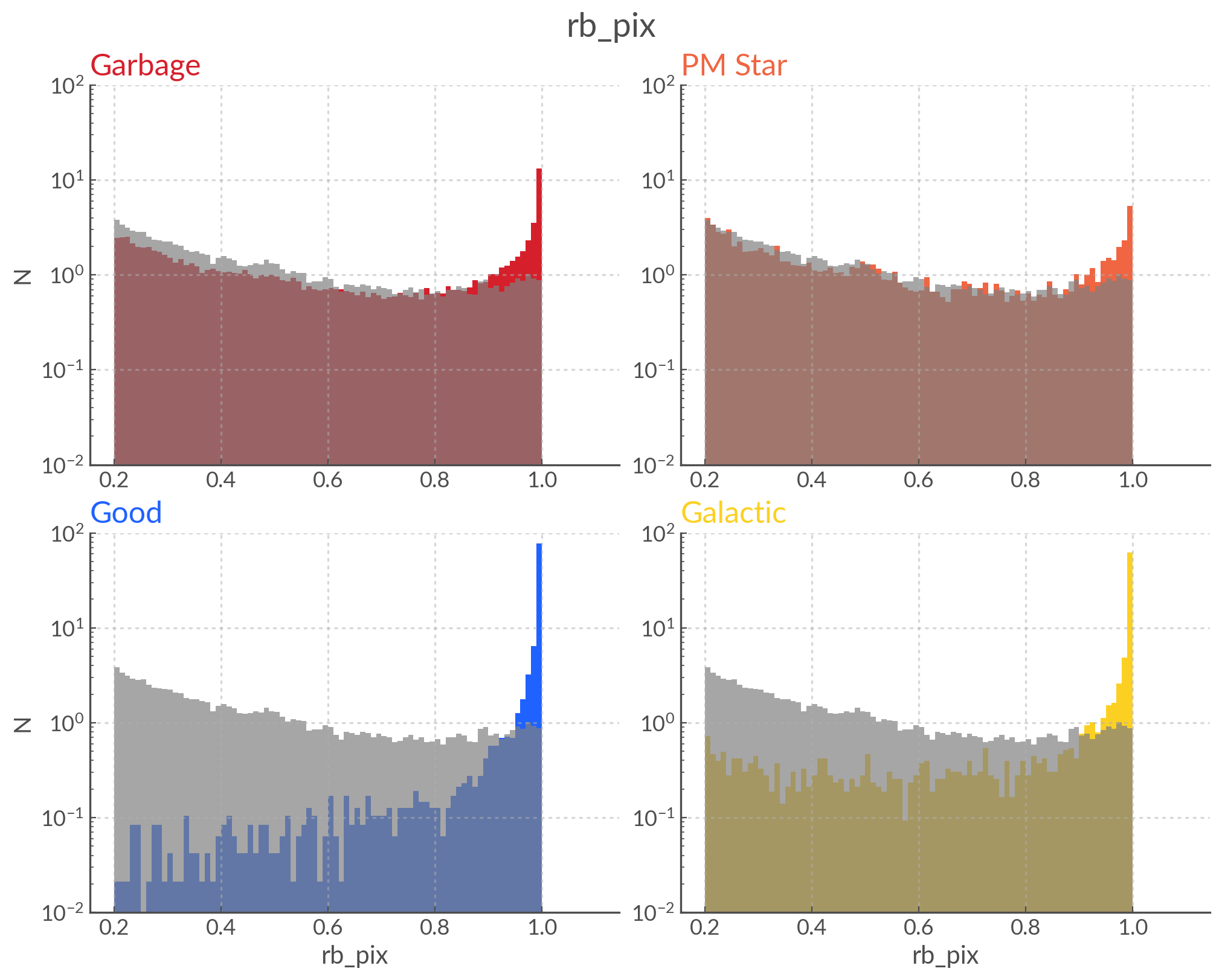}

\caption{Real Bogus score distribution splits by alert types. The Auto-garbage alerts are plotted in grey over each plot. Note that we logged the x-axis for better visualization. The features given to the scoring algorithms are not logged. Also note that the {\tt rb\_pix} feature distribution starts at 0.2 because we only use data that passed all previous up-stream cuts. 
\label{fig:rbpix_distrib}}
\end{figure*}

The real/bogus score from the CNN ({\tt rb\_pix} feature) is crucial predicting for the real and galactic scores (see Section \ref{sec:models}). 
In Figure \ref{fig:rbpix_distrib} we can see that the lower tail of the distribution is what distinguishes the ``Good" (and ``Galactic") alerts from the Bogus alerts (``Garbage" and ``PM"). 
Another notable characteristic highlighted in the figure is that the ``Auto-garbage" alerts do not have a pronounced spike in {\tt rb\_pix} at and round 1.
This is because alerts with a very high {\tt rb\_pix} value are unlikely to be meet the auto-garbaging policies as their overall VRA score will not be sufficiently low. 

\subsubsection{Right Ascension and Declination}

\begin{figure*}[ht!]
\centering
\includegraphics[width=7cm]{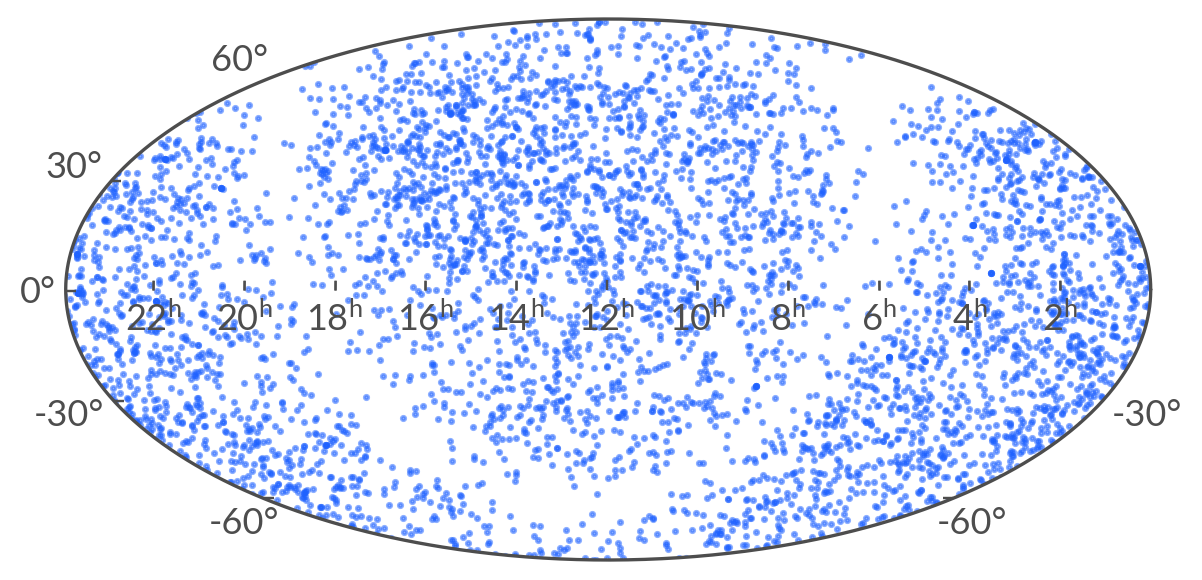}
\includegraphics[width=7cm]{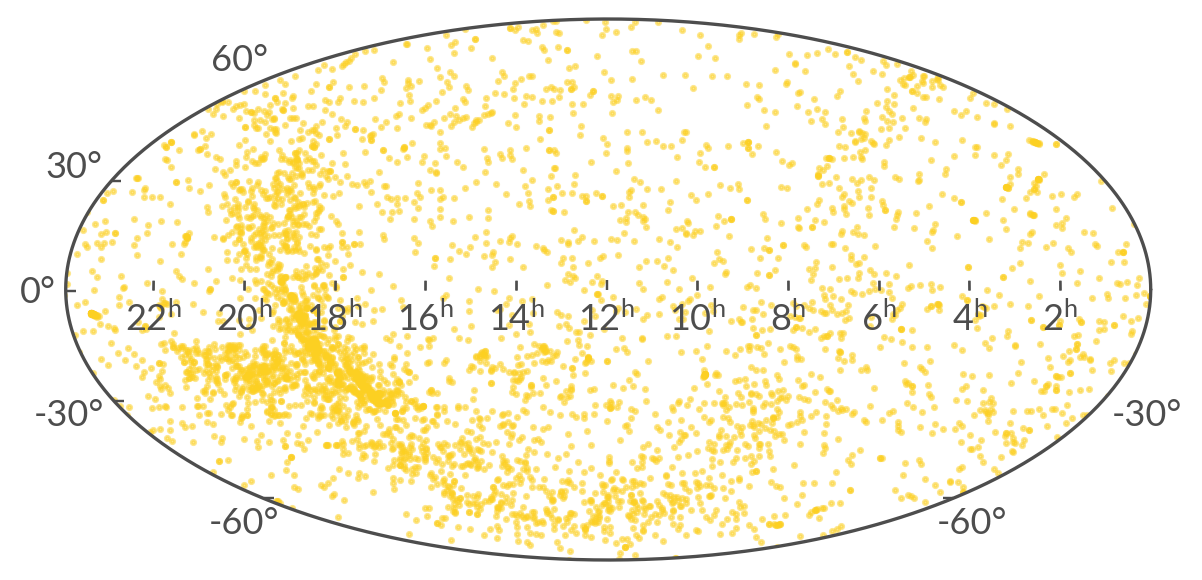}
\includegraphics[width=7cm]{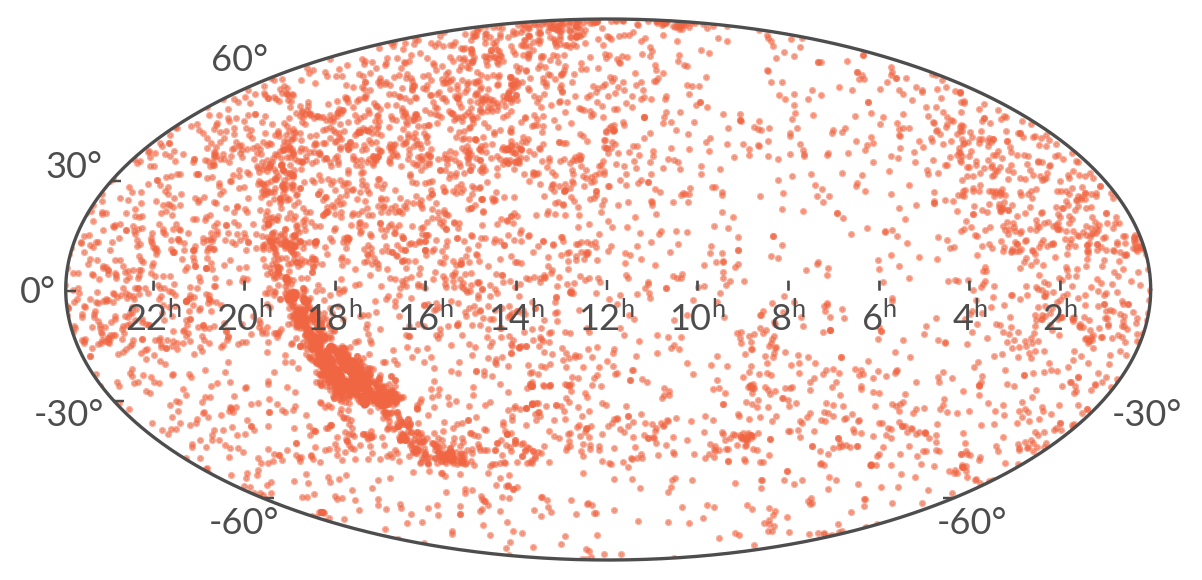}
\includegraphics[width=7cm]{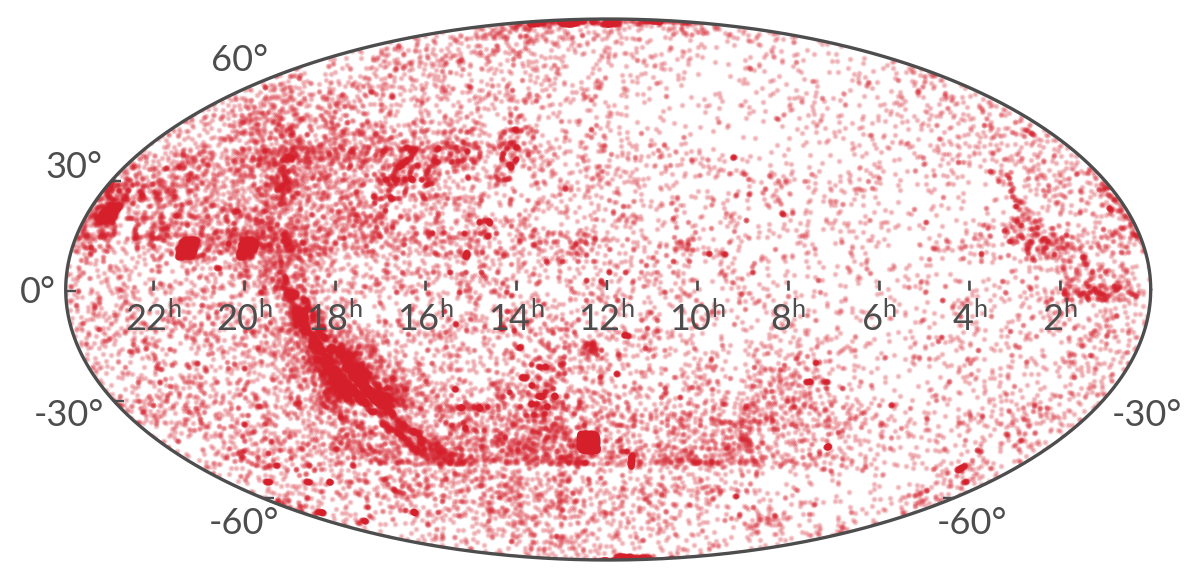}
\includegraphics[width=7cm]{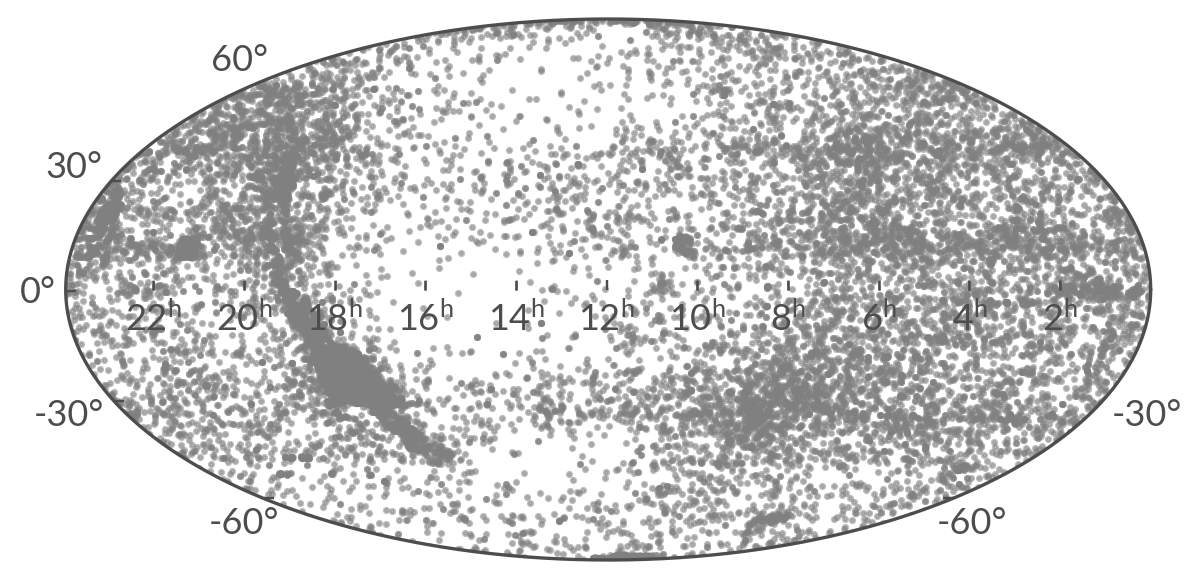}
\caption{Right Ascension (RA) and Declination (Dec) distribution of our data split by types: ``Good" (blue), ``Galactic" (yellow), ``PM" (orange), ``Garbage" (red), ``Auto-garbage" (grey).
\label{fig:RADEC}}
\end{figure*}

In Figure \ref{fig:RADEC} we show the RA and Dec distribution of our labeled data. 
We can see that the density of PM and Garbage objects is affected by the start of VRA operations in August 2024, after which fewer objects were labeled in this categories.
The alerts that would have been labeled as such by human scanners were for the most part auto-garbaged, which is noticeable in the auto-garbage distribution which is visibly denser for RAs observed after August 2024.

We can also see in the ``Garbage" and ``Auto-garbage" maps the effects of the ATLAS tiling pattern. 
This is a consequence of the higher incidence of bogus alerts on the edges of the ATLAS field of view. 
Although we have masks to reject alerts from regions that are known to cause problems, these masks are not perfect.
Since the tiling pattern is not visible in the map of the ``Good" alerts (as notably empty lines or patches of sky) we do not think this is will be an issue, and since introducing the VRA in August 2024 we have not noticed a pattern of missed objects associated with tiling.
Since we will continue to monitor potential misses (see Section \ref{sec:monitoring}) we will assess whether a tiling pattern emerges in our missed transients. 
If so we can mitigate this by resampling the ``Garbage" and ``Auto-garbage"  across the x,y position on the detectors (balancing the sample across Ra and Dec would erase the important correlation with the galactic plane). 

In the ``Garbage" and ``Auto-garbage"  (and PM to a lesser extent) distributions we can also see the Dec limits of the Northern and Southern units and where they overlap. 
As with the tiling pattern we do not think this is an issue but we will monitor in the long term.  


\subsubsection{Additional on-sky localization features}
\begin{figure*}[ht!]
\centering
\includegraphics[width=8.5cm]{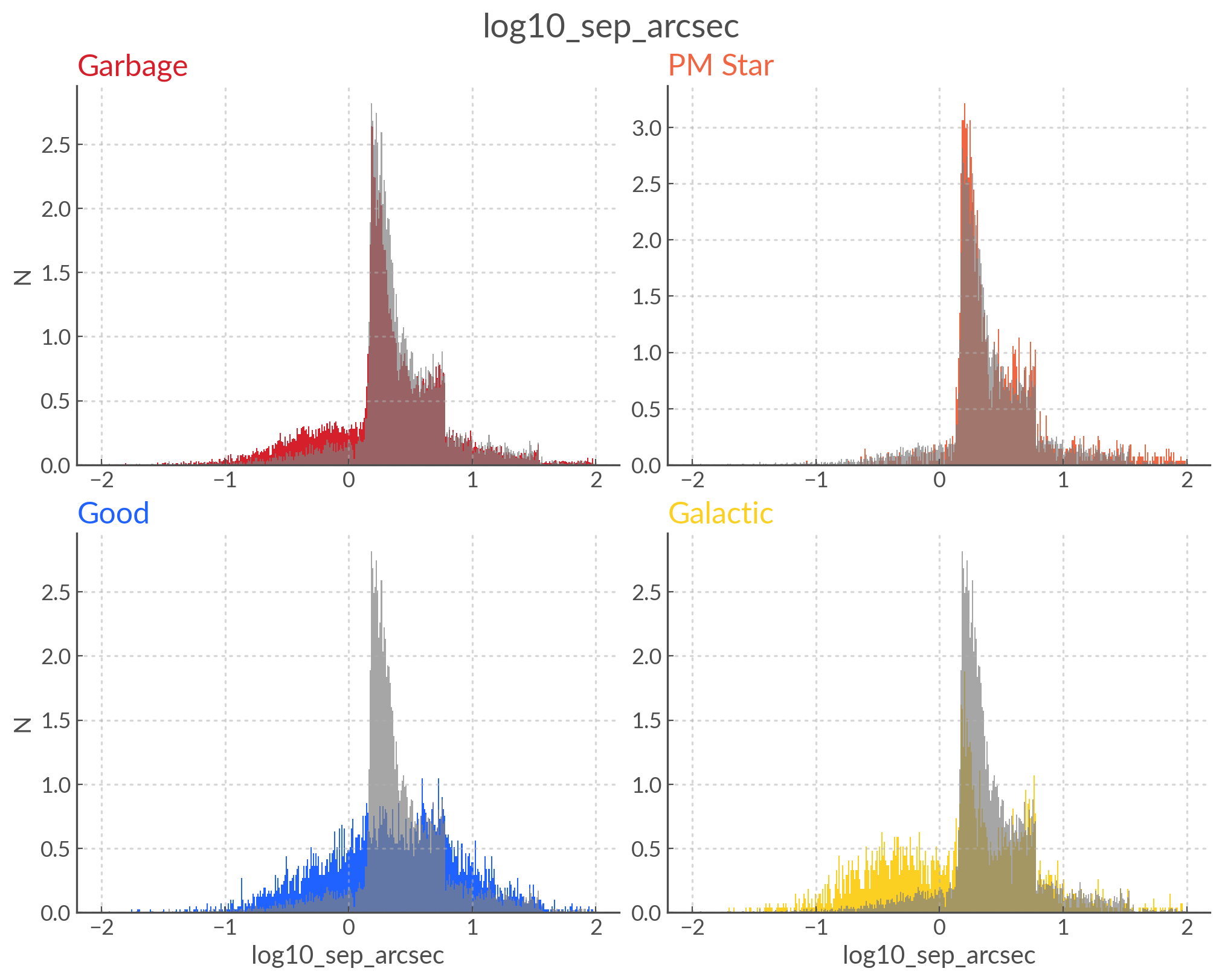}
\includegraphics[width=8.5cm]{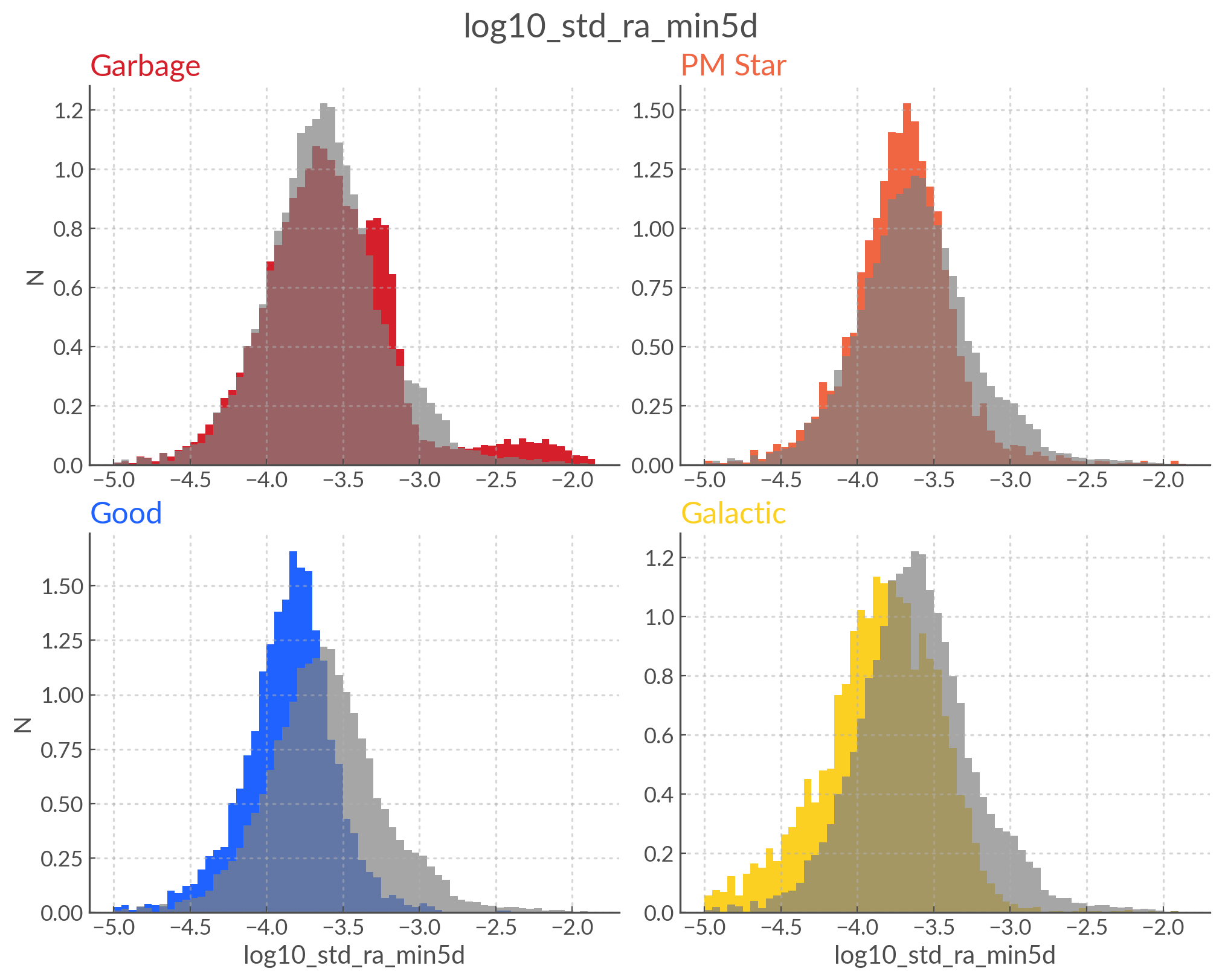}
\includegraphics[width=8.5cm]{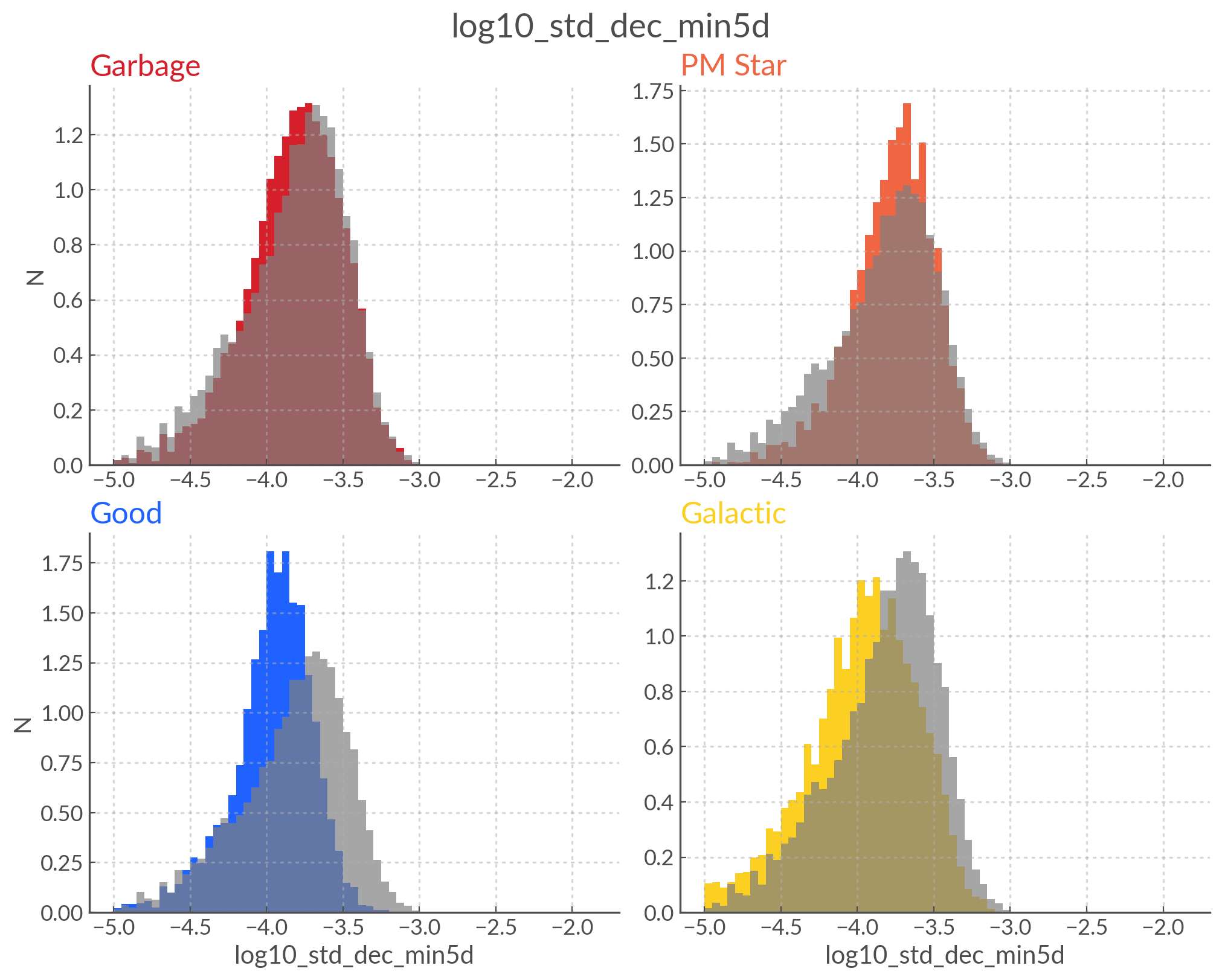}
\caption{Additional positional feature distributions. From left to right: The separation between the alert and the most nearby cross-matching catalog source; the standard deviation of the RA and Dec localizations of all detections recorded. All these features were logged (base 10). We plot separately the labels given by human scanners and show the Auto-garbage label distribution in grey over-top. The Auto-garbaged alerts are nearly exclusively Garbage and Proper Motion stars, which is reflected in how their distributions overlap the other four categories. 
\label{fig:radecscatter_separation}}
\end{figure*}

There are three other features related to the on-sky localization of each alerts (see Figure \ref{fig:radecscatter_separation}): the scatter in RA and Dec for the detections related to this alert, and the separation of the alert from the catalog source it is associated with (in arc seconds). The latter is provided by {\sc Sherlock} \citep{Young_sherlock_2023}.
All these features were logged (base 10) to increase their dynamic range. 

The separation can also be None when there is no viable catalog cross-match. 
Null values are given as such in our models since the chosen algorithm natively handles Null values by reserving one bin of the histogram for them. 
This is relevant particularly in the case of the separation feature as the Null value there represents what {\sc Sherlock} would flag as an Orphan detection. 

\subsubsection{Redshift}
\begin{figure*}[ht!]
\centering
\includegraphics[width=8.5cm]{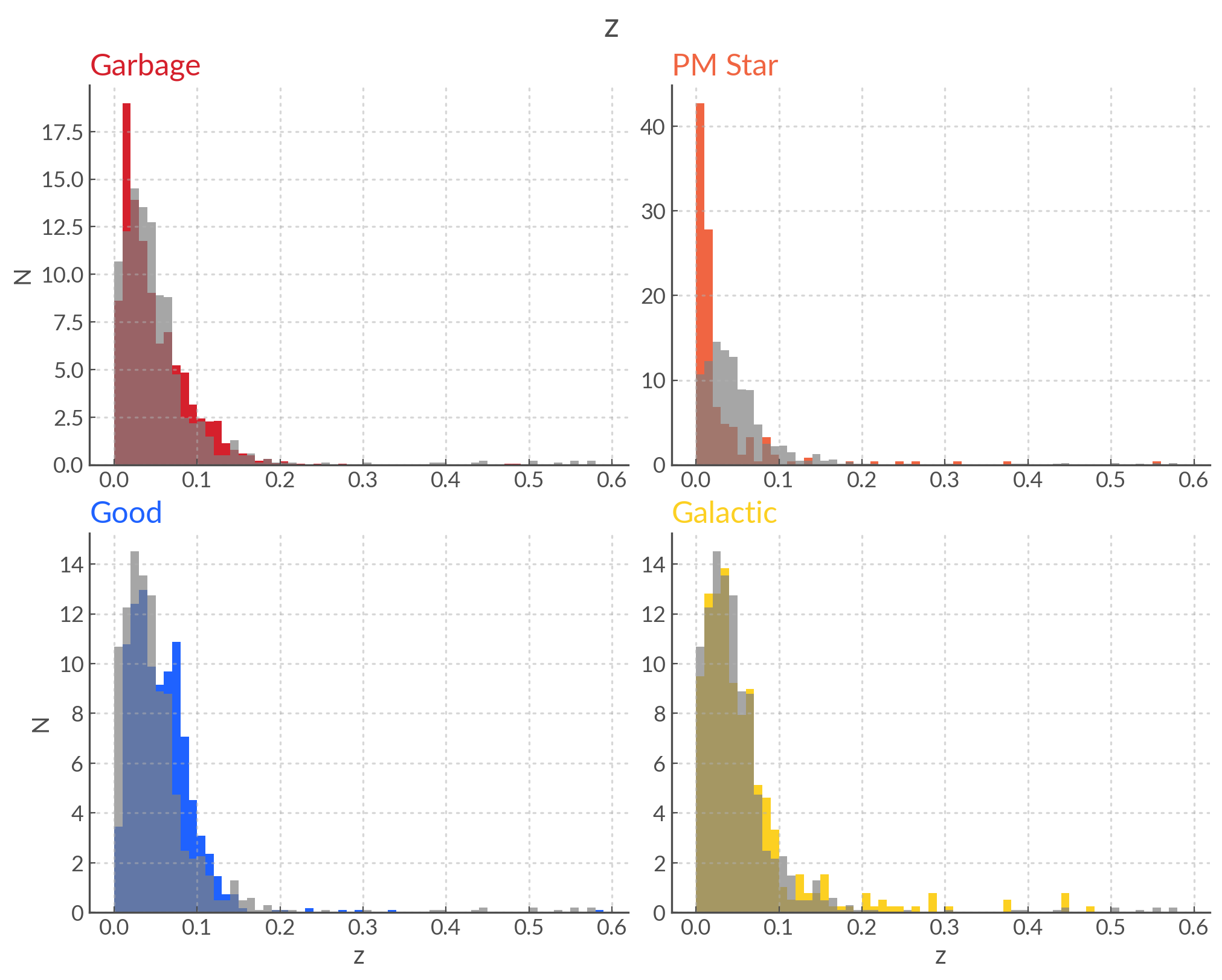}
\includegraphics[width=8.5cm]{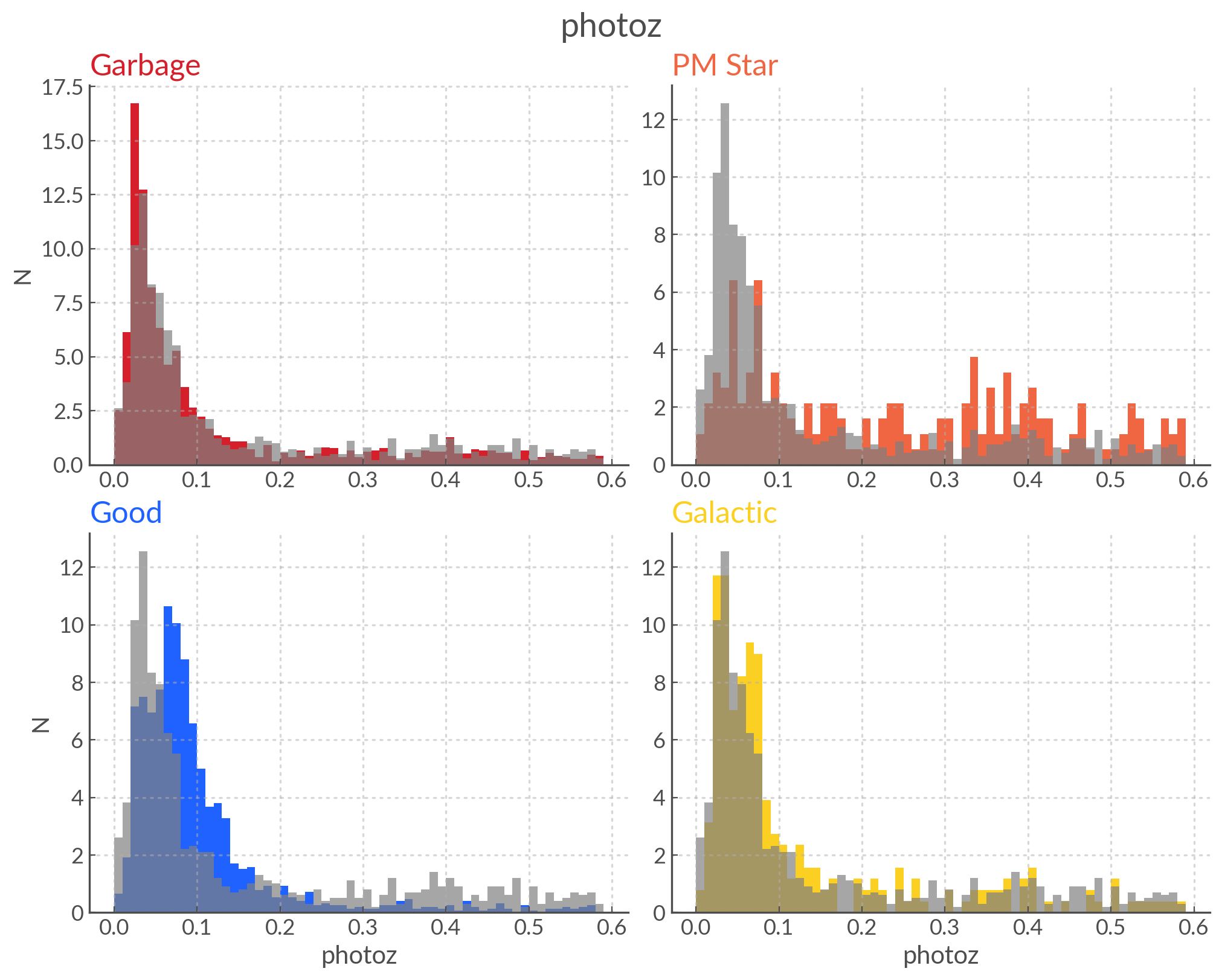}
\caption{Distribution of the redshift measurements across our alert types. We separate the spectroscopic (z) and photometric redshifts (photo z). We plot separately the labels given by human scanners and show the Auto-garbage label distribution in grey over-top.
\label{fig:redshift}}
\end{figure*}

The redshift information is known for 4,737 alerts or 13.7\% of our data set (2905 have a spectroscopic redshift, 2748 have a photometric redshift). 
This is provided by {\sc Sherlock} \citep{Young_sherlock_2023} if there is an associated catalog source and that source has a known redshift. 
There are 916 sources for which both the spectroscopic and photometric redshift are known. In Figure \ref{fig:redshift} we show the features distribution separated by alert type. 
We have not taken the extra step of combining spectroscopic and photometric redshift into a single column because we want to keep these two sources of information distinct. 
The spectroscopic redshift is much more reliable that the photometric redshift and although we cannot easily and formally ``tell" the models, by keeping these two pieces of information separate the decision trees can learn to use one over the other.

\subsubsection{Galactic Extinction: $E(B-V)$}
\begin{figure*}[ht!]
\centering
\includegraphics[width=8.5cm]{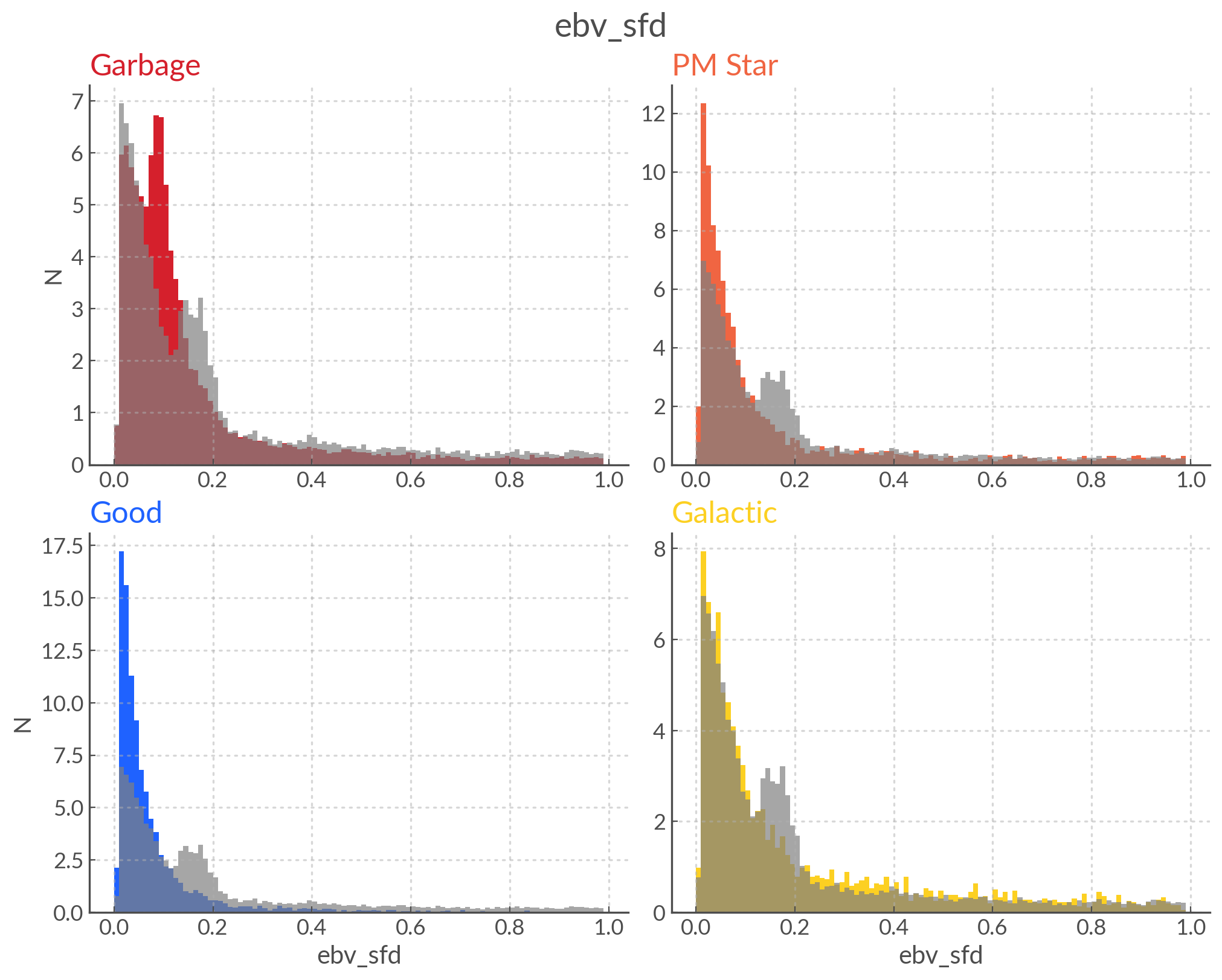}
\caption{ Distribution of the $E(B-V)$ feature for our different alert types. We plot separately the labels given by human scanners and show the Auto-garbage label distribution in grey over-top. 
\label{fig:ebv}}
\end{figure*}

The Galactic extinction feature is calculated using the {\tt dustmaps} python package by \cite{green2018} and selecting the \cite{SFD98} extinction maps.
In Figure \ref{fig:ebv} we show the $E(B-V)$ distribution separated by alert type. 
Unlike previous plots where the auto-garbage distribution showed a behaviour very similar to the garbage and proper motion labels, in this case there is a secondary peak around $E(B-V)$ values of 0.2 that is not shown in any of the other distributions. 
We interpret this behavior as follows:
There is a secondary peak in the Garbage $E(B-V)$ distribution around 0.1, however the fraction of Good and Galactic (therefore Real) objects with $E(B-V)$ $\approx 0.1$ is still significant. 
Therefore the Auto-garbage distribution does not exactly follow that of the Garbage alerts. 
The secondary peak remains but has a lower amplitude and moves to $\approx 0.2$ where the fraction of Good and Galactic alerts is lower. 

\subsection{Day 1 Light curve features}
\subsubsection{Long term history}
\begin{figure*}[ht!]
\centering
\includegraphics[width=8.5cm]{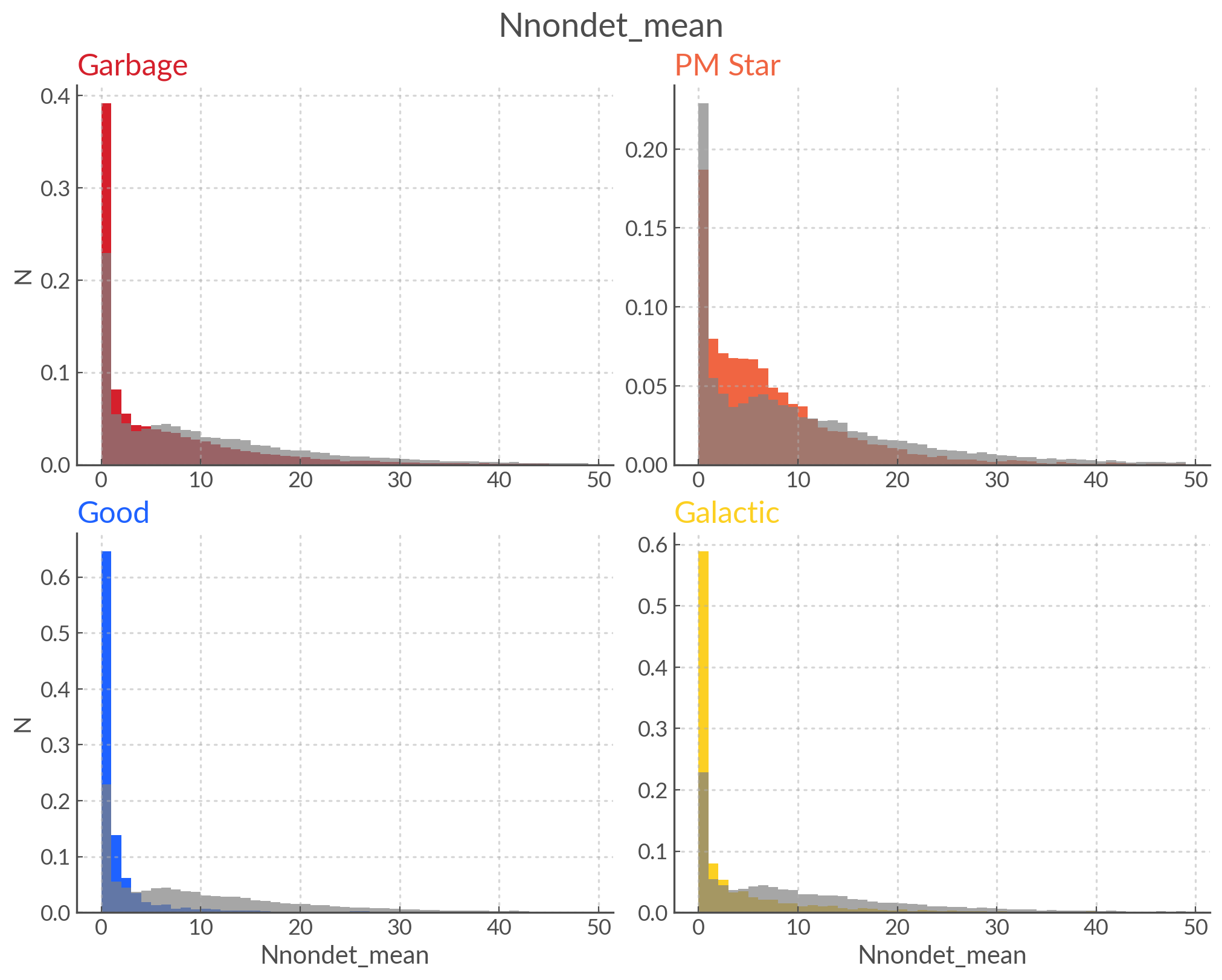}
\includegraphics[width=8.5cm]{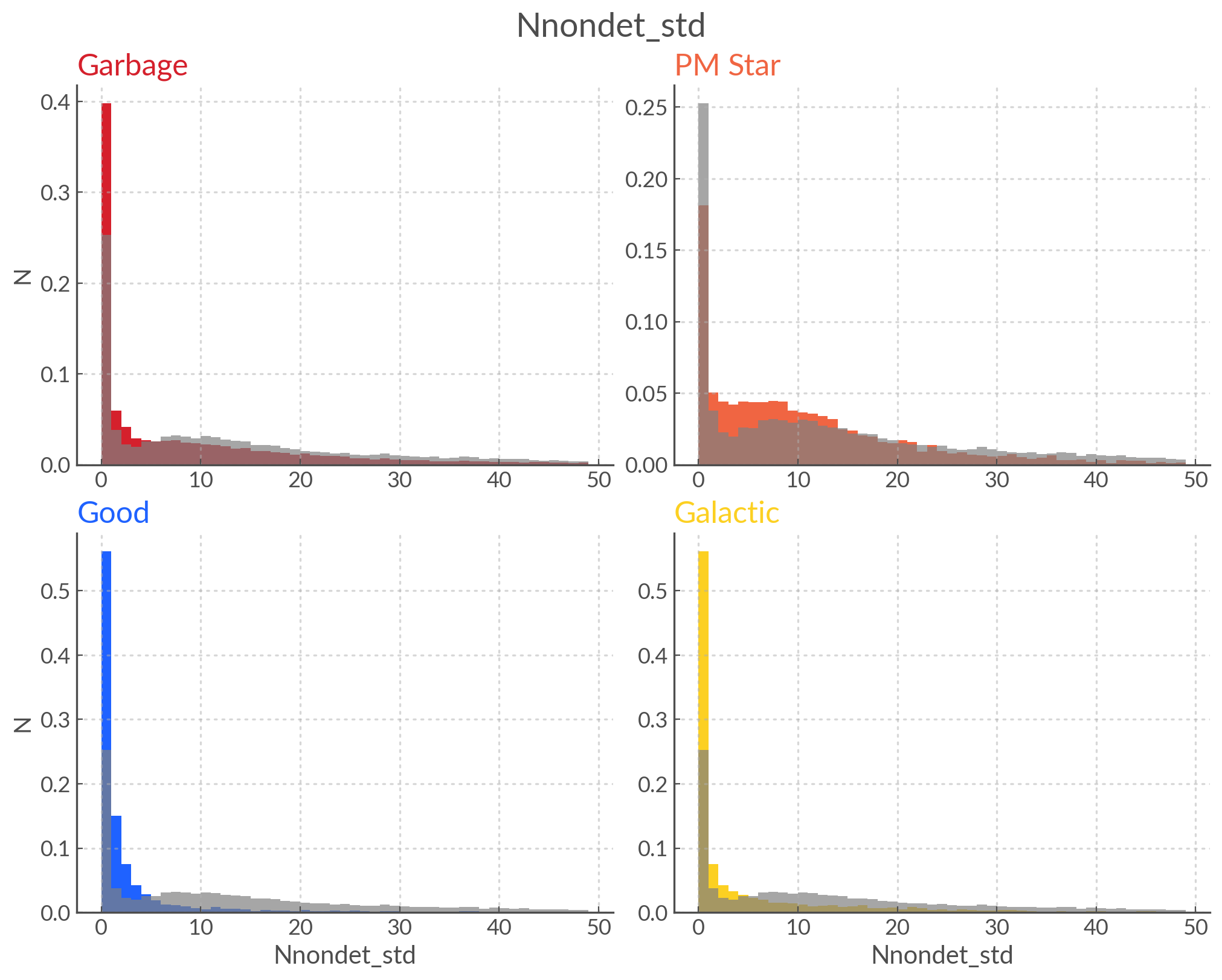}
\includegraphics[width=8.5cm]{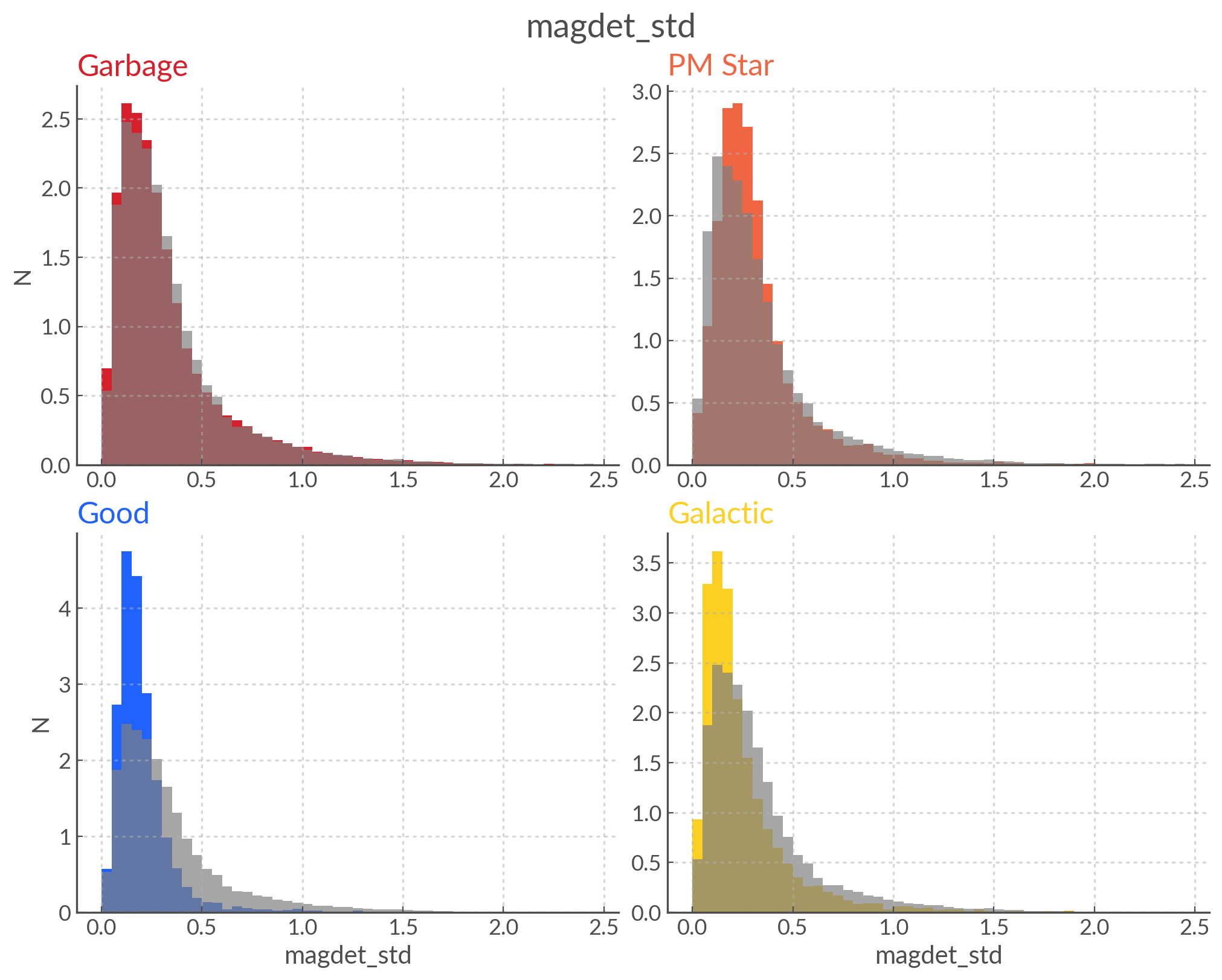}
\caption{Long term history (from -100 days with respect to first alert) features, from left to right: The mean and standard deviation of the number of non detections between each detection, and the standard deviation of the magnitude values of these detections. We plot separately the labels given by human scanners and show the Auto-garbage label distribution in grey over-top. 
\label{fig:long_lc_hist}}
\end{figure*}

For a description of how these features are calculated see Section \ref{sec:data} and Figure \ref{fig:lc_history}.
In Figure \ref{fig:long_lc_hist} we show the distributions of the three long term history features. 

\subsubsection{Short term history}
\begin{figure*}[ht!]
\centering
\includegraphics[width=8cm]{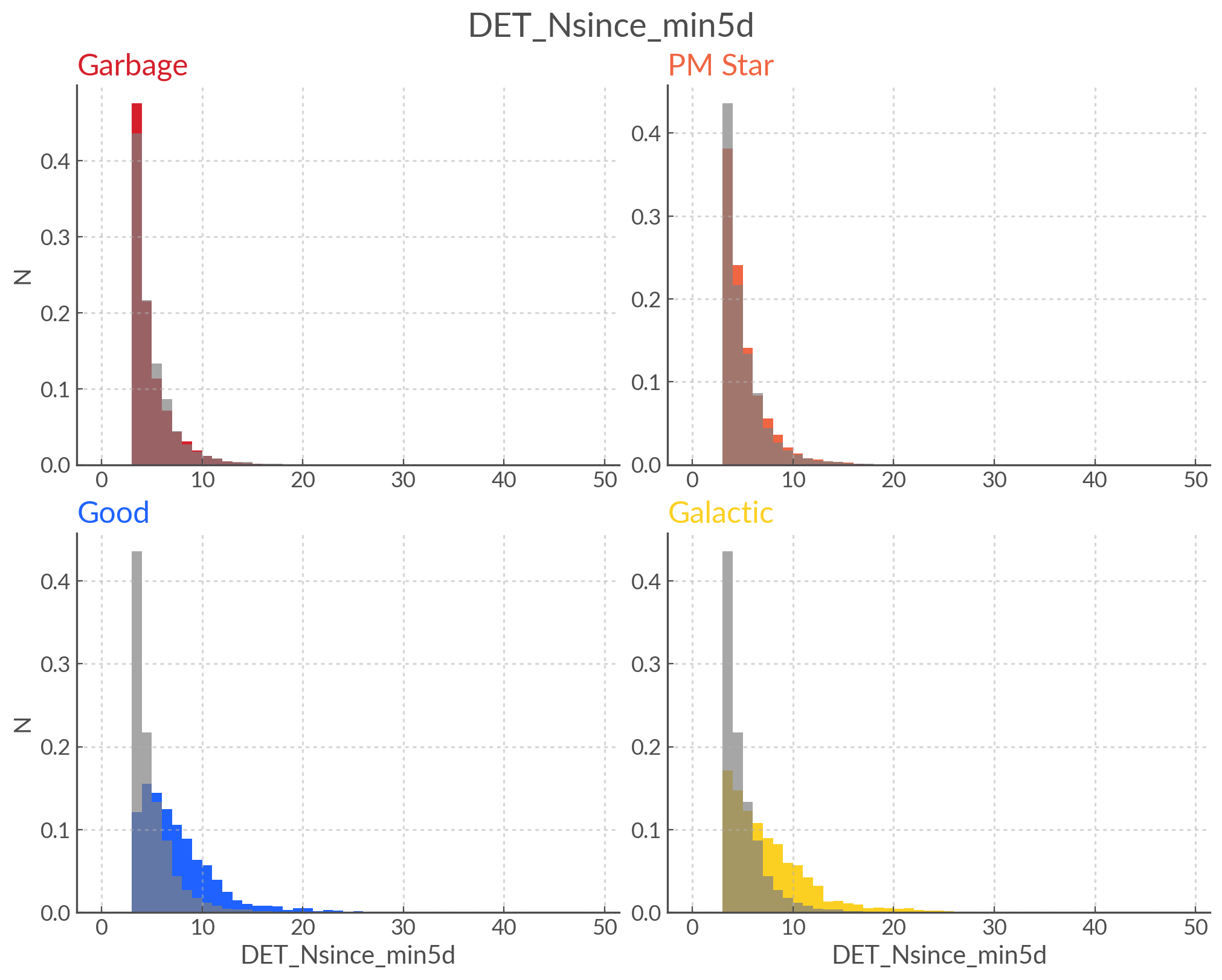}
\includegraphics[width=8cm]{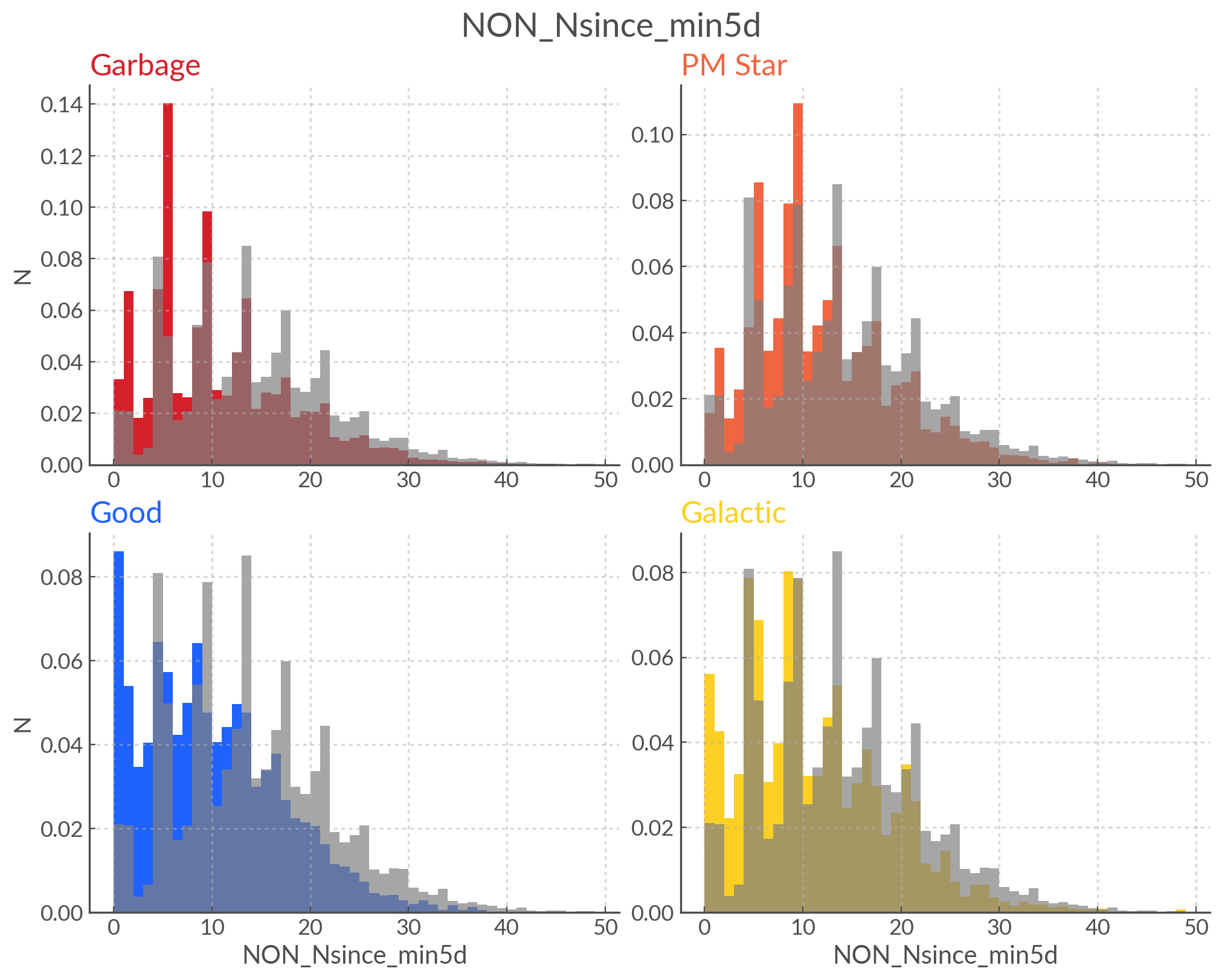}
\caption{Short term history features. We plot separately the labels given by human scanners and show the Auto-garbage label distribution in grey over-top. 
\label{fig:short_lc_history}}
\end{figure*}

See Section \ref{sec:data} for description and motivation. 
We show the distribution of the two short term light curve history features in Figure \ref{fig:short_lc_history}.

\newpage
\subsection{Day 1 features of AT2024lwd}
We only show the features that are most out of distribution for a ``Good" alert.

\begin{figure*}
\centering
\includegraphics[width=8cm]{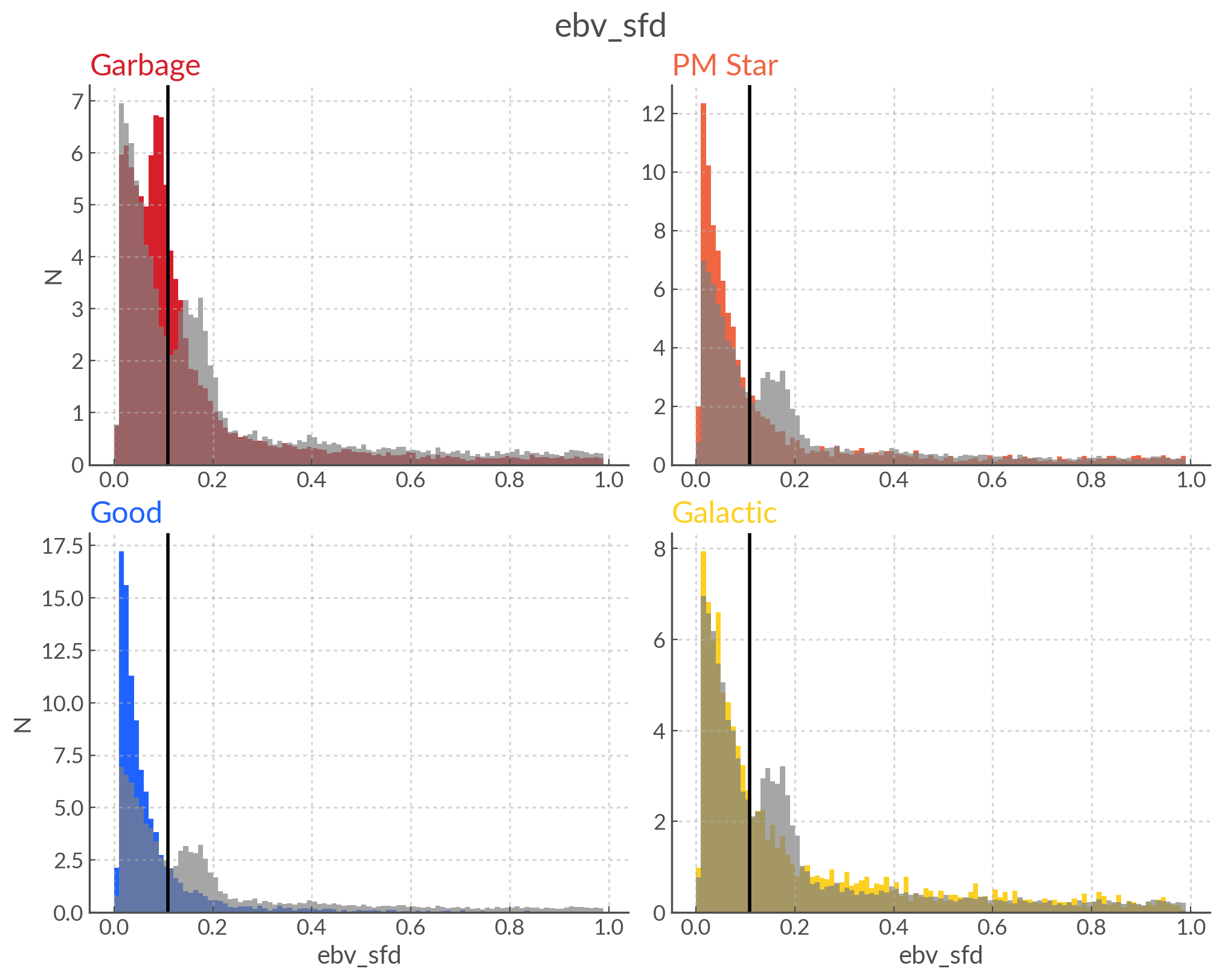}
\includegraphics[width=8cm]{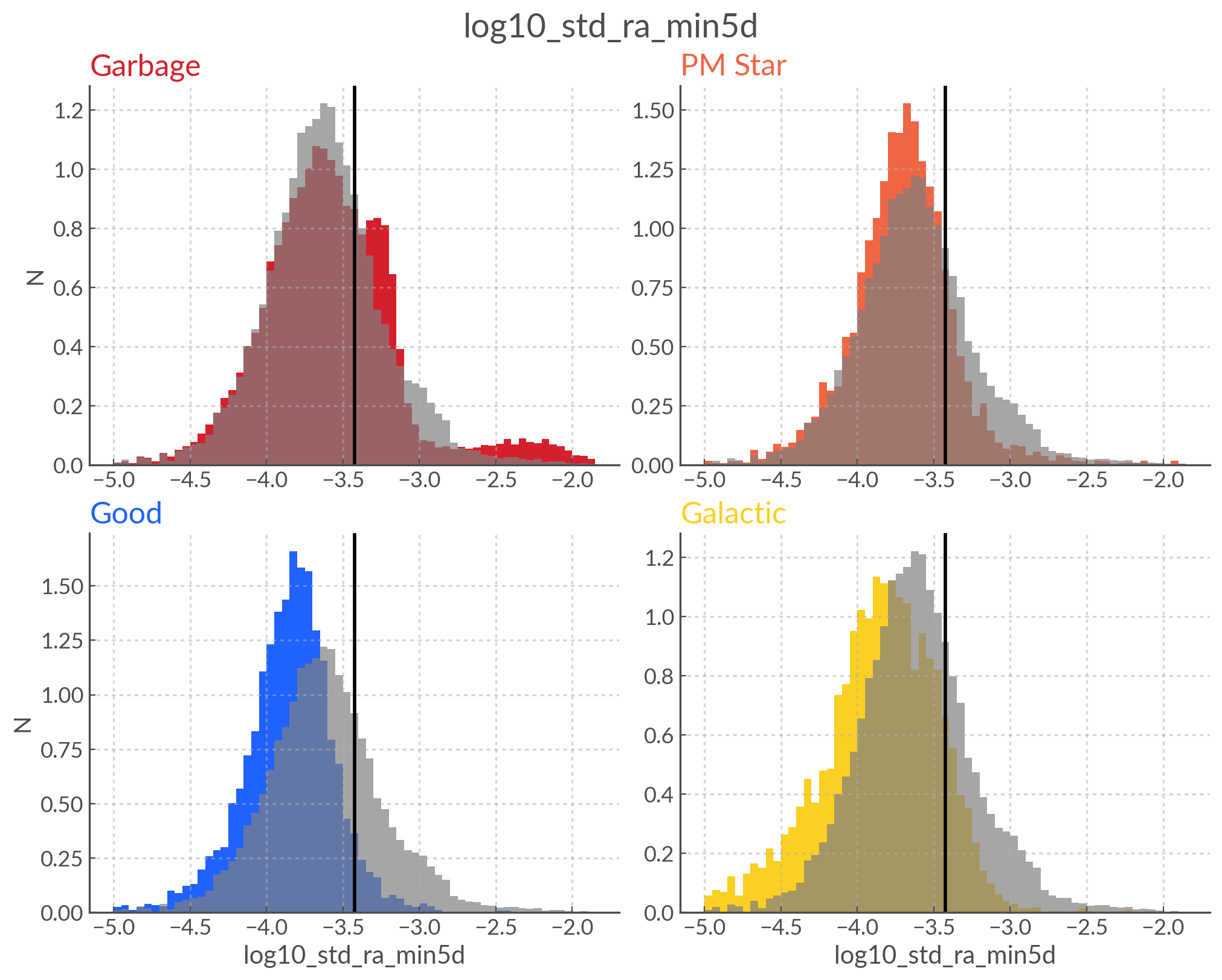}
\includegraphics[width=8cm]{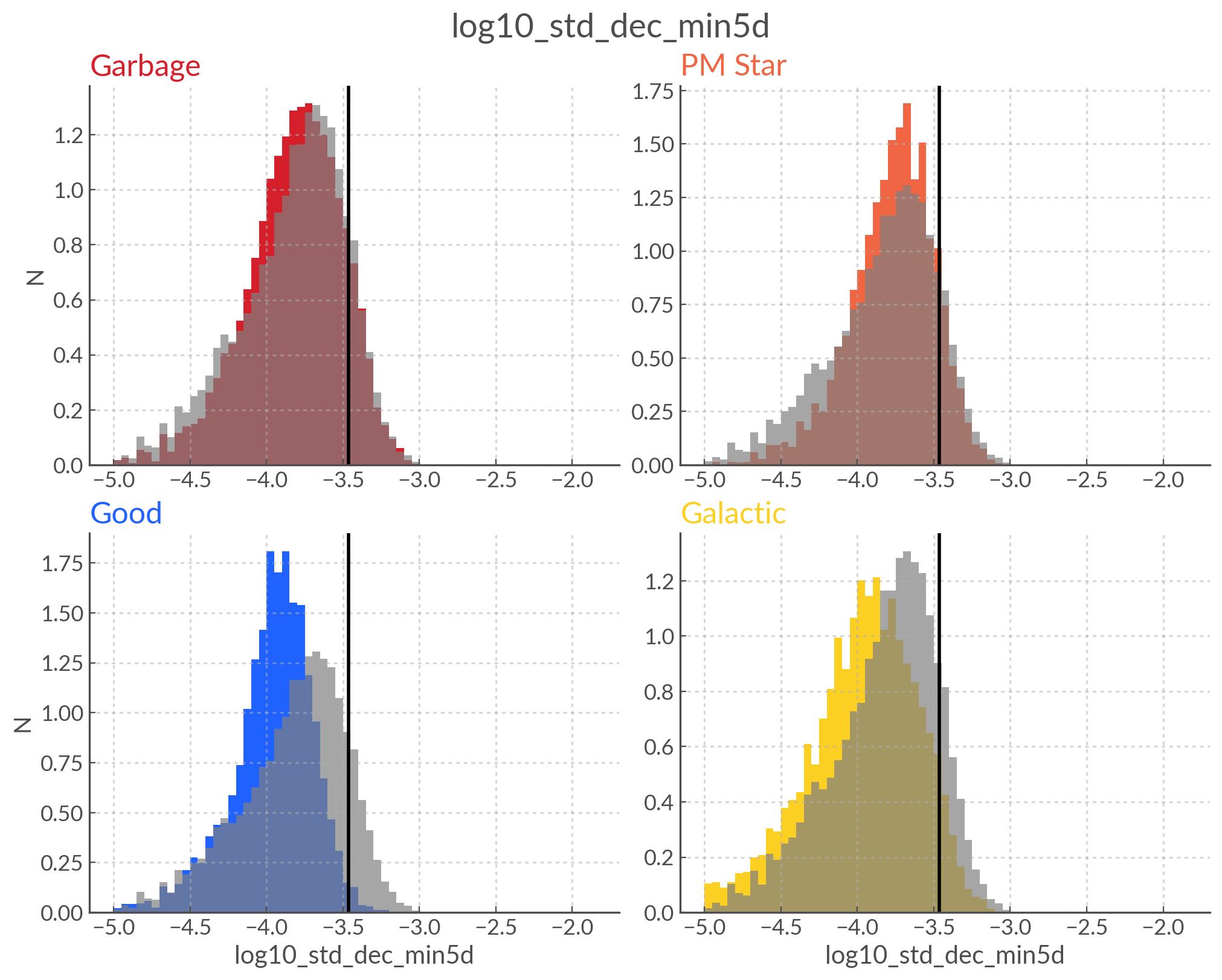}
\includegraphics[width=8cm]{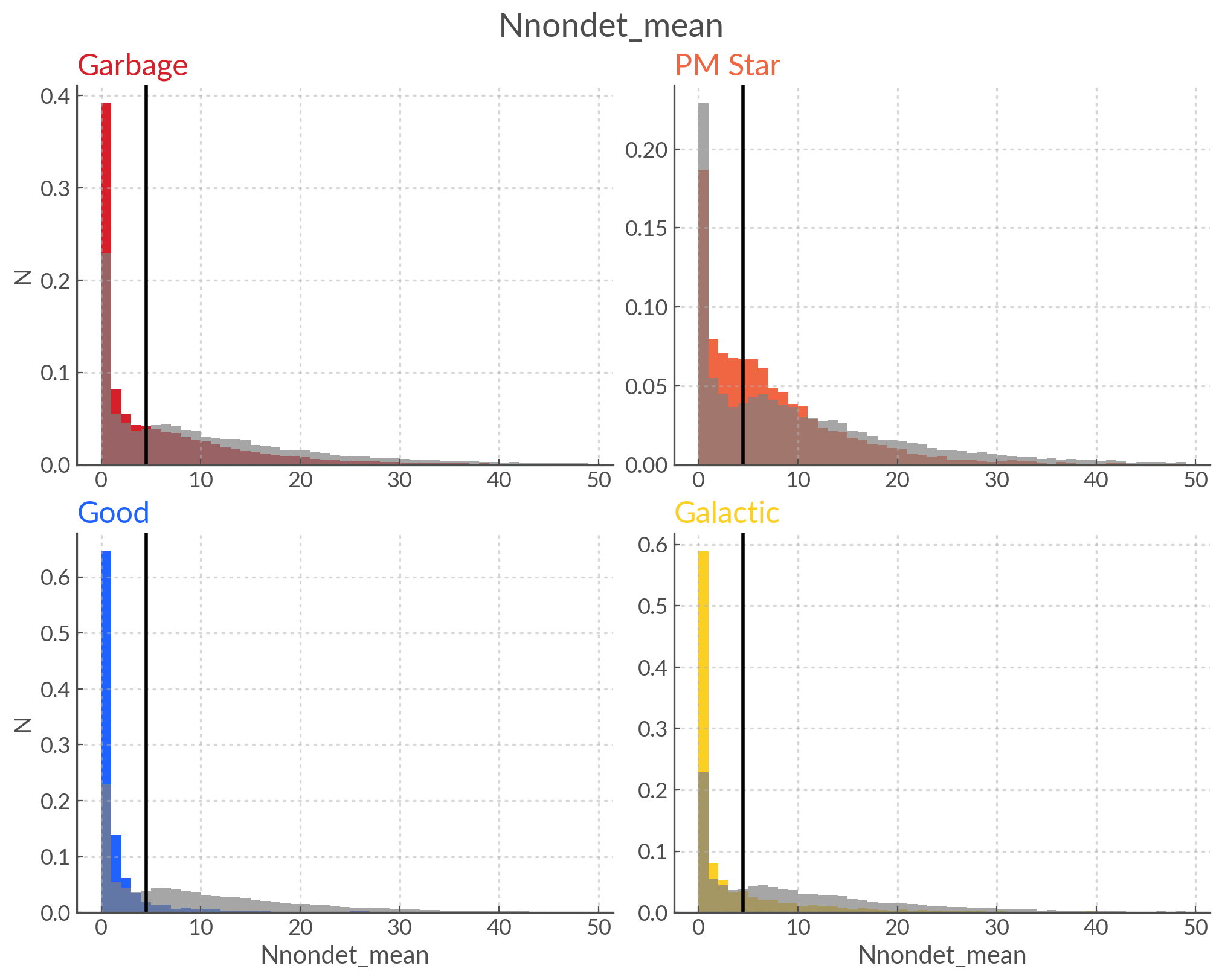}
\includegraphics[width=8cm]{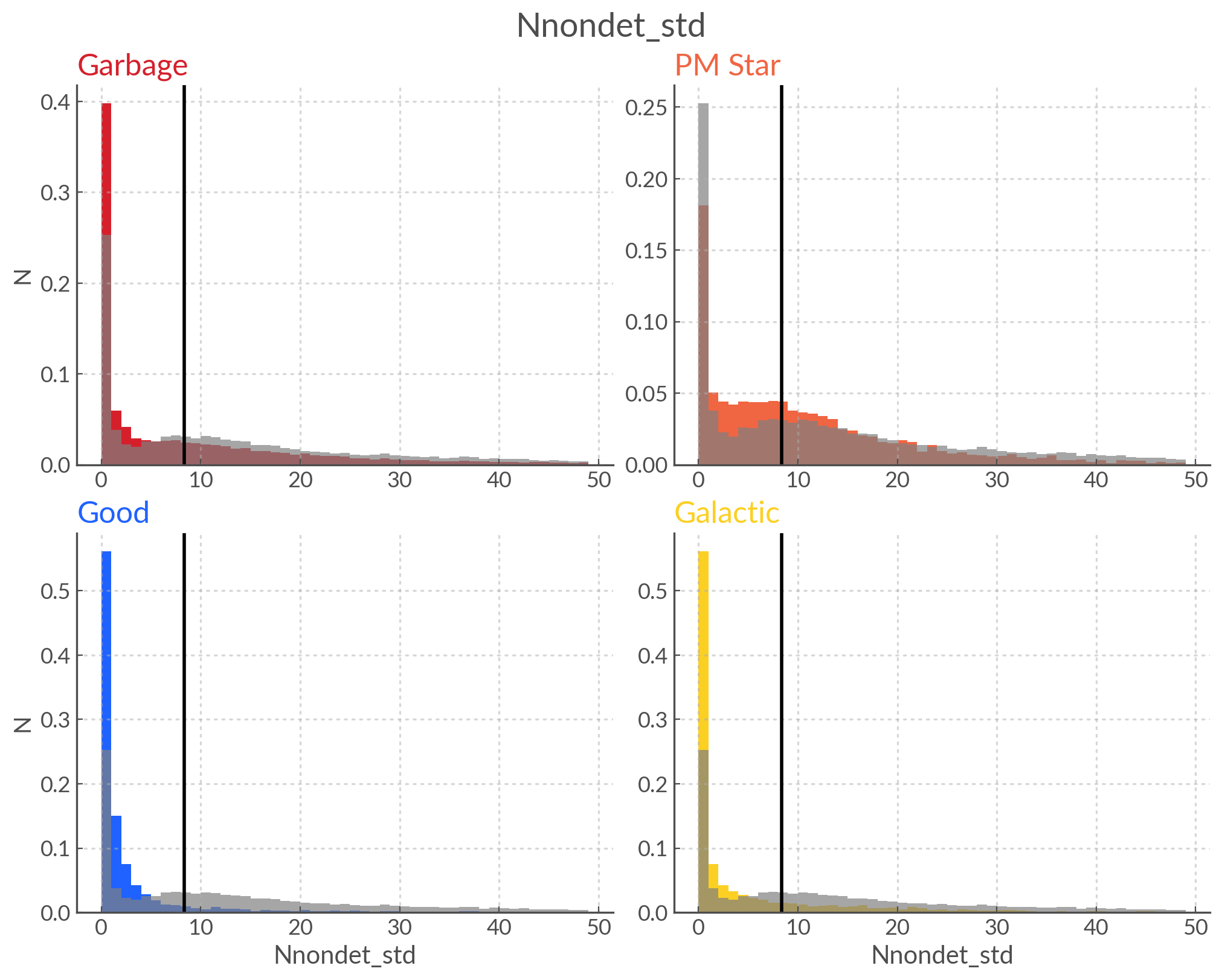}
\caption{The five day 1 features of AT2024lwd whose values are anomalous for a ``Good" object, shown on top of the distributions for these features split by alert types. The grey distribution is the ``Auto-Garbage" which is superimposed over the human-vetted labels. 
\label{fig:24lwd}}
\end{figure*}

\newpage
\newpage
\section{On features and metrics, bias and expertise}
\label{sec:imkeepingitstephen}
As neural networks and transformers have gained in popularity, a commonly used justification for choosing these methods is that they do not require feature extraction.
The details of why this is an advantage are most often not discussed and it has become common to view not needing to extract or engineer features as an automatic advantage.

A first argument for the use of ``feature-free" models is that they allow us to remove a step (or steps) of data processing that can be computationally expensive.
What is then omitted however is that a new form of processing is required to make astronomical data usable to neural networks or transformers. 
To cite a recent example, \cite{bad} studied how visual transformers can be used for photometric classification, stating in their aims that these methods could ``classify photometric light curves without the need for feature extraction or multi-band preprocessing".
On further inspection the use of visual transformers requires some significant pre-processing steps such as turning multi-band light-curves into images, this is a done via matplotlib in their implementation.
Lightcurve feature extraction as done in, for example, the VRA or in BTSbot \citep{btsbot} is much more lightweight than the generation of images of lightcurves to be parsed to a visual transformer, especially in the context of a large data stream such as LSST.
Feature-based models can therefore have a lesser data-processing burden than ``feature-free" methods, whether they be bespoke neural-networks or pre-trained large scale models\footnote{sometimes referred to as ``foundational models"}.

Another argument encountered is that feature extraction is a biased formed of pre-processing, or that an algorithm which handles raw data ``extracts by itself the best feature representation for a given problem" \citep{pasquet2019}. 
In a context where it would be untractable for a human to extract meaningful features that can be interpreted by an algorithm, either due to the level of abstraction required or the large quantity and diversity of data  (e.g. computer vision), this statement holds true.
But sometimes the underlying sentiment is that ``raw" data is unbiased (or less biased), and that processing performed by a human inevitably taints these data with unwanted bias.
We will take it as granted that the reader agrees that no data is unbiased, and focus on the discussion surrounding human intervention in data processing and to what degree ``human bias" is an issue. 

Here we understand the term bias to mean a representation of data that is unrepresentative of the characteristics of interest. 
\textbf{There are three key areas where bias can be introduced: data collection, data abstraction/pre-processing, metrics choice/interpretation.} 

An example of biased data collection can be found for example in a kilonova transient classifer reported to have 95\% precision \citep{liang2023} but whose training set contains ``contaminant" transients (type Ia, Ib/c, II, SLSN-I) which omit the main contaminants we can expect in a real life setting (shock-break out of CCSNe and cataclysmic variables). 
An example of biased data abstraction is the use of inadequate statistics, such as using the mean of a sewed distribution which would be biased by the tails (a common everyday example is household income: in 2022 the mean UK income was 39,328 whereas the median was 32,349 \citealt{ukincome}).
Finally an example of poor metric choice can be taken from a medical imaging methodology paper where the authors showed that CNNs trained to detect tumours were biased against finding small tumours (the ultimate goal being early detection) because their performance metric was based on the number of cancerous pixels detected in each image \citep{reinke2024}. 

\textit{Removing bias can therefore not be achieved by removing a single step of the development process.}
Feature engineering is a form of data abstraction, even if we assumed that delegating all the data abstraction to the algorithm removes bias from this step (it does not), bias can still be an issue at the data collection level and when creating and evaluating metrics. 
Bias cannot be eliminated but it can be mitigated and disclosed.

We will take this discussion further and posit that wherever bias can be introduced so can \textit{expertise} or domain knowledge. AI for Science professionals can and should take advantage of this. 

At the data collection stage, domain knowledge is required to chose training sets that span the \textit{full set of characteristics} we expect our models to encounter in production. 
A training set having different properties from a production data set is sometimes referred to as ``data-drift". This can occur when live data properties slowly change over time from the initial training, but a discrepancy between training and live data can occur as soon as a model is put into the world if the choice of training data is not informed by expert knowledge of the real-life setting.

At the data abstraction stage, expertise can be imbued in the models through feature engineering and feature choice. 
Even when using methods that can be feature-free, such as CNNs or RNNs, recent successful examples of automation  \citep{btsbot, sheng2024, gupta2025} make use of features or metadata to provide more reliable results. 
These additional features provide context to the images that researchers know are relevant because of their domain knowledge. 
It is also worth noting once again that these features can be computationally inexpensive, such as the maximum magnitude and date at maximum computed by BTSBot and the VRA, or the lightning-bolt method recently proposed to capture the shock break out peak and main peak of some core collapse supernovae \citep{crawford2025}. 
The sole use of raw data when other information is available should be well justified, as it is not obvious that feature-free models are faster and less biased.

Domain knowledge is also essential in designing and choosing metrics that capture the science problems we are addressing. 
In the case of the VRA, which is designed to help rank eyeball lists to look for extra-galactic transients, we assess model performance by measuring the recall at rank k where recall focuses on extra-galactic transients in our sample. 
This provides a metric that is directly informative, compared to an accuracy, recall or precision score, which would tell us very little about the value of the ranking\footnote{It also makes the models \textit{biased} towards the fast recovery of extra-galactic transients, at the expense of galactic transients, but that is a design choice that is known and reported.}.

Finally another area where expertise is essential is in benchmarking and assessing how the models compare to the state-of-the-art, including non machine learning solutions. 
This benchmarking step is not systematically shown in the astro-ML literature, or the comparisons are limited to other machine learning proof-of-concepts rather than the currently used methods.
Currently it is common-place for models to be deemed ``smarter" when they are more complex (e.g. \citealt{crawford2025}), which we suspect is the result of the way new advances in ML and AI have been communicated to the general public\footnote{and since astronomers are not primarily AI professionals we \textit{are} the general public in that context} in the last few years.
The reason we would urge science professional to stay away from these terms is that `\textit{`smart" and ``intelligent" are not descriptive and they are not value neutral.}
``Smarter" is better than ``less smart". 
Yet, more complex models are not necessarily better. 
They have the potential to capture more complex data abstractions, which may be completely irrelevant to a given use-case, leading to the creation of bots that, at best, are unnecessarily complex to understand for future team members or users, and unnecessarily difficult to maintain; at worst, they generalise poorly (overfit) and introduce uninterpretable layers of data processing that future generations of scientists will have to wrestle with.
Additionally, different machine learning methods were developed and specialized for different uses; choosing a model that is best suited to one's given type (and volume) of data is preferable to choosing a model that is more complex but built for a different data type, or used in settings where the amount of data available for training is far superior to the data available in our field. 

Overall, we hope to remind our colleagues that larger machine learning models and larger data sets rarely mitigate the effects of unrepresentative data, poor design and irrelevant metrics; we refer the reader to \cite{huppenkothen2023} for an extended discussion of machine-learning best practices in astronomy.


\bibliography{vra}{}

\begin{thebibliography}{}
\expandafter\ifx\csname natexlab\endcsname\relax\def\natexlab#1{#1}\fi
\providecommand{\url}[1]{\href{#1}{#1}}
\providecommand{\dodoi}[1]{doi:~\href{http://doi.org/#1}{\nolinkurl{#1}}}
\providecommand{\doeprint}[1]{\href{http://ascl.net/#1}{\nolinkurl{http://ascl.net/#1}}}
\providecommand{\doarXiv}[1]{\href{https://arxiv.org/abs/#1}{\nolinkurl{https://arxiv.org/abs/#1}}}

\bibitem[{Alwosheel {et~al.}(2018)Alwosheel, {van Cranenburgh}, \& Chorus}]{ALWOSHEEL2018167}
Alwosheel, A., {van Cranenburgh}, S., \& Chorus, C.~G. 2018, Journal of Choice Modelling, 28, 167, \dodoi{https://doi.org/10.1016/j.jocm.2018.07.002}

\bibitem[{{Bellm} {et~al.}(2019){Bellm}, {Kulkarni}, {Graham}, {Dekany}, {Smith}, {Riddle}, {Masci}, {Helou}, {Prince}, {Adams}, {Barbarino}, {Barlow}, {Bauer}, {Beck}, {Belicki}, {Biswas}, {Blagorodnova}, {Bodewits}, {Bolin}, {Brinnel}, {Brooke}, {Bue}, {Bulla}, {Burruss}, {Cenko}, {Chang}, {Connolly}, {Coughlin}, {Cromer}, {Cunningham}, {De}, {Delacroix}, {Desai}, {Duev}, {Eadie}, {Farnham}, {Feeney}, {Feindt}, {Flynn}, {Franckowiak}, {Frederick}, {Fremling}, {Gal-Yam}, {Gezari}, {Giomi}, {Goldstein}, {Golkhou}, {Goobar}, {Groom}, {Hacopians}, {Hale}, {Henning}, {Ho}, {Hover}, {Howell}, {Hung}, {Huppenkothen}, {Imel}, {Ip}, {Ivezi{\'c}}, {Jackson}, {Jones}, {Juric}, {Kasliwal}, {Kaspi}, {Kaye}, {Kelley}, {Kowalski}, {Kramer}, {Kupfer}, {Landry}, {Laher}, {Lee}, {Lin}, {Lin}, {Lunnan}, {Giomi}, {Mahabal}, {Mao}, {Miller}, {Monkewitz}, {Murphy}, {Ngeow}, {Nordin}, {Nugent}, {Ofek}, {Patterson}, {Penprase}, {Porter}, {Rauch}, {Rebbapragada}, {Reiley}, {Rigault}, {Rodriguez}, {van Roestel}, {Rusholme}, {van
  Santen}, {Schulze}, {Shupe}, {Singer}, {Soumagnac}, {Stein}, {Surace}, {Sollerman}, {Szkody}, {Taddia}, {Terek}, {Van Sistine}, {van Velzen}, {Vestrand}, {Walters}, {Ward}, {Ye}, {Yu}, {Yan}, \& {Zolkower}}]{bellm2019}
{Bellm}, E.~C., {Kulkarni}, S.~R., {Graham}, M.~J., {et~al.} 2019, \pasp, 131, 018002, \dodoi{10.1088/1538-3873/aaecbe}

\bibitem[{Buitinck {et~al.}(2013)Buitinck, Louppe, Blondel, Pedregosa, Mueller, Grisel, Niculae, Prettenhofer, Gramfort, Grobler, Layton, VanderPlas, Joly, Holt, \& Varoquaux}]{sklearn_api}
Buitinck, L., Louppe, G., Blondel, M., {et~al.} 2013, in ECML PKDD Workshop: Languages for Data Mining and Machine Learning, 108--122

\bibitem[{{Cenko} {et~al.}(2013){Cenko}, {Kulkarni}, {Horesh}, {Corsi}, {Fox}, {Carpenter}, {Frail}, {Nugent}, {Perley}, {Gruber}, {Gal-Yam}, {Groot}, {Hallinan}, {Ofek}, {Rau}, {MacLeod}, {Miller}, {Bloom}, {Filippenko}, {Kasliwal}, {Law}, {Morgan}, {Polishook}, {Poznanski}, {Quimby}, {Sesar}, {Shen}, {Silverman}, \& {Sternberg}}]{2013ApJ...769..130C}
{Cenko}, S.~B., {Kulkarni}, S.~R., {Horesh}, A., {et~al.} 2013, \apj, 769, 130, \dodoi{10.1088/0004-637X/769/2/130}

\bibitem[{{Chambers} {et~al.}(2016){Chambers}, {Magnier}, {Metcalfe}, {Flewelling}, {Huber}, {Waters}, {Denneau}, {Draper}, {Farrow}, {Finkbeiner}, {Holmberg}, {Koppenhoefer}, {Price}, {Saglia}, {Schlafly}, {Smartt}, {Sweeney}, {Wainscoat}, {Burgett}, {Grav}, {Heasley}, {Hodapp}, {Jedicke}, {Kaiser}, {Kudritzki}, {Luppino}, {Lupton}, {Monet}, {Morgan}, {Onaka}, {Stubbs}, {Tonry}, {Banados}, {Bell}, {Bender}, {Bernard}, {Botticella}, {Casertano}, {Chastel}, {Chen}, {Chen}, {Cole}, {Deacon}, {Frenk}, {Fitzsimmons}, {Gezari}, {Goessl}, {Goggia}, {Goldman}, {Grebel}, {Hambly}, {Hasinger}, {Heavens}, {Heckman}, {Henderson}, {Henning}, {Holman}, {Hopp}, {Ip}, {Isani}, {Keyes}, {Koekemoer}, {Kotak}, {Long}, {Lucey}, {Liu}, {Martin}, {McLean}, {Morganson}, {Murphy}, {Nieto-Santisteban}, {Norberg}, {Peacock}, {Pier}, {Postman}, {Primak}, {Rae}, {Rest}, {Riess}, {Riffeser}, {Rix}, {Roser}, {Schilbach}, {Schultz}, {Scolnic}, {Szalay}, {Seitz}, {Shiao}, {Small}, {Smith}, {Soderblom}, {Taylor}, {Thakar}, {Thiel},
  {Thilker}, {Urata}, {Valenti}, {Walter}, {Watters}, {Werner}, {White}, {Wood-Vasey}, \& {Wyse}}]{Chambers2016}
{Chambers}, K.~C., {Magnier}, E.~A., {Metcalfe}, N., {et~al.} 2016, ArXiv e-prints.
\newblock \doarXiv{1612.05560}

\bibitem[{{Crawford} {et~al.}(2025){Crawford}, {Pritchard}, {Modjaz}, {Pellegrino}, {Kumar}, \& {Baer-Way}}]{crawford2025}
{Crawford}, A., {Pritchard}, T.~A., {Modjaz}, M., {et~al.} 2025, arXiv e-prints, arXiv:2503.03735, \dodoi{10.48550/arXiv.2503.03735}

\bibitem[{{Dom{\'\i}nguez S{\'a}nchez} {et~al.}(2019){Dom{\'\i}nguez S{\'a}nchez}, {Huertas-Company}, {Bernardi}, {Kaviraj}, {Fischer}, {Abbott}, {Abdalla}, {Annis}, {Avila}, {Brooks}, {Buckley-Geer}, {Carnero Rosell}, {Carrasco Kind}, {Carretero}, {Cunha}, {D'Andrea}, {da Costa}, {Davis}, {De Vicente}, {Doel}, {Evrard}, {Fosalba}, {Frieman}, {Garc{\'\i}a-Bellido}, {Gaztanaga}, {Gerdes}, {Gruen}, {Gruendl}, {Gschwend}, {Gutierrez}, {Hartley}, {Hollowood}, {Honscheid}, {Hoyle}, {James}, {Kuehn}, {Kuropatkin}, {Lahav}, {Maia}, {March}, {Melchior}, {Menanteau}, {Miquel}, {Nord}, {Plazas}, {Sanchez}, {Scarpine}, {Schindler}, {Schubnell}, {Smith}, {Smith}, {Soares-Santos}, {Sobreira}, {Suchyta}, {Swanson}, {Tarle}, {Thomas}, {Walker}, \& {Zuntz}}]{domingezSanchez2019}
{Dom{\'\i}nguez S{\'a}nchez}, H., {Huertas-Company}, M., {Bernardi}, M., {et~al.} 2019, \mnras, 484, 93, \dodoi{10.1093/mnras/sty3497}

\bibitem[{{Dyer} {et~al.}(2024){Dyer}, {Ackley}, {Jim{\'e}nez-Ibarra}, {Lyman}, {Ulaczyk}, {Steeghs}, {Galloway}, {Dhillon}, {O'Brien}, {Ramsay}, {Noysena}, {Kotak}, {Breton}, {Nuttall}, {Pall{\'e}}, {Pollacco}, {Killestein}, {Kumar}, {O'Neill}, {Kelsey}, {Godson}, \& {Jarvis}}]{dyer2024}
{Dyer}, M.~J., {Ackley}, K., {Jim{\'e}nez-Ibarra}, F., {et~al.} 2024, in Society of Photo-Optical Instrumentation Engineers (SPIE) Conference Series, Vol. 13094, Ground-based and Airborne Telescopes X, ed. H.~K. {Marshall}, J.~{Spyromilio}, \& T.~{Usuda}, 130941X, \dodoi{10.1117/12.3018305}

\bibitem[{{Erasmus} {et~al.}(2024{\natexlab{a}}){Erasmus}, {Steele}, {Piascik}, {Bates}, {Mottram}, {Rosie}, {van Gend}, {Geen}, {Pretorius}, {Potter}, {Loubser}, {Koorts}, {Gajjar}, {Titus}, {Worters}, {Sickafoose}, {Chandra}, {O'Connor}, {Matlala}, {Crook-Mansour}, {Ranjbar}, {Smith}, {Jermak}, {Abiodun}, \& {Egbo}}]{erasmus2024}
{Erasmus}, N., {Steele}, I.~A., {Piascik}, A.~S., {et~al.} 2024{\natexlab{a}}, Journal of Astronomical Telescopes, Instruments, and Systems, 10, 025005, \dodoi{10.1117/1.JATIS.10.2.025005}

\bibitem[{{Erasmus} {et~al.}(2024{\natexlab{b}}){Erasmus}, {Potter}, {van Gend}, {Loubser}, {Rosie}, {Titus}, {Chandra}, {Worters}, {Gajjar}, {Hlakola}, \& {Julie}}]{erasmus2024spie}
{Erasmus}, N., {Potter}, S.~B., {van Gend}, C. H.~D.~R., {et~al.} 2024{\natexlab{b}}, in Society of Photo-Optical Instrumentation Engineers (SPIE) Conference Series, Vol. 13096, Ground-based and Airborne Instrumentation for Astronomy X, ed. J.~J. {Bryant}, K.~{Motohara}, \& J.~R.~D. {Vernet}, 130968K, \dodoi{10.1117/12.3015250}

\bibitem[{Friedman(2001)}]{friedman2001}
Friedman, J.~H. 2001, The Annals of Statistics, 29, 1189 , \dodoi{10.1214/aos/1013203451}

\bibitem[{{Gagliano} {et~al.}(2023){Gagliano}, {Contardo}, {Foreman-Mackey}, {Malz}, \& {Aleo}}]{gagliano2023}
{Gagliano}, A., {Contardo}, G., {Foreman-Mackey}, D., {Malz}, A.~I., \& {Aleo}, P.~D. 2023, \apj, 954, 6, \dodoi{10.3847/1538-4357/ace326}

\bibitem[{{Gal-Yam}(2019)}]{galyam2019}
{Gal-Yam}, A. 2019, \araa, 57, 305, \dodoi{10.1146/annurev-astro-081817-051819}

\bibitem[{{Gal-Yam}(2021)}]{tns2021}
{Gal-Yam}, A. 2021, in American Astronomical Society Meeting Abstracts, Vol. 237, American Astronomical Society Meeting Abstracts, 423.05

\bibitem[{{Gezari}(2021)}]{gezari2021}
{Gezari}, S. 2021, \araa, 59, 21, \dodoi{10.1146/annurev-astro-111720-030029}

\bibitem[{{Gillanders} {et~al.}(2024){Gillanders}, {Rhodes}, {Srivastav}, {Carotenuto}, {Bright}, {Huber}, {Stevance}, {Smartt}, {Chambers}, {Chen}, {Fender}, {Andersson}, {Cooper}, {Jonker}, {Cowie}, {de Boer}, {Erasmus}, {Fulton}, {Gao}, {Herman}, {Lin}, {Lowe}, {Magnier}, {Miao}, {Minguez}, {Moore}, {Ngeow}, {Nicholl}, {Pan}, {Pignata}, {Rest}, {Sheng}, {Smith}, {Smith}, {Tonry}, {Wainscoat}, {Weston}, {Yang}, \& {Young}}]{gillanders2024}
{Gillanders}, J.~H., {Rhodes}, L., {Srivastav}, S., {et~al.} 2024, \apjl, 969, L14, \dodoi{10.3847/2041-8213/ad55cd}

\bibitem[{{Green}(2018)}]{green2018}
{Green}, G. 2018, The Journal of Open Source Software, 3, 695, \dodoi{10.21105/joss.00695}

\bibitem[{{Groot} {et~al.}(2024){Groot}, {Bloemen}, {Vreeswijk}, {van Roestel}, {Jonker}, {Nelemans}, {Klein-Wolt}, {Lepoole}, {Pieterse}, {Rodenhuis}, {Boland}, {Haverkorn}, {Aerts}, {Bakker}, {Balster}, {Bekema}, {Dijkstra}, {Dolron}, {Elswijk}, {van Elteren}, {Engels}, {Fokker}, {de Haan}, {Hahn}, {ter Horst}, {Lesman}, {Kragt}, {Morren}, {Nillissen}, {Pessemier}, {Raskin}, {de Rijke}, {Scheers}, {Schuil}, {Timmer}, {Antunes Amaral}, {Arancibia-Rojas}, {Arcavi}, {Blagorodnova}, {Biswas}, {Breton}, {Dawson}, {Dayal}, {De Wet}, {Duffy}, {Faris}, {Fausnaugh}, {Gal-Yam}, {Geier}, {Horesh}, {Johnston}, {Katusiime}, {Kelley}, {Kosakowski}, {Kupfer}, {Leloudas}, {Levan}, {Modiano}, {Mogawana}, {Munday}, {Paice}, {Patat}, {Pelisoli}, {Ramsay}, {Ranaivomanana}, {Ruiz-Carmona}, {Schaffenroth}, {Scaringi}, {Stoppa}, {Street}, {Tranin}, {Uzundag}, {Valenti}, {Veresvarska}, {Vuc̆kovi{\'c}}, {Wichern}, {Wijers}, {Wijnands}, \& {Zimmerman}}]{blackgem}
{Groot}, P.~J., {Bloemen}, S., {Vreeswijk}, P.~M., {et~al.} 2024, \pasp, 136, 115003, \dodoi{10.1088/1538-3873/ad8b6a}

\bibitem[{{Gupta} \& {Muthukrishna}(2025)}]{gupta2025}
{Gupta}, R., \& {Muthukrishna}, D. 2025, arXiv e-prints, arXiv:2502.18558, \dodoi{10.48550/arXiv.2502.18558}

\bibitem[{Harris {et~al.}(2020)Harris, Millman, van~der Walt, Gommers, Virtanen, Cournapeau, Wieser, Taylor, Berg, Smith, Kern, Picus, Hoyer, van Kerkwijk, Brett, Haldane, del R{\'{i}}o, Wiebe, Peterson, G{\'{e}}rard-Marchant, Sheppard, Reddy, Weckesser, Abbasi, Gohlke, \& Oliphant}]{harris2020array}
Harris, C.~R., Millman, K.~J., van~der Walt, S.~J., {et~al.} 2020, Nature, 585, 357, \dodoi{10.1038/s41586-020-2649-2}

\bibitem[{{Heinze} {et~al.}(2021){Heinze}, {Denneau}, {Tonry}, {Smartt}, {Erasmus}, {Fitzsimmons}, {Robinson}, {Weiland}, {Flewelling}, {Stalder}, {Rest}, \& {Young}}]{Heinze2021}
{Heinze}, A.~N., {Denneau}, L., {Tonry}, J.~L., {et~al.} 2021, \psj, 2, 12, \dodoi{10.3847/PSJ/abd325}

\bibitem[{Hunter(2007)}]{Hunter:2007}
Hunter, J.~D. 2007, Computing in Science \& Engineering, 9, 90, \dodoi{10.1109/MCSE.2007.55}

\bibitem[{{Huppenkothen} {et~al.}(2023){Huppenkothen}, {Ntampaka}, {Ho}, {Fouesneau}, {Nord}, {Peek}, {Walmsley}, {Wu}, {Avestruz}, {Buck}, {Brescia}, {Finkbeiner}, {Goulding}, {Kacprzak}, {Melchior}, {Pasquato}, {Ramachandra}, {Ting}, {van de Ven}, {Villar}, {Villar}, \& {Zinger}}]{huppenkothen2023}
{Huppenkothen}, D., {Ntampaka}, M., {Ho}, M., {et~al.} 2023, arXiv e-prints, arXiv:2310.12528, \dodoi{10.48550/arXiv.2310.12528}

\bibitem[{{Kaiser} {et~al.}(2002){Kaiser}, {Aussel}, {Burke}, {Boesgaard}, {Chambers}, {Chun}, {Heasley}, {Hodapp}, {Hunt}, {Jedicke}, {Jewitt}, {Kudritzki}, {Luppino}, {Maberry}, {Magnier}, {Monet}, {Onaka}, {Pickles}, {Rhoads}, {Simon}, {Szalay}, {Szapudi}, {Tholen}, {Tonry}, {Waterson}, \& {Wick}}]{keiser2002}
{Kaiser}, N., {Aussel}, H., {Burke}, B.~E., {et~al.} 2002, in Society of Photo-Optical Instrumentation Engineers (SPIE) Conference Series, Vol. 4836, Survey and Other Telescope Technologies and Discoveries, ed. J.~A. {Tyson} \& S.~{Wolff}, 154--164, \dodoi{10.1117/12.457365}

\bibitem[{Ke {et~al.}(2017)Ke, Meng, Finley, Wang, Chen, Ma, Ye, \& Liu}]{lightGMB}
Ke, G., Meng, Q., Finley, T., {et~al.} 2017, in Advances in Neural Information Processing Systems, ed. I.~Guyon, U.~V. Luxburg, S.~Bengio, H.~Wallach, R.~Fergus, S.~Vishwanathan, \& R.~Garnett, Vol.~30 (Curran Associates, Inc.).
\newblock \url{https://proceedings.neurips.cc/paper_files/paper/2017/file/6449f44a102fde848669bdd9eb6b76fa-Paper.pdf}

\bibitem[{{Killestein} {et~al.}(2021){Killestein}, {Lyman}, {Steeghs}, {Ackley}, {Dyer}, {Ulaczyk}, {Cutter}, {Mong}, {Galloway}, {Dhillon}, {O'Brien}, {Ramsay}, {Poshyachinda}, {Kotak}, {Breton}, {Nuttall}, {Pall{\'e}}, {Pollacco}, {Thrane}, {Aukkaravittayapun}, {Awiphan}, {Burhanudin}, {Chote}, {Chrimes}, {Daw}, {Duffy}, {Eyles-Ferris}, {Gompertz}, {Heikkil{\"a}}, {Irawati}, {Kennedy}, {Levan}, {Littlefair}, {Makrygianni}, {Mata S{\'a}nchez}, {Mattila}, {Maund}, {McCormac}, {Mkrtichian}, {Mullaney}, {Rol}, {Sawangwit}, {Stanway}, {Starling}, {Str{\o}m}, {Tooke}, {Wiersema}, \& {Williams}}]{killenstein2021}
{Killestein}, T.~L., {Lyman}, J., {Steeghs}, D., {et~al.} 2021, \mnras, 503, 4838, \dodoi{10.1093/mnras/stab633}

\bibitem[{{Killestein} {et~al.}(2024){Killestein}, {Kelsey}, {Wickens}, {Nuttall}, {Lyman}, {Krawczyk}, {Ackley}, {Dyer}, {Jim{\'e}nez-Ibarra}, {Ulaczyk}, {O'Neill}, {Kumar}, {Steeghs}, {Galloway}, {Dhillon}, {O'Brien}, {Ramsay}, {Noysena}, {Kotak}, {Breton}, {Pall{\'e}}, {Pollacco}, {Awiphan}, {Belkin}, {Chote}, {Clark}, {Coppejans}, {Duffy}, {Eyles-Ferris}, {Godson}, {Gompertz}, {Graur}, {Irawati}, {Jarvis}, {Julakanti}, {Kennedy}, {Kuncarayakti}, {Levan}, {Littlefair}, {Magee}, {Mandhai}, {Mata S{\'a}nchez}, {Mattila}, {McCormac}, {Mullaney}, {Munday}, {Patel}, {Pursiainen}, {Rana}, {Sawangwit}, {Stanway}, {Starling}, {Warwick}, \& {Wiersema}}]{killenstein2024}
{Killestein}, T.~L., {Kelsey}, L., {Wickens}, E., {et~al.} 2024, \mnras, 533, 2113, \dodoi{10.1093/mnras/stae1817}

\bibitem[{{Law} {et~al.}(2009){Law}, {Kulkarni}, {Dekany}, {Ofek}, {Quimby}, {Nugent}, {Surace}, {Grillmair}, {Bloom}, {Kasliwal}, {Bildsten}, {Brown}, {Cenko}, {Ciardi}, {Croner}, {Djorgovski}, {van Eyken}, {Filippenko}, {Fox}, {Gal-Yam}, {Hale}, {Hamam}, {Helou}, {Henning}, {Howell}, {Jacobsen}, {Laher}, {Mattingly}, {McKenna}, {Pickles}, {Poznanski}, {Rahmer}, {Rau}, {Rosing}, {Shara}, {Smith}, {Starr}, {Sullivan}, {Velur}, {Walters}, \& {Zolkower}}]{law2009}
{Law}, N.~M., {Kulkarni}, S.~R., {Dekany}, R.~G., {et~al.} 2009, \pasp, 121, 1395, \dodoi{10.1086/648598}

\bibitem[{{Liang} {et~al.}(2023){Liang}, {Liu}, {Lei}, \& {Zhao}}]{liang2023}
{Liang}, R., {Liu}, Z., {Lei}, L., \& {Zhao}, W. 2023, Universe, 10, 10, \dodoi{10.3390/universe10010010}

\bibitem[{{McElfresh} {et~al.}(2023){McElfresh}, {Khandagale}, {Valverde}, {Prasad C}, {Feuer}, {Hegde}, {Ramakrishnan}, {Goldblum}, \& {White}}]{2023arXiv230502997M}
{McElfresh}, D., {Khandagale}, S., {Valverde}, J., {et~al.} 2023, arXiv e-prints, arXiv:2305.02997, \dodoi{10.48550/arXiv.2305.02997}

\bibitem[{{M{\"o}ller} {et~al.}(2021){M{\"o}ller}, {Peloton}, {Ishida}, {Arnault}, {Bachelet}, {Blaineau}, {Boutigny}, {Chauhan}, {Gangler}, {Hernandez}, {Hrivnac}, {Leoni}, {Leroy}, {Moniez}, {Pateyron}, {Ramparison}, {Turpin}, {Ansari}, {Allam}, {Bajat}, {Biswas}, {Boucaud}, {Bregeon}, {Campagne}, {Cohen-Tanugi}, {Coleiro}, {Dornic}, {Fouchez}, {Godet}, {Gris}, {Karpov}, {Nebot Gomez-Moran}, {Neveu}, {Plaszczynski}, {Savchenko}, \& {Webb}}]{fink2021}
{M{\"o}ller}, A., {Peloton}, J., {Ishida}, E. E.~O., {et~al.} 2021, \mnras, 501, 3272, \dodoi{10.1093/mnras/staa3602}

\bibitem[{{Moreno-Cartagena} {et~al.}(2025){Moreno-Cartagena}, {Protopapas}, {Cabrera-Vives}, {C{\'a}diz-Leyton}, {Becker}, \& {Donoso-Oliva}}]{bad}
{Moreno-Cartagena}, D., {Protopapas}, P., {Cabrera-Vives}, G., {et~al.} 2025, arXiv e-prints, arXiv:2502.20479, \dodoi{10.48550/arXiv.2502.20479}

\bibitem[{{Muthukrishna} {et~al.}(2019){Muthukrishna}, {Narayan}, {Mandel}, {Biswas}, \& {Hlo{\v{z}}ek}}]{muthukrishna2019}
{Muthukrishna}, D., {Narayan}, G., {Mandel}, K.~S., {Biswas}, R., \& {Hlo{\v{z}}ek}, R. 2019, \pasp, 131, 118002, \dodoi{10.1088/1538-3873/ab1609}

\bibitem[{ONS-UK(2022)}]{ukincome}
ONS-UK. 2022, {Office of National Statistics}, \url{https://www.ons.gov.uk/peoplepopulationandcommunity/personalandhouseholdfinances/incomeandwealth/bulletins/householddisposableincomeandinequality/financialyearending2022}

\bibitem[{pandas~development team(2020)}]{reback2020pandas}
pandas~development team, T. 2020, pandas-dev/pandas: Pandas, latest,  Zenodo, \dodoi{10.5281/zenodo.3509134}

\bibitem[{{Pasquet} {et~al.}(2019){Pasquet}, {Pasquet}, {Chaumont}, \& {Fouchez}}]{pasquet2019}
{Pasquet}, J., {Pasquet}, J., {Chaumont}, M., \& {Fouchez}, D. 2019, \aap, 627, A21, \dodoi{10.1051/0004-6361/201834473}

\bibitem[{Pedregosa {et~al.}(2011)Pedregosa, Varoquaux, Gramfort, Michel, Thirion, Grisel, Blondel, Prettenhofer, Weiss, Dubourg, Vanderplas, Passos, Cournapeau, Brucher, Perrot, \& Duchesnay}]{scikit-learn}
Pedregosa, F., Varoquaux, G., Gramfort, A., {et~al.} 2011, Journal of Machine Learning Research, 12, 2825

\bibitem[{{Perley} {et~al.}(2020){Perley}, {Fremling}, {Sollerman}, {Miller}, {Dahiwale}, {Sharma}, {Bellm}, {Biswas}, {Brink}, {Bruch}, {De}, {Dekany}, {Drake}, {Duev}, {Filippenko}, {Gal-Yam}, {Goobar}, {Graham}, {Graham}, {Ho}, {Irani}, {Kasliwal}, {Kim}, {Kulkarni}, {Mahabal}, {Masci}, {Modak}, {Neill}, {Nordin}, {Riddle}, {Soumagnac}, {Strotjohann}, {Schulze}, {Taggart}, {Tzanidakis}, {Walters}, \& {Yan}}]{perley2020}
{Perley}, D.~A., {Fremling}, C., {Sollerman}, J., {et~al.} 2020, \apj, 904, 35, \dodoi{10.3847/1538-4357/abbd98}

\bibitem[{{Perley} {et~al.}(2025){Perley}, {Ho}, {Fausnaugh}, {Lamb}, {Kasliwal}, {Ahumada}, {Anand}, {Andreoni}, {Bellm}, {Bhalerao}, {Bolin}, {Brink}, {Burns}, {Cenko}, {Corsi}, {Filippenko}, {Frederiks}, {Goldstein}, {Hamburg}, {Jayaraman}, {Jonker}, {Kool}, {Kulkarni}, {Kumar}, {Laher}, {Levan}, {Lysenko}, {Perley}, {Ricker}, {Riddle}, {Ridnaia}, {Rusholme}, {Smith}, {Svinkin}, {Ulanov}, {Vanderspek}, {Waratkar}, \& {Yao}}]{2025MNRAS.537.2362P}
{Perley}, D.~A., {Ho}, A. Y.~Q., {Fausnaugh}, M., {et~al.} 2025, \mnras, 537, 2362, \dodoi{10.1093/mnras/staf125}

\bibitem[{{Potter} {et~al.}(2024){Potter}, {Erasmus}, {van Gend}, {Chandra}, {Worters}, {Hlakola}, \& {Julie}}]{potter2024}
{Potter}, S.~B., {Erasmus}, N., {van Gend}, C. H.~D.~R., {et~al.} 2024, in Society of Photo-Optical Instrumentation Engineers (SPIE) Conference Series, Vol. 13098, Observatory Operations: Strategies, Processes, and Systems X, ed. C.~R. {Benn}, A.~{Chrysostomou}, \& L.~J. {Storrie-Lombardi}, 130980Y, \dodoi{10.1117/12.3018154}

\bibitem[{{Prentice} {et~al.}(2018){Prentice}, {Maguire}, {Smartt}, {Magee}, {Schady}, {Sim}, {Chen}, {Clark}, {Colin}, {Fulton}, {McBrien}, {O'Neill}, {Smith}, {Ashall}, {Chambers}, {Denneau}, {Flewelling}, {Heinze}, {Holoien}, {Huber}, {Kochanek}, {Mazzali}, {Prieto}, {Rest}, {Shappee}, {Stalder}, {Stanek}, {Stritzinger}, {Thompson}, \& {Tonry}}]{2018ApJ...865L...3P}
{Prentice}, S.~J., {Maguire}, K., {Smartt}, S.~J., {et~al.} 2018, \apjl, 865, L3, \dodoi{10.3847/2041-8213/aadd90}

\bibitem[{{Radhakrishnan Santhakumari} {et~al.}(2024){Radhakrishnan Santhakumari}, {Battaini}, {Di Filippo}, {Di Rosa}, {Cabona}, {Claudi}, {Lessio}, {Dima}, {Young}, {Landoni}, {Colapietro}, {D'Orsi}, {Aliverti}, {Genoni}, {Munari}, {Zanmar Sanchez}, {Vitali}, {Ricci}, {Schipani}, {Campana}, {Achren}, {Araiza-Duran}, {Arcavi}, {Baruffolo}, {Ben-Ami}, {Bitchkovsky}, {Brucalassi}, {Bruch}, {Capasso}, {Cappellaro}, {Cosentino}, {D'Alessio}, {D'Avanzo}, {Della Valle}, {Di Benedetto}, {Gal-Yam}, {Hernandez Diaz}, {Hershko}, {Kotilainen}, {Kuncarayakti}, {Li Causi}, {Marafatto}, {Martinetti}, {Marty}, {Mattila}, {Micciche}, {Nicotra}, {Oggioni}, {Perez Ventura}, {Pariani}, {Pignata}, {Rappaport}, {Riva}, {Rubin}, {Salasnich}, {Savarese}, {Scuderi}, {Smartt}, \& {Stritzinger}}]{soxs}
{Radhakrishnan Santhakumari}, K.~K., {Battaini}, F., {Di Filippo}, S., {et~al.} 2024, arXiv e-prints, arXiv:2407.17288, \dodoi{10.48550/arXiv.2407.17288}

\bibitem[{{Rehemtulla} {et~al.}(2024){Rehemtulla}, {Miller}, {Jegou Du Laz}, {Coughlin}, {Fremling}, {Perley}, {Qin}, {Sollerman}, {Mahabal}, {Laher}, {Riddle}, {Rusholme}, \& {Kulkarni}}]{btsbot}
{Rehemtulla}, N., {Miller}, A.~A., {Jegou Du Laz}, T., {et~al.} 2024, \apj, 972, 7, \dodoi{10.3847/1538-4357/ad5666}

\bibitem[{{Reinke} \& et~al.(2024)}]{reinke2024}
{Reinke}, A., \& et~al. 2024, Nature Methods, 21, \dodoi{doi:10.1038/s41592-023-02150-0}

\bibitem[{{Schlegel} {et~al.}(1998){Schlegel}, {Finkbeiner}, \& {Davis}}]{SFD98}
{Schlegel}, D.~J., {Finkbeiner}, D.~P., \& {Davis}, M. 1998, \apj, 500, 525, \dodoi{10.1086/305772}

\bibitem[{{Shappee} {et~al.}(2014){Shappee}, {Prieto}, {Grupe}, {Kochanek}, {Stanek}, {De Rosa}, {Mathur}, {Zu}, {Peterson}, {Pogge}, {Komossa}, {Im}, {Jencson}, {Holoien}, {Basu}, {Beacom}, {Szczygie{\l}}, {Brimacombe}, {Adams}, {Campillay}, {Choi}, {Contreras}, {Dietrich}, {Dubberley}, {Elphick}, {Foale}, {Giustini}, {Gonzalez}, {Hawkins}, {Howell}, {Hsiao}, {Koss}, {Leighly}, {Morrell}, {Mudd}, {Mullins}, {Nugent}, {Parrent}, {Phillips}, {Pojmanski}, {Rosing}, {Ross}, {Sand}, {Terndrup}, {Valenti}, {Walker}, \& {Yoon}}]{shappee2014}
{Shappee}, B.~J., {Prieto}, J.~L., {Grupe}, D., {et~al.} 2014, \apj, 788, 48, \dodoi{10.1088/0004-637X/788/1/48}

\bibitem[{{Sheng} {et~al.}(2024){Sheng}, {Nicholl}, {Smith}, {Young}, {Williams}, {Stevance}, {Smartt}, {Srivastav}, \& {Moore}}]{sheng2024}
{Sheng}, X., {Nicholl}, M., {Smith}, K.~W., {et~al.} 2024, \mnras, 531, 2474, \dodoi{10.1093/mnras/stae1253}

\bibitem[{{Smartt} {et~al.}(2015){Smartt}, {Valenti}, {Fraser}, {Inserra}, {Young}, {Sullivan}, {Pastorello}, {Benetti}, {Gal-Yam}, {Knapic}, {Molinaro}, {Smareglia}, {Smith}, {Taubenberger}, {Yaron}, {Anderson}, {Ashall}, {Balland}, {Baltay}, {Barbarino}, {Bauer}, {Baumont}, {Bersier}, {Blagorodnova}, {Bongard}, {Botticella}, {Bufano}, {Bulla}, {Cappellaro}, {Campbell}, {Cellier-Holzem}, {Chen}, {Childress}, {Clocchiatti}, {Contreras}, {Dall'Ora}, {Danziger}, {de Jaeger}, {De Cia}, {Della Valle}, {Dennefeld}, {Elias-Rosa}, {Elman}, {Feindt}, {Fleury}, {Gall}, {Gonzalez-Gaitan}, {Galbany}, {Morales Garoffolo}, {Greggio}, {Guillou}, {Hachinger}, {Hadjiyska}, {Hage}, {Hillebrandt}, {Hodgkin}, {Hsiao}, {James}, {Jerkstrand}, {Kangas}, {Kankare}, {Kotak}, {Kromer}, {Kuncarayakti}, {Leloudas}, {Lundqvist}, {Lyman}, {Hook}, {Maguire}, {Manulis}, {Margheim}, {Mattila}, {Maund}, {Mazzali}, {McCrum}, {McKinnon}, {Moreno-Raya}, {Nicholl}, {Nugent}, {Pain}, {Pignata}, {Phillips}, {Polshaw}, {Pumo}, {Rabinowitz},
  {Reilly}, {Romero-Ca{\~n}izales}, {Scalzo}, {Schmidt}, {Schulze}, {Sim}, {Sollerman}, {Taddia}, {Tartaglia}, {Terreran}, {Tomasella}, {Turatto}, {Walker}, {Walton}, {Wyrzykowski}, {Yuan}, \& {Zampieri}}]{smartt2015}
{Smartt}, S.~J., {Valenti}, S., {Fraser}, M., {et~al.} 2015, \aap, 579, A40, \dodoi{10.1051/0004-6361/201425237}

\bibitem[{{Smartt} {et~al.}(2018){Smartt}, {Clark}, {Smith}, {McBrien}, {Maguire}, {O'Neil}, {Fulton}, {Magee}, {Prentice}, {Colin}, {Tonry}, {Denneau}, {Stalder}, {Heinze}, {Weiland}, {Flewelling}, \& {Rest}}]{2018ATel11727....1S}
{Smartt}, S.~J., {Clark}, P., {Smith}, K.~W., {et~al.} 2018, The Astronomer's Telegram, 11727, 1

\bibitem[{{Smith} {et~al.}(2020){Smith}, {Smartt}, {Young}, {Tonry}, {Denneau}, {Flewelling}, {Heinze}, {Weiland}, {Stalder}, {Rest}, {Stubbs}, {Anderson}, {Chen}, {Clark}, {Do}, {F{\"o}rster}, {Fulton}, {Gillanders}, {McBrien}, {O'Neill}, {Srivastav}, \& {Wright}}]{smith2020}
{Smith}, K.~W., {Smartt}, S.~J., {Young}, D.~R., {et~al.} 2020, \pasp, 132, 085002, \dodoi{10.1088/1538-3873/ab936e}

\bibitem[{{Stalder} {et~al.}(2017){Stalder}, {Tonry}, {Smartt}, {Coughlin}, {Chambers}, {Stubbs}, {Chen}, {Kankare}, {Smith}, {Denneau}, {Sherstyuk}, {Heinze}, {Weiland}, {Rest}, {Young}, {Huber}, {Flewelling}, {Lowe}, {Magnier}, {Schultz}, {Waters}, {Wainscoat}, {Willman}, {Wright}, {Chu}, {Sanders}, {Inserra}, {Maguire}, \& {Kotak}}]{2017ApJ...850..149S}
{Stalder}, B., {Tonry}, J., {Smartt}, S.~J., {et~al.} 2017, \apj, 850, 149, \dodoi{10.3847/1538-4357/aa95c1}

\bibitem[{{Steeghs} {et~al.}(2022){Steeghs}, {Galloway}, {Ackley}, {Dyer}, {Lyman}, {Ulaczyk}, {Cutter}, {Mong}, {Dhillon}, {O'Brien}, {Ramsay}, {Poshyachinda}, {Kotak}, {Nuttall}, {Pall{\'e}}, {Breton}, {Pollacco}, {Thrane}, {Aukkaravittayapun}, {Awiphan}, {Burhanudin}, {Chote}, {Chrimes}, {Daw}, {Duffy}, {Eyles-Ferris}, {Gompertz}, {Heikkil{\"a}}, {Irawati}, {Kennedy}, {Killestein}, {Kuncarayakti}, {Levan}, {Littlefair}, {Makrygianni}, {Marsh}, {Mata-Sanchez}, {Mattila}, {Maund}, {McCormac}, {Mkrtichian}, {Mullaney}, {Noysena}, {Patel}, {Rol}, {Sawangwit}, {Stanway}, {Starling}, {Str{\o}m}, {Tooke}, {West}, {White}, \& {Wiersema}}]{steeghs2022}
{Steeghs}, D., {Galloway}, D.~K., {Ackley}, K., {et~al.} 2022, \mnras, 511, 2405, \dodoi{10.1093/mnras/stac013}

\bibitem[{{Steele} {et~al.}(2004){Steele}, {Smith}, {Rees}, {Baker}, {Bates}, {Bode}, {Bowman}, {Carter}, {Etherton}, {Ford}, {Fraser}, {Gomboc}, {Lett}, {Mansfield}, {Marchant}, {Medrano-Cerda}, {Mottram}, {Raback}, {Scott}, {Tomlinson}, \& {Zamanov}}]{steele2004}
{Steele}, I.~A., {Smith}, R.~J., {Rees}, P.~C., {et~al.} 2004, in Society of Photo-Optical Instrumentation Engineers (SPIE) Conference Series, Vol. 5489, Ground-based Telescopes, ed. J.~M. {Oschmann}, Jr., 679--692, \dodoi{10.1117/12.551456}

\bibitem[{Stevance(2025{\natexlab{a}})}]{stevance_2025_14944209}
Stevance, H. 2025{\natexlab{a}}, ATLAS VRA Technical Manual,  Zenodo, \dodoi{10.5281/zenodo.14944208}

\bibitem[{Stevance(2025{\natexlab{b}})}]{stevance_2025_14906192}
---. 2025{\natexlab{b}}, ATLAS VRA v1 - Training Data and Code,  Zenodo, \dodoi{10.5281/zenodo.15195392}

\bibitem[{Stevance \& Smith(2025)}]{heloise_2025_14983116}
Stevance, H., \& Smith, K. 2025, HeloiseS/atlasvras: VRA 1.1, v1.1,  Zenodo, \dodoi{10.5281/zenodo.14363396}

\bibitem[{{Stevance} {et~al.}(2025){Stevance}, {Leland}, \& {Smith}}]{atlaspapiclient}
{Stevance}, H.~F., {Leland}, J., \& {Smith}, K.~W. 2025, arXiv e-prints, arXiv:2506.06403.
\newblock \doarXiv{2506.06403}

\bibitem[{{The PLAsTiCC team} {et~al.}(2018){The PLAsTiCC team}, {Allam}, {Bahmanyar}, {Biswas}, {Dai}, {Galbany}, {Hlo{\v{z}}ek}, {Ishida}, {Jha}, {Jones}, {Kessler}, {Lochner}, {Mahabal}, {Malz}, {Mandel}, {Mart{\'\i}nez-Galarza}, {McEwen}, {Muthukrishna}, {Narayan}, {Peiris}, {Peters}, {Ponder}, {Setzer}, {The LSST Dark Energy Science Collaboration}, {LSST Transients}, \& {Variable Stars Science Collaboration}}]{plasticc}
{The PLAsTiCC team}, {Allam}, Jr., T., {Bahmanyar}, A., {et~al.} 2018, arXiv e-prints, arXiv:1810.00001, \dodoi{10.48550/arXiv.1810.00001}

\bibitem[{{Tonry} {et~al.}(2018){Tonry}, {Denneau}, {Heinze}, {Stalder}, {Smith}, {Smartt}, {Stubbs}, {Weiland}, \& {Rest}}]{tonry2018}
{Tonry}, J.~L., {Denneau}, L., {Heinze}, A.~N., {et~al.} 2018, \pasp, 130, 064505, \dodoi{10.1088/1538-3873/aabadf}

\bibitem[{{Vernet} {et~al.}(2011){Vernet}, {Dekker}, {D'Odorico}, {Kaper}, {Kjaergaard}, {Hammer}, {Randich}, {Zerbi}, {Groot}, {Hjorth}, {Guinouard}, {Navarro}, {Adolfse}, {Albers}, {Amans}, {Andersen}, {Andersen}, {Binetruy}, {Bristow}, {Castillo}, {Chemla}, {Christensen}, {Conconi}, {Conzelmann}, {Dam}, {de Caprio}, {de Ugarte Postigo}, {Delabre}, {di Marcantonio}, {Downing}, {Elswijk}, {Finger}, {Fischer}, {Flores}, {Fran{\c{c}}ois}, {Goldoni}, {Guglielmi}, {Haigron}, {Hanenburg}, {Hendriks}, {Horrobin}, {Horville}, {Jessen}, {Kerber}, {Kern}, {Kiekebusch}, {Kleszcz}, {Klougart}, {Kragt}, {Larsen}, {Lizon}, {Lucuix}, {Mainieri}, {Manuputy}, {Martayan}, {Mason}, {Mazzoleni}, {Michaelsen}, {Modigliani}, {Moehler}, {M{\o}ller}, {Norup S{\o}rensen}, {N{\o}rregaard}, {P{\'e}roux}, {Patat}, {Pena}, {Pragt}, {Reinero}, {Rigal}, {Riva}, {Roelfsema}, {Royer}, {Sacco}, {Santin}, {Schoenmaker}, {Spano}, {Sweers}, {Ter Horst}, {Tintori}, {Tromp}, {van Dael}, {van der Vliet}, {Venema}, {Vidali}, {Vinther}, {Vola},
  {Winters}, {Wistisen}, {Wulterkens}, \& {Zacchei}}]{vernet2011}
{Vernet}, J., {Dekker}, H., {D'Odorico}, S., {et~al.} 2011, \aap, 536, A105, \dodoi{10.1051/0004-6361/201117752}

\bibitem[{{W}es {M}c{K}inney(2010)}]{mckinney-proc-scipy-2010}
{W}es {M}c{K}inney. 2010, in {P}roceedings of the 9th {P}ython in {S}cience {C}onference, ed. {S}t\'efan van~der {W}alt \& {J}arrod {M}illman, 56 -- 61, \dodoi{10.25080/Majora-92bf1922-00a}

\bibitem[{{Weston} {et~al.}(2024){Weston}, {Smith}, {Smartt}, {Tonry}, \& {Stevance}}]{weston2024}
{Weston}, J.~G., {Smith}, K.~W., {Smartt}, S.~J., {Tonry}, J.~L., \& {Stevance}, H.~F. 2024, RAS Techniques and Instruments, 3, 385, \dodoi{10.1093/rasti/rzae027}

\bibitem[{{Williams} {et~al.}(2024{\natexlab{a}}){Williams}, {Francis}, {Lawrence}, {Sloan}, {Smartt}, {Smith}, \& {Young}}]{lasair}
{Williams}, R.~D., {Francis}, G.~P., {Lawrence}, A., {et~al.} 2024{\natexlab{a}}, RAS Techniques and Instruments, 3, 362, \dodoi{10.1093/rasti/rzae024}

\bibitem[{{Williams} {et~al.}(2024{\natexlab{b}}){Williams}, {Francis}, {Lawrence}, {Sloan}, {Smartt}, {Smith}, \& {Young}}]{lasair2024}
---. 2024{\natexlab{b}}, RAS Techniques and Instruments, 3, 362, \dodoi{10.1093/rasti/rzae024}

\bibitem[{{Worters} {et~al.}(2016){Worters}, {O'Connor}, {Carter}, {Loubser}, {Fourie}, {Sickafoose}, \& {Swanevelder}}]{worters2016}
{Worters}, H.~L., {O'Connor}, J.~E., {Carter}, D.~B., {et~al.} 2016, in Society of Photo-Optical Instrumentation Engineers (SPIE) Conference Series, Vol. 9908, Ground-based and Airborne Instrumentation for Astronomy VI, ed. C.~J. {Evans}, L.~{Simard}, \& H.~{Takami}, 99083Y, \dodoi{10.1117/12.2231636}

\bibitem[{Young(2023)}]{Young_sherlock_2023}
Young, D.~R. 2023, {Sherlock. Contextual classification of astronomical transient sources}, \dodoi{10.5281/zenodo.8038057}

\end{thebibliography}
\bibliographystyle{aasjournal}



\end{document}